%% file: gpdff.tex
\pdfoutput=1
\documentclass[11pt,twocolumn]{article}

\usepackage[english]{babel}
\usepackage{graphicx}
\usepackage{amsmath}

\usepackage{color}

\usepackage{rotating}

\usepackage{url}

\usepackage[linktocpage]{hyperref}

\allowdisplaybreaks[3]

\newcommand{\ms}{\mskip 1.5mu}

\newcommand{\gev}{\operatorname{GeV}}
\newcommand{\fm}{\operatorname{fm}}

\newcommand{\gsim}{\raisebox{-4pt}{%
    $\,\stackrel{\textstyle >}{\sim}\,$}}

\newcommand{\dd}{\mathrm{d}}

\newcommand{\tvec}[1]{\boldsymbol{#1}}

\newcommand{\vbs}{\boldsymbol{b}{}}

\newcommand{\tab}[1]{table~\protect\ref{#1}}
\newcommand{\tabs}[1]{tables~\protect\ref{#1}}
\newcommand{\fig}[1]{figure~\protect\ref{#1}}
\newcommand{\figs}[1]{figures~\protect\ref{#1}}
\newcommand{\sect}[1]{section~\protect\ref{#1}}
\newcommand{\sects}[1]{sections~\protect\ref{#1}}
\newcommand{\app}[1]{appendix~\protect\ref{#1}}

\begin{document}

\title{%
\raggedleft{\normalsize DESY 13-025} \\[0.5em]
\centering\textbf{\LARGE Nucleon form factors, \\[0.2em]
  generalized parton distributions \\[0.2em]
  and quark angular momentum}} 

\author{Markus Diehl \\[0.4em]
  \textit{\normalsize Deutsches Elektronen-Synchroton DESY, 22603 Hamburg,
    Germany} \\[1em]
Peter Kroll \\[0.4em]
\textit{\normalsize Fachbereich Physik, Universit\"at Wuppertal, 42097
  Wuppertal, Germany} \\[-0.1em]
\textit{\normalsize Institut f\"ur Theoretische Physik, Universit\"at
  Regensburg, 93040 Regensburg, Germany}
}

\date{}

\onecolumn

\maketitle

\begin{center}
\textbf{Abstract}
\end{center}

\parbox{0.9\textwidth}{%
  We extract the individual contributions from $u$ and $d$ quarks to the
  Dirac and Pauli form factors of the proton, after a critical examination
  of the available measurements of electromagnetic nucleon form factors.
  From this data we determine generalized parton distributions for valence
  quarks, assuming a particular form for their functional dependence.  The
  result allows us to study various aspects of nucleon structure in the
  valence region.  In particular, we evaluate Ji's sum rule and estimate
  the total angular momentum carried by valence quarks at the scale $\mu =
  2 \gev$ to be $J_v^u = 0.230^{+ 0.009}_{- 0.024}$ and $J_v^d = -0.004^{+
    0.010}_{- 0.016}$.}

\newpage

\tableofcontents

\newpage

\twocolumn

\input{intro}
\input{basics}
\input{data}
\input{strange}
\input{interpol}
\input{fitgpd}
\input{properties}
\input{applications}

\input{summary}

%%%%%%%%%%%%%%%%%%%%%%%%%%%%%%%%%%%%%%%%%%%%%%%%%%%%%%%%%%%%%%%%%%

\section*{Acknowledgments}

We gratefully thank
John Arrington for discussions about the proton form factors and for
providing numerical values for several of his results,
Mariaelena Boglione for providing numerical values of the extraction
\cite{Anselmino:2012aa} of the Sivers function,
Daniela Rohe for correspondence about the neutron form factor results 
\cite{Becker:1999tw,Golak:2000nt},
and Thorsten Feldmann for his collaboration at an early stage of this
project.

%%%%%%%%%%%%%%%%%%%%%%%%%%%%%%%%%%%%%%%%%%%%%%%%%%%%%%%%%%%%%%%%%%

\appendix

\input{appendices}

%%%%%%%%%%%%%%%%%%%%%%%%%%%%%%%%%%%%%%%%%%%%%%%%%%%%%%%%%%%%%%%%%%

\onecolumn

% following three lines create entry in table of contents
\clearpage
\phantomsection

\end{document}

%% file: intro.tex
\section{Introduction}
\label{sec:intro}

Together with parton distributions, electromagnetic form factors are among
the most important quantities that provide information about the internal
structure of the nucleon.  Their experimental determination has entered
the realm of precision physics.  Generalized parton distributions (GPDs)
combine and enlarge the different types of information contained in
ordinary parton densities (PDFs) and form factors, but they remain much
less well known experimentally.  After pioneering measurements at DESY and
Jefferson Lab, the upcoming energy upgrade at Jefferson Lab will
significantly advance the determination of GPDs in the valence quark
region, whereas measurements at COMPASS will explore the region of sea
quarks and gluons with momentum fractions between $10^{-2}$ and $10^{-1}$.
For reviews of the many facets of GPDs we refer to
\cite{Goeke:2001tz,Diehl:2003ny,Belitsky:2005qn,Boffi:2007yc}.

GPDs can be extracted from hard exclusive processes like deeply virtual
Compton scattering and meson production.  In complement, one can constrain
the GPDs for valence quarks indirectly via the sum rules that connect them
with electromagnetic form factors.  This requires an ansatz for the
functional form of the GPDs and in this sense is intrinsically model
dependent, but on the other hand it can reach values of the invariant
momentum transfer $t$ much larger than what can conceivably be measured in
hard exclusive scattering.  We performed such an indirect determination
some time ago \cite{DFJK4}.  Since then, there have been significant
improvements in the experimental determination of the electromagnetic form
factors, and we find it timely to investigate how this progress impacts on
the extraction of GPDs and of important quantities such as the total
angular momentum carried by quarks in the proton.  This is the purpose of
the present work.

In \sect{sec:basics} we briefly recall some essentials about form factors
and GPDs and introduce our notation.  A critical discussion of the form
factor data used in our analysis is given in \sect{sec:data-sel}, where we
also provide a simple and precise parameterization of the selected data.
In \sect{sec:strange} we estimate the contribution of strange quarks to
the form factors.  Section~\ref{sec:interpol} describes how we combine the
experimental results on proton and neutron form factors in order to
extract Dirac and Pauli form factors for individual quark flavors, which
are most closely connected with GPDs.  Our fit of the GPDs to the form
factor data, including a number of variants that allow us to investigate
systematic uncertainties, is described in \sects{sec:gpd-fits} and
\ref{sec:fit-properties}.  In particular, we evaluate Ji's sum rule and
thus obtain an estimate for the total angular momentum carried by $u$ and
by $d$ quarks minus the corresponding contribution from antiquarks.  Using
our extracted GPDs, we explore in \sect{sec:applications} a number of
further connections, namely the axial form factor, wide-angle Compton
scattering, chromodynamic lensing and GPDs at nonzero skewness.  We
summarize our findings in \sect{sec:sum} and list various numerical
results in two appendices.

%% file: basics.tex
\section{Basics and notation}
\label{sec:basics}

To begin with, let us recall some basics about the electromagnetic form
factors of the nucleon.  Experimental results are typically expressed in
terms of the Sachs form factors $G_M^p(t)$, $G_E^p(t)$ and $G_M^n(t)$,
$G_E^n(t)$, where $t$ is the squared momentum transfer to the proton.
Several measurements determine the ratio of electric and magnetic form
factors, which is commonly written as
\begin{equation}
  \label{GE-GM-ratio}
R^i(t) = \mu_i \ms G_E^i(t) / G_M^i(t)
\end{equation}
with $i=p,n$ for the proton and the neutron.  The magnetic moments $\mu_p$
and $\mu_n$ normalize this ratio to unity at $t=0$.

For convenience the magnetic form factors are often divided by the
conventional dipole form
\begin{equation}
\label{GDipole}
G _{\text{dipole}}^{i}(t) = 
  \frac{\mu_{i}}{\bigl[ 1 - t /(0.71 \gev{\!}^2) \bigr]{}^2_{\phantom{1}}}
\end{equation}
with $i=p,n$.  Plotting this ratio allows one to discern details in the
data over a wide range of $t$, since the ratios $G_M^i
/G_{\text{dipole}}^i$ show only a mild variation, unlike the form factors
themselves.

The Dirac and Pauli form factors, $F_1^i$ and $F_2^i$, are related to the
Sachs form factors by
\begin{align}
  \label{Sachs-Dirac}
G_M^i &= F_1^i + F_2^i \,,
&
G_E^i &= F_1^i + \frac{t}{4 m^2}\, F_2^i \,,
\end{align}
where $m$ is the nucleon mass and again $i=p,n$.  One can further
decompose
\begin{align}
\label{eq:decomposition}
F_i^{p} &= e_u\ms F_i^{u} + e_d\ms F_i^{d} + e_s\ms F_i^{s} \,,
\nonumber\\
F_i^{n} &= e_u\ms F_i^{d} + e_d\ms F_i^{u} + e_s\ms F_i^{s} \,, 
\end{align}
where $F^{q}_i$ denotes the contribution from quark flavor $q$ to the form
factor $F^{p}_i$ of the proton.  Here $i=1,2$ and $e_q$ is the electric
charge of the quark in units of the positron charge.  It is instructive to
rewrite \eqref{eq:decomposition} as
\begin{align}
\label{inverse-dec}
2 F_i^{p} + F_i^n &= F_i^u - F_i^s \,,
\nonumber \\
2 F_i^{n} + F_i^p &= F_i^d - F_i^s \,.
\end{align}
To the extent that the strangeness contributions $F_1^s$ and $F_2^s$ can
be neglected, one can hence reconstruct the form factors for $u$ and $d$
quarks from the electromagnetic form factors alone.  We will return to the
issue of strangeness form factors in \sect{sec:strange}.
For brevity we will refer to the set of $F_i^q$ as ``flavor form factors''
in this work.  We will also use self-explaining abbreviations
\begin{align}
F_i^{u-s} &= F_i^u - F_i^s \,,
&
F_i^{u+d} &= F_i^u + F_i^d
\end{align}
etc.\ for linear combinations of these form factors.

The flavor form factors can be written in terms of GPDs at zero skewness.
For each quark flavor we have the sum rules
\begin{align}
  \label{eq:SumRule}
F_1^q(t) &= \int_{0}^1 \dd x\, H_v^q(x,t) \,,
\nonumber \\
F_2^q(t) &= \int_{0}^1 \dd x\, E_v^q(x,t)
\end{align}
with
\begin{align}
  \label{valence-GPDs}
H_v^q(x,t) &= H^q(x,0,t) + H^q(-x,0,t) \,,
\nonumber \\
E_v^q(x,t) &= E^q(x,0,t) + E^q(-x,0,t) \,,
\end{align}
where $H^q(x,\xi,t)$ and $E^q(x,\xi,t)$ denote the proton GPDs for
unpolarized quarks of flavor $q$ in the standard notation
\cite{Diehl:2003ny}.  The combinations \eqref{valence-GPDs} correspond to
the difference of quarks and antiquarks (as it must be for the
electromagnetic form factors) and in this sense can be called ``valence
GPDs''.  For positive $x$ one recovers the usual quark and antiquark
densities as $H^q(x,0,0) = q(x)$ and $H^q(-x,0,0) = - \bar{q}(x)$.

We will also need the combination
\begin{align}
  \label{valence-Htilde}
\widetilde{H}^q_v(x,t) &= \widetilde{H}^q(x,0,t)
                        - \widetilde{H}^q(-x,0,t)
\end{align}
for the difference of longitudinally polarized quarks and antiquarks, as
well as the antiquark distributions
\begin{align}
  \label{antiquark-GPDs}
H^{\bar{q}}(x,t) &= - H^q(-x,0,t) \,,
\nonumber \\[0.1em]
E^{\bar{q}}(x,t) &= - E^q(-x,0,t) \,,
\nonumber \\
\smash{\widetilde{H}}^{\bar{q}}(x,t) 
  &= \phantom{-} \smash{\widetilde{H}^q(-x,0,t)} \,.
\end{align}
With these definitions, the isovector axial form factor of the nucleon can
be written as
\begin{multline}
F_A(t) = \int_0^1 \dd x\,
 \bigl[ \widetilde{H}^u_v(x,t)-\widetilde{H}^d_v(x,t)\bigr]
\\[0.1em]
+ 2 \int_0^1 \dd x\, \bigl[ \widetilde{H}^{\bar u}(x,t)
                          - \widetilde{H}^{\bar d}(x,t) \bigr] \,.
  \label{eq:FA}
\end{multline}
The sea quark contribution does not drop out in this sum rule since the
axial form factor has positive charge parity and thus corresponds to the
sum and not the difference of quark and antiquark contributions.  The
value of $F_A$ at $t=0$, the axial charge, is well known from
$\beta$-decay experiments.

%% file: data.tex
\section{Data selection}
\label{sec:data-sel}

The determination of the electromagnetic nucleon form factors has not only
a long history but remains at the forefront of experimental research.
With quoted uncertainties typically in the percent region, the consistency
between different measurements and the control of the theory underlying
them have become nontrivial issues, as we shall see.  In this section we
discuss the selection of data used in our subsequent analysis and point
out open problems and discrepancies between data sets.
Earlier overviews and discussions of form factor data can be found in
\cite{Kelly:2002if,Friedrich:2003iz,Gao:2004zi}, and for $G_E^n$ also in
\cite{Plaster:2005cx}.

\begin{table}
\begin{center}
\renewcommand{\arraystretch}{1.2}
\begin{tabular}{ccc}
\hline
        & $-t \, [\gev^2]$ & references \\
\hline
$G_M^p$ & ~0.017 -- 31.2~ & \protect\cite{Arrington:2007ux} \\
\cline{2-3}
$R^p$   & ~0.069 -- 0.138 & \protect\cite{Arrington:2007ux} \\
        & ~0.246 -- 8.49~ & \protect\cite{Milbrath:1997de} --
                            \protect\cite{Zhan:2011ji} \\
\cline{2-3}
$G_M^n$ & ~0.071 -- 0.235 & \protect\cite{Anklin:1998ae,Kubon:2001rj} \\
        & ~~~0.1 -- 4.77~ & \protect\cite{Anklin:1994ae} --
                            \protect\cite{Anderson:2006jp} \\
\cline{2-3}
$R^n$   & ~0.142 -- 3.41~ & \protect\cite{Herberg:1999ud} --
                            \protect\cite{Riordan:2010id} \\
\cline{2-3}
$G_E^n$ & 0.0389 -- 1.644 & \protect\cite{Schiavilla:2001qe} \\
$r^2_{En}$ & 0 & \protect\cite{Beringer:1900zz} \\
\hline
\end{tabular}
\end{center}
\caption{\label{tab:data-overview} Overview of our default data set.  More
  information about the data on $R^p$, $G_M^n$ and $R^n$ is given in
  \tabs{tab:Rp-data}, \ref{tab:GMn-data} and \ref{tab:Rn-data}.}
\end{table}

A synopsis of the default data set that we use in later sections is given
in \tab{tab:data-overview}.  Several of these data do not have separated
statistical and systematic errors.  To have a uniform treatment, we add
those errors in quadrature for the data sets where they are available.

%%%%%%%%%%%%%%%

\subsection{Proton form factors}
\label{sec:proton-data}

One of the main observables for the extraction of the proton form factors
is the unpolarized elastic $ep$ cross section, from which $G_M^p$ and
$G_E^p$ may be obtained by a Rosenbluth separation.  About a decade ago it
has become evident that the effects of two-photon exchange are substantial
in this extraction method \cite{Guichon:2003qm}, especially for $G_E^p$
but at the precision level also for $G_M^p$.  Two-photon exchange must
hence be described accurately, which is nontrivial because it involves the
proton structure in a way that is even more complex than for the
one-photon exchange term from which one wants to extract the form factors.
For a review of this subject we refer to \cite{Arrington:2011dn}.

A second method uses the correlation between the polarizations of the beam
electron and either the proton target or the scattered proton, i.e.\ the
processes $\vec{p} \ms (\vec{e},e'p)$ or $p (\vec{e},e'\vec{p} \, )$.
This is typically found to be less sensitive to two-photon effects.
However, it gives access only to the ratio $R^p$, so that information from
the $ep$ cross section remains indispensable for the separate
determination of $G_M^p$ and $G_E^p$.

There exist several global analyses that combine Rosenbluth separation and
polarization data, plus in some cases the ratio of $e^+p$ and $e^-p$ cross
sections.  Most of them include a parameterization of two-photon exchange
terms, whose parameters are fitted to data.  This approach has been taken
e.g.\ in
\cite{Arrington:2004ae,Alberico:2008sz,Qattan:2011ke,Qattan:2012zf}.  By
contrast, the analysis in \cite{Arrington:2007ux} uses two-photon exchange
terms calculated in dynamical models, including an estimate of their
uncertainties.  A comparative discussion of the different methods can be
found in \cite{Qattan:2011ke}.

For our data set we use the $G_M^p$ results from the analysis
\cite{Arrington:2007ux} of Arrington, Melnitchouk and Tjon (hereafter
referred to as AMT 07).  This set covers the $t$ range from $0.007$ to
$31.2 \gev^2$.  We omit the two data points with the lowest $t$ values,
which have huge errors and would not affect any our fits (but spoil the
legibility of plots), and remain with data in the range from $0.017$ to
$31.2 \gev^2$.

\begin{figure}[t]
\begin{center}
\includegraphics[width=0.45\textwidth,%
viewport=70 50 395 295]{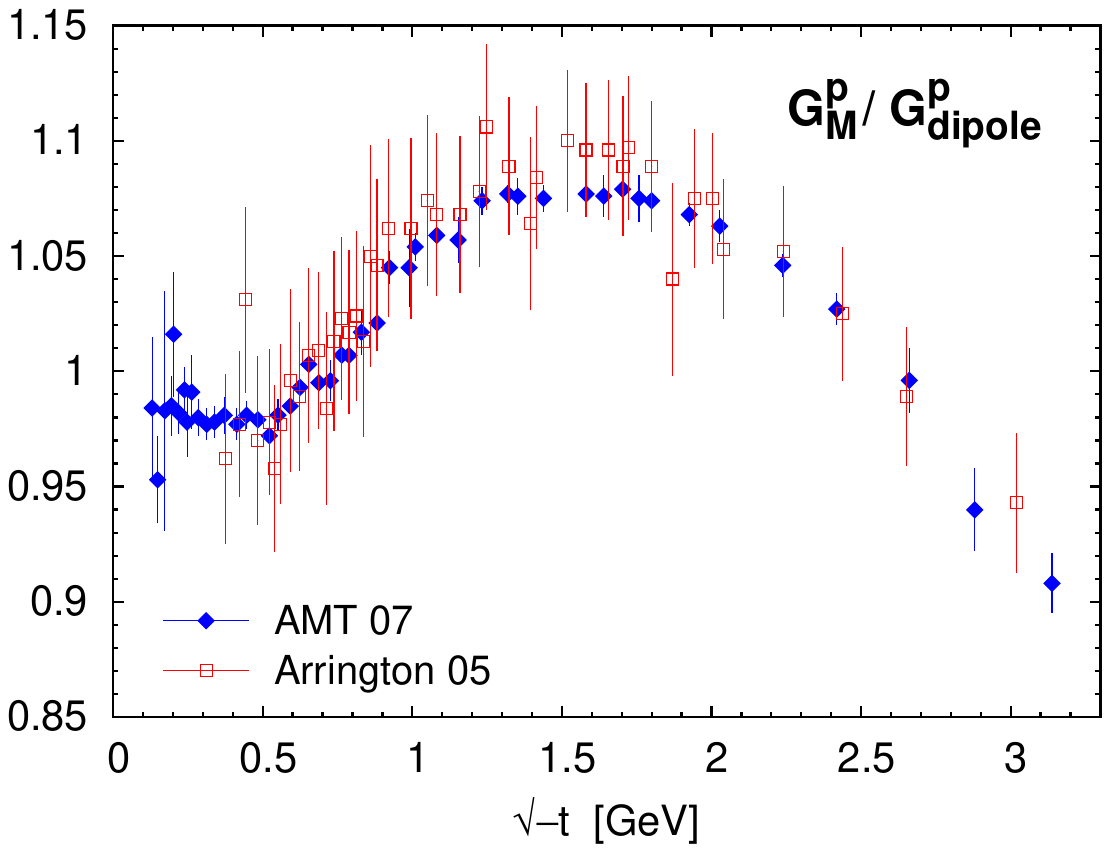} \\[1em]
\includegraphics[width=0.45\textwidth,%
viewport=70 50 395 295]{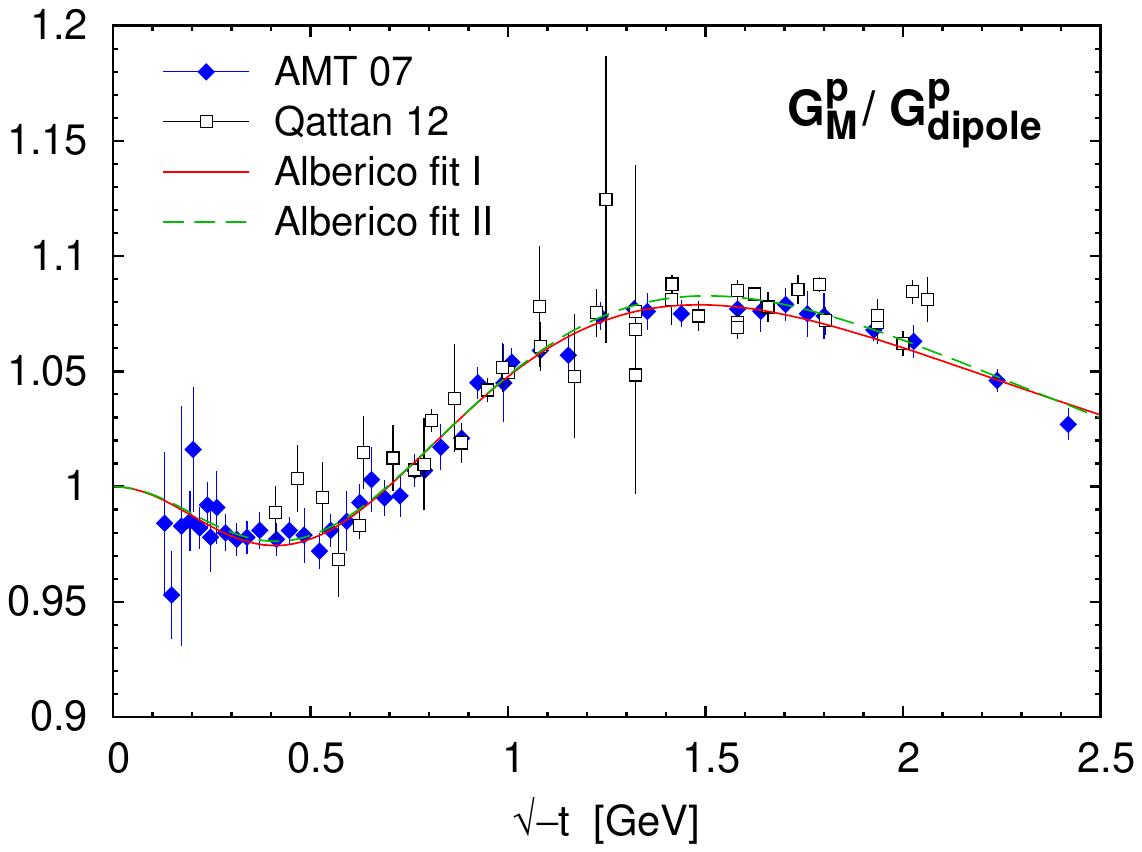}
\end{center}
\caption{\label{fig:GMp-data} Data for $G_M^p$ from the global analyses
  AMT 07 \protect\cite{Arrington:2007ux}, Arrington 05
  \protect\cite{Arrington:2004ae} and Qattan 12
  \protect\cite{Qattan:2012zf}, along with two fits from the global
  analysis by Alberico et al.\ \protect\cite{Alberico:2008sz}.  The form
  factor is divided by the dipole parameterization
  \protect\eqref{GDipole}.}
\end{figure}

A comparison of the results for $G_M^p$ extracted in
\cite{Arrington:2007ux,Arrington:2004ae,Qattan:2012zf} is shown in
\fig{fig:GMp-data}, as well as two fits proposed in
\cite{Alberico:2008sz}.  Here and in the following we plot form factors
and their ratios against $\sqrt{-t}$ rather than against the more common
variable $-t$, which permits a clearer view of the data at low $t$, where
the density of measurements is highest.  We observe that the error bars of
AMT 07 are significantly smaller than those of \cite{Arrington:2004ae} and
\cite{Qattan:2012zf}, which reflects that in
\cite{Arrington:2004ae,Qattan:2012zf} the size of two-photon exchange
effects was fitted to the data rather than provided as an external, albeit
model dependent, input.  Given the good agreement between AMT 07 and the
other analyses, including the central fit curves of
\cite{Alberico:2008sz}, we think that the AMT 07 data may be taken as a
representative of the current best knowledge of $G_M^p$.
We will later cross check our results by using instead the values for
$G_M^p$ obtained in \cite{Arrington:2004ae} (Arrington 05), see
\sect{sec:var-fits}.

%%%%%%

Let us now turn to the electric proton form factor.  Rather than the
values from the above global fits, we select in this case results for the
ratio $R^p$ measured with the recoil polarization method or with a
polarized proton target, i.e.\ with $p (\vec{e},e'\vec{p} \, )$ or
$\vec{p} \ms (\vec{e},e'p)$.  The recoil polarization data cover a range
of $-t$ from $0.246$ to $8.49 \gev^2$, and the polarized proton data from
$0.162$ to $1.51 \gev^2$.
To have some coverage at lower $-t$, we use in addition the values of
$R^p$ with $-t = 0.069, 0.098$ and $0.138 \gev^2$ from the analysis of AMT
07 \cite{Arrington:2007ux}.

\begin{table*}
\begin{center}
\renewcommand{\arraystretch}{1.2}
\begin{tabular}{lcllll}
\hline
\multicolumn{6}{c}{$R^p = \mu_p G_E^p /G_M^p$} \\
\hline
process & $-t \, [\gev^2]$ & \multicolumn{2}{c}{reference} & facility
& remarks \\
\hline
$p (\vec{e},e'\vec{p} \, )$
& ~0.38 -- 0.50~ & Milbrath 99   & \cite{Milbrath:1997de}   & MIT Bates \\
& 0.373 -- 0.441 & Pospischil 01 & \cite{Pospischil:2001pp} & MAMI A1 \\
& ~0.32 -- 1.76~ & Gayou 01      & \cite{Gayou:2001qt}      & JLab Hall A \\
& ~3.50 -- 5.54~ & Gayou 02      & \cite{Gayou:2001qd}      & JLab Hall A \\
& ~0.49 -- 3.47~ & Punjabi 05    & \cite{Punjabi:2005wq}    & JLab Hall A \\
& 1.13           & MacLachlan 06 & \cite{MacLachlan:2006vw} & JLab Hall C \\
& ~5.17 -- 8.49~ & Puckett 10    & \cite{Puckett:2010ac}    & JLab Hall C
\\ \cline{2-6}
& 0.8, 1.3       & Paolone 10    & \cite{Paolone:2010qc}    & JLab Hall A \\
& 0.246 -- 0.474 & Ron 11        & \cite{Ron:2011rd}        & JLab Hall A
& updates \cite{Ron:2007vr} \\
& 0.298 -- 0.695 & Zhan 11       & \cite{Zhan:2011ji}       & JLab Hall A \\
\hline
$\vec{p} \ms (\vec{e},e'p)$
& 1.51           & Jones 06      & \cite{Jones:2006kf}      & JLab Hall C \\
& 0.162 -- 0.591 & Crawford 07   & \cite{Crawford:2006rz}   & MIT Bates \\
\hline
\end{tabular}
\end{center}
\caption{\label{tab:Rp-data} Measurements of $R^p$ obtained with recoil
  polarization or with a polarized target (last two rows). }
\end{table*}

\begin{figure}[t]
\begin{center}
\includegraphics[width=0.45\textwidth,%
viewport=70 50 395 295]{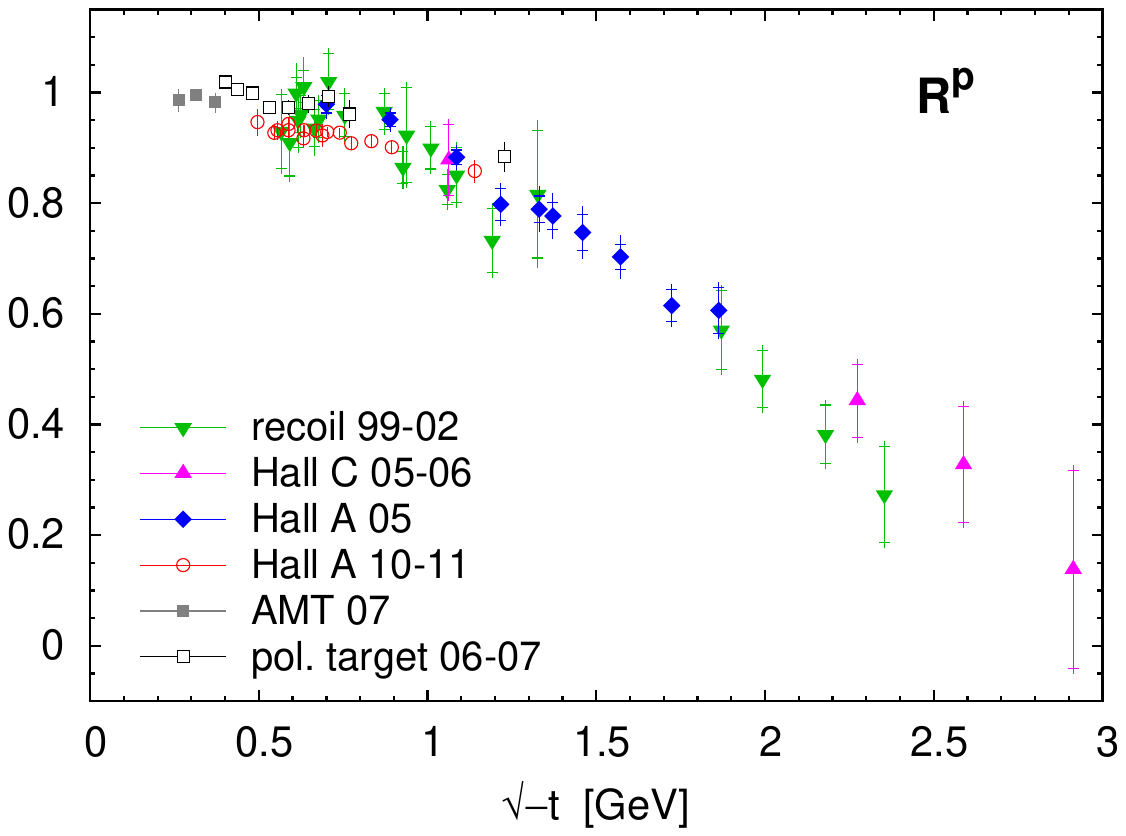} \\[0.5em]
\includegraphics[width=0.45\textwidth,%
viewport=70 50 395 295]{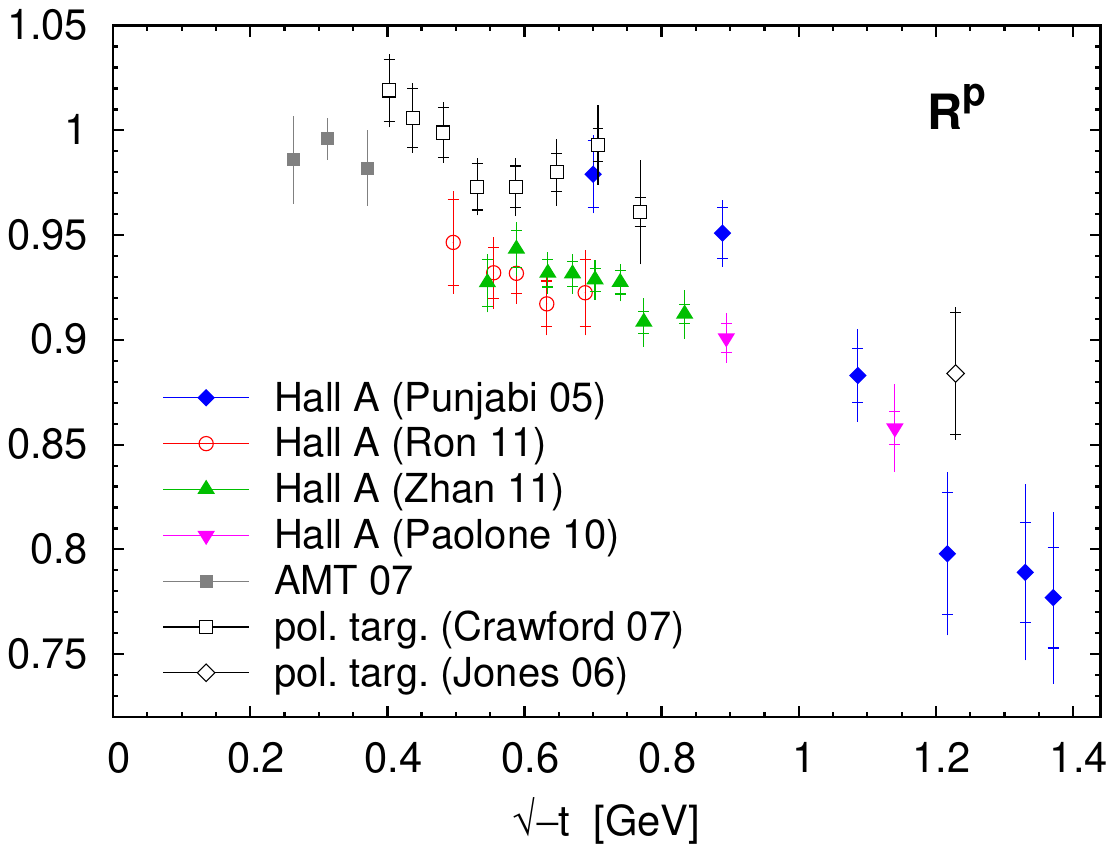}
\end{center}

\vspace{-0.5em}

\caption{\label{fig:Rp-data} Results for $R^p$ from the data sets in
  \tab{tab:Rp-data}.  The polarized target data (empty white squares or
  diamonds) are not used in our default data set.  Inner error bars
  indicate statistical uncertainties when those are given separately.}
\end{figure}

An overview of the data sets is given in \tab{tab:Rp-data}, and the
corresponding values of $R^p$ are shown in \fig{fig:Rp-data}.
We see that for $\sqrt{-t}$ between $0.5$ and $0.9 \gev$ the recent data
from JLab Hall A (Paolone 10, Ron 11 and Zhan 11) show a clear systematic
discrepancy with earlier data.  Let us emphasize that this discrepancy
concerns not only the polarized target data from MIT Bates (Crawford 07)
but also the two data points of the Hall A recoil polarization measurement
of Punjabi 05 in this $t$ range, as well as older recoil polarization data
(which have however relatively large errors).  We also note a discrepancy
at $\sqrt{-t} \approx 1.2 \gev$ between polarized target data (Jones 06)
and recoil polarization (Punjabi 05), which is however less significant at
the scale of the errors.  To the best of our knowledge, the origin of
these discrepancies is currently not understood.  We sincerely hope that
future investigations will clarify the experimental situation, which is
quite unsatisfactory as it stands.

Including all of the above data in least-square fits would not be useful
in our opinion, since without modifications this fitting method is not
designed to cope with manifestly inconsistent data sets.  Lacking better
criteria, we have chosen to include the more recent measurements in our
default data set, which is composed of all recoil polarization data in
\tab{tab:Rp-data} and the three points of AMT 07 mentioned above.  Note
that this set still includes a tension in the data, since we have not
removed the two points of Punjabi 05 with $\sqrt{-t}$ below $0.9 \gev$.
In \sect{sec:var-fits} we will investigate as an alternative the data set
obtained by removing the measurements of Paolone 10, Ron 11 and Zhan 11
and by adding the polarized target data of Jones 06 and Crawford 07.

%%%%%%%%%%%%%%%

\subsection{Neutron form factors}
\label{sec:neutron-data}

\begin{figure}[t]
\begin{center}
\includegraphics[width=0.45\textwidth,%
viewport=70 50 395 295]{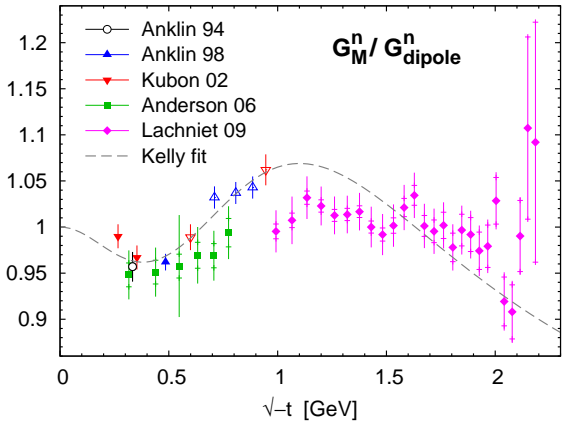}
\end{center}
\caption{\label{fig:GMn-data}  Data for $G_M^n$ divided by the dipole
  parameterization \protect\eqref{GDipole}.  The data points with open
  triangles (Anklin 98 and Kubon 02) are not used in our fits.  Inner
  error bars indicate statistical uncertainties.}
\end{figure}

The magnetic form factor of the neutron can be determined from the cross
section ratio of the quasi-elastic processes $d(e,e'n)$ and $d(e,e'p)$,
and also from scattering polarized electrons on polarized
${}^3\mathrm{He}$.  The relevant data sets are listed in
\tab{tab:GMn-data} and shown in \fig{fig:GMn-data}.  We do not use results
of the MIT Bates measurement \cite{Gao:1998nn,Gao:1994ud} of
$\smash{{}^3\overrightarrow{\mathrm{He}}} (\vec{e}, e')$, which has rather
large errors and would not influence our analysis.

Looking at the deuterium measurements at $\sqrt{-t}$ just below $1 \gev$,
we observe an abrupt change between the data of Anklin 98 and Kubon 02 on
one hand and the Hall B data (Lachniet 09) on the other.  We do not
consider this a physically plausible behavior and suspect a consistency
problem between these measurements (which were using the same process),
but we have not found any discussion of this issue in the literature.  We
also observe that the two data points of Anklin 98 for $\sqrt{-t}$ between
$0.7$ and $0.8 \gev$ are systematically higher than the data of Anderson
07, which connect smoothly with those of Hall B.  Since the Anderson 07
points are extracted from ${}^3\mathrm{He}$ and thus have entirely
different theoretical and systematic uncertainties than the Hall B
measurement, we consider this as a strong argument in favor of the more
recent data.  We hence decided to discard the points with $-t = 0.504,
0.652$ and $0.784 \gev^2$ from Anklin 98 and the points with $-t = 0.359$
and $0.894 \gev^2$ from Kubon 02.  We keep, however, the point with $-t =
0.235 \gev^2$ from Anklin 98 and the points with $-t = 0.071$ and $0.125
\gev^2$ from Kubon 02, which are consistent with other data in the same
$t$ region.

We remark that the global fit by Kelly \cite{Kelly:2004hm} used the full
data sets of Anklin 98 and Kubon 02, whereas the results of Anderson 07
and Lachniet 09 were not available at the time of that fit.  As a
consequence, the Kelly parameterization for $G_M^n$ faithfully reproduces
the older data but is in clear conflict with the newer ones, as seen in
\fig{fig:GMn-data}.

\begin{table*}[t]
\begin{center}
\renewcommand{\arraystretch}{1.2}
\begin{tabular}{lcllll}
\hline
\multicolumn{6}{c}{$G_M^n$} \\
\hline
process & $-t \, [\gev^2]$ & \multicolumn{2}{c}{reference} & facility & remarks \\
\hline
$d(e,e'n)$, $d(e,e'p)$
& 0.111 & Anklin 94 & \cite{Anklin:1994ae} & NIKHEF, PSI & \\
& 0.235 -- 0.784 & Anklin 98   & \cite{Anklin:1998ae} & MAMI, PSI
  & used partially \\
& 0.071 -- 0.894 & Kubon 02    & \cite{Kubon:2001rj}  & MAMI, PSI
  & used partially \\
& 0.985 -- 4.773 & Lachniet 09 & \cite{Lachniet:2008qf} & JLab Hall B \\
\hline
${}^3\overrightarrow{\mathrm{He}} (\vec{e}, e')$ \rule{0pt}{2.9ex}
& 0.1 -- 0.6     & Anderson 07 \hspace{-1em}
  & \cite{Anderson:2006jp} & JLab Hall A
  & updates \cite{Xu:2000xw,Xu:2002xc} \\
\hline
\end{tabular}
\end{center}
\caption{\label{tab:GMn-data} Data sets for the determination of $G_M^n$.}
\end{table*}

The electric neutron form factor $G_E^n$ has been extracted from a series
of polarization measurements using deuterium or ${}^{3}\mathrm{He}$, which
are compiled in \tab{tab:Rn-data}.  In the approximation of vanishing
final state interactions, these experiments directly measure $G_E^n
/G_M^n$ and obtain $G_E^n$ by using a value for $G_M^n$ as external input.
This is problematic since the used values are not always in good agreement
with the current experimental determination of $G_M^n$, especially if they
are computed from the dipole parameterization \eqref{GDipole} or from the
Kelly parameterization \cite{Kelly:2004hm}.  To circumvent this bias, we
use $R^n$ as input for our data selection; values of $G_E^n$ then result
from our selection and analysis of the $G_M^n$ data.  The papers
\cite{Plaster:2005cx,Geis:2008aa,Riordan:2010id} directly quote results
for $R^n$, whereas in the other cases we have calculated $R^n$ from the
quoted results for $G_E^n$, using the $G_M^n$ values and their errors as
specified in the experimental papers.
% (For \cite{Passchier:1999cj} no error on $G_M^n$ was assumed.)
The resulting values and errors of $R^n$ are given in \app{app:tables} for
reference.

\begin{table*}
\begin{center}
\renewcommand{\arraystretch}{1.2}
\begin{tabular}{lcllll}
\hline
\multicolumn{6}{c}{$R^n = \mu_n G_E^n /G_M^n$} \\
\hline
process & $-t \, [\gev^2]$ & \multicolumn{2}{c}{reference} & facility & remarks \\
\hline
$d (\vec{e},e'\vec{n})p$
& 0.15, 0.34     & Herberg 99 & \cite{Herberg:1999ud} & MAMI A1 \\
& ~~0.3 -- 0.79~ & Glazier 05 & \cite{Glazier:2004ny} & MAMI A1
   & updates \cite{Ostrick:1999xa} \\
& 0.447 -- 1.45~ & Plaster 06 & \cite{Plaster:2005cx} & Jlab Hall C
   & updates \cite{Madey:2003av} \\
\hline
$\vec{d}(\vec{e},en)p$ \rule{0pt}{2.8ex}
& 0.21~          & Passchier 99 & \cite{Passchier:1999cj} & NIKHEF \\
& 0.495          & Zhu 01       & \cite{Zhu:2001md} & JLab Hall C \\
& 0.5, 1.0       & Warren 04    & \cite{Warren:2003ma} & JLab Hall C \\
& 0.142 -- 0.415 & Geis 08      & \cite{Geis:2008aa} & MIT Bates \\
\hline
${}^3\overrightarrow{\mathrm{He}} (\vec{e}, e'n)$ \rule{0pt}{2.9ex}
& 0.67~          & Bermuth 03  & \cite{Bermuth:2003qh} & MAMI A1 
    & updates \cite{Rohe:1999sh} \\
& 0.35~          & Rohe 05     & \cite{Rohe:2005pc} & MAMI A3 
    & updates \cite{Becker:1999tw,Golak:2000nt} \\
& ~1.72 -- 3.41~ & Riordan 10  & \cite{Riordan:2010id} & JLab Hall A \\
\hline
\end{tabular}
\end{center}
\caption{\label{tab:Rn-data} Data sets for the determination of the
  neutron form factor ratio $R^n$.}
\end{table*}

\begin{figure}[t]
\begin{center}
\includegraphics[width=0.45\textwidth,%
viewport=70 50 395 295]{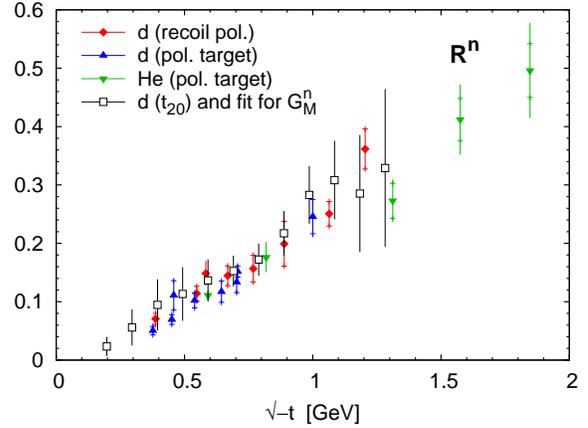}
\end{center}
\caption{\label{fig:Rn-data} Data for $R^n$.  The open (white) symbols
  correspond to data for $G_E^n$ together with $G_M^n$ taken from the fit
  described in \protect\sect{sec:power-law-fit}.  Inner error bars indicate
  statistical uncertainties.}
\end{figure}

As an independent source of information we take the values of $G_E^n$
extracted from the deuteron quadrupole form factor in Schiavilla 01
\cite{Schiavilla:2001qe}, which uses as input the measurement of the
polarization observable $t_{20}$ and provides data points with $-t$ from
$0.0389$ to $1.644 \gev^2$.  These data have relatively large errors and
are shown in \fig{fig:data-vs-interpol} below.  Taking the fit for $G_M^n$
that will be described in \sect{sec:power-law-fit}, we can compute $R^n$
from these data.  As can be seen in \fig{fig:Rn-data}, the result agrees
well with $R^n$ extracted from the the measurements in \tab{tab:Rn-data}.

We note that the smallest value of $\sqrt{-t}$ for which we have a precise
determination of $R^n$ (from Herberg 99 and Geis 08) is about $0.4 \gev$.
It is therefore highly welcome that entirely independent information about
the small-$t$ behavior of $G_E^n$ is provided by the squared charge radius
of the neutron, which is defined by
\begin{equation}
  \label{neutron-radius}
r^2_{nE} = 6 \, \frac{\dd G_E^{n}}{\dd t} \bigg|_{t=0}
\end{equation}
and can be measured in scattering a neutron beam off the shell electrons
in a nuclear target.  We use the value quoted in the 2012 Review of
Particle Physics \cite{Beringer:1900zz},
\begin{equation}
  \label{r2nE-value}
r^2_{nE} = - ( 0.1161 \pm 0.0022 ) \fm^2 \,,
\end{equation}
which has been stable since its first listing in the 2002 edition of the
same review.  In principle we might have also included in our data
selection the electric charge radius of the proton as determined from
atomic physics, but the present controversy concerning the value of this
radius prevented us from following this path.  We will briefly discuss
this in \sect{sec:charge-radii}.

%%%%%%%%%%%%%%%%%%%%%%%%%%%%%%%%%%%%%%%%%%%%%%%%%%

\subsection{A note on the dipole parameterization}
\label{sec:dipole-par}

When plotting $G_M^p$ and $G_M^n$ we follow common practice and divide by
the dipole parameterization
\begin{equation}
\label{general-dipole}
F(t) = \frac{F(0)}{\bigl( 1
            - t / M^2_{\text{dip}} \bigr){}^2_{\phantom{1}}}
\end{equation}
of these form factors.  As already emphasized, this is a matter of pure
convenience, with the value $M^2_{\text{dip}} = 0.71 \gev^2$ fixed by
convention.  As is evident from \figs{fig:GMp-data} and
\ref{fig:GMn-data}, neither $G_M^p$ nor $G_M^{n\phantom{p}}$ is
particularly well described by this parameterization.

\begin{figure*}[t]
\begin{center}
\includegraphics[width=0.45\textwidth,%
viewport=70 50 395 295]{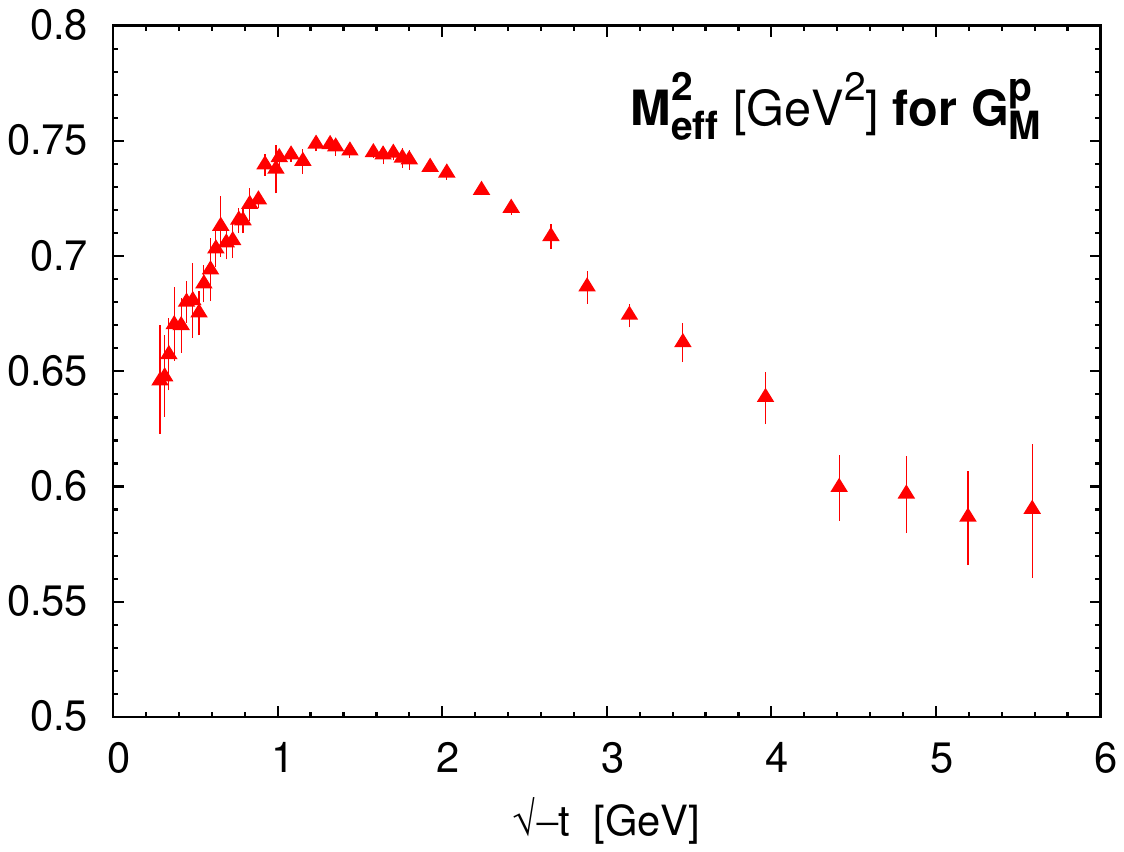}
\hspace{1.5em}
\includegraphics[width=0.45\textwidth,%
viewport=70 50 395 295]{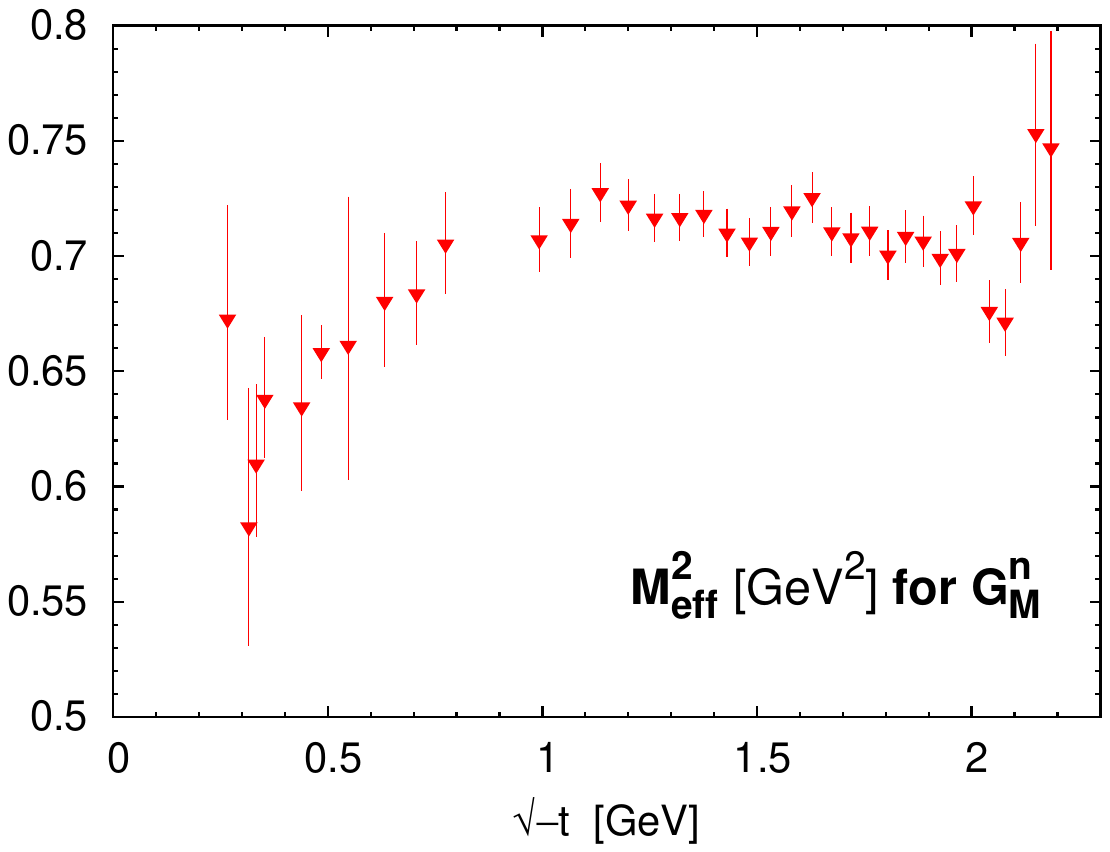}
\end{center}
\caption{\label{fig:GMp-GMn-dipmass} Effective dipole masses as defined in
  \protect\eqref{eff-dip-mass} for $G_M^p$ and $G_M^{n\protect\phantom{p}}$,
  evaluated from our data set.}
\end{figure*}

One may ask whether a good description of these form factors can be
achieved with a different value of the dipole mass, at least in a certain
range of $t$.  To investigate this question, we define the effective
dipole mass
\begin{equation}
\label{eff-dip-mass}
M_{\text{eff}}^2(t) = \frac{-t}{\sqrt{F(0)/F(t) \rule{0pt}{1.9ex}} - 1} \,,
\end{equation}
which is $t$ independent and equal to the dipole mass for a form factor
with the shape \eqref{general-dipole}.  From the plots of this quantity in
\fig{fig:GMp-GMn-dipmass} we see that within their currently known
precision, neither of the magnetic form factors is well represented by a
dipole form in any interval that starts at $t=0$.  Somewhat amusingly, the
dipole law with its conventional mass value approximately describes
$G_M^n$ in the region of $\sqrt{-t}$ between 1 and $2 \gev$, as is already
visible in \fig{fig:GMn-data}.

%%%%%%%%%%%%%%%%%%%%%%%%%%%%%%%%%%%%%%%%%%%%%%%%%%

\subsection{A global form factor fit}
\label{sec:power-law-fit}

Although the main goal of our study is the determination of the GPDs
$H_v^q$ and $E_v^q$ from the form factors with the help of the sum rules
\eqref{eq:SumRule} (see \sect{sec:gpd-fits}), there is some interest in
having a simple parameterization of the form factor data.  This
parameterization can serve as a baseline for comparison with our GPD fits,
for the interpolation of data or for the convenient evaluation of cross
sections.  An example for the latter is the evaluation of the
Bethe-Heitler process, which plays a special role in connection with
deeply virtual Compton scattering.

\begin{table*}[t]
\begin{center}
\renewcommand{\arraystretch}{1.2}
\begin{tabular}{lcccc}
\hline
      & $a_{iq}+b_{iq}$& $b_{iq}$ & $p_{iq}$ & $q_{iq}-p_{iq}$ \\  
\hline
$F_1^{u-s}$ & $3.116\pm0.102$& $1.122\pm 0.101$& $0.347\pm 0.061$& $1.527\pm 0.110$ \\
$F_1^{d-s}$ & $3.184   \pm   0.103$& $1.638   \pm   0.242$& $0.278   \pm   0.119$ &  $3.5$ \\
$F_2^{u-s}$ & $3.192   \pm   0.060$& $0.122   \pm   0.034$& $1.812   \pm
0.110$ &  $2.5$ \\
$F_2^{d-s}$ & $3.478   \pm   0.149$& $1.649   \pm   0.250$& $0.296   \pm
0.100$ & $2.5$ \\
\hline
\end{tabular}
\end{center}
\caption{\label{tab:global-power-fit-pars} The parameters of the global fit 
  \protect\eqref{global-power-fit-fct} to our default data
  set.  Parameters without quoted errors are kept fixed in the fit.}
%\end{table*} 

\vspace{1em}

%\begin{table*}
\begin{center}
\renewcommand{\arraystretch}{1.2}
\begin{tabular}{lcccccccc}
\hline
  & total & \multicolumn{2}{c}{--- $G_M^p$ ---} & $R^p$
          & $G_M^n$ & $R^n$ & $G_E^n$ & $r^2_{nE}$ \\
\hline
$\chi^2_{\text{min}}$
  & $122.3$ & $28.8$ &  $1.8$ & $52.7$ & $20.4$ & $15.3$ & $3.4$ & $0.0$ \\
data points & $178$ & $48$ & $6$ & $54$ & $36$ & $21$ & $12$ & $1$ \\
\hline
\end{tabular}
\end{center}
\caption{\label{tab:global-power-fit-chi} Total and partial values of
  $\chi^2_{\text{min}}$ for the global power law fit specified by
  \protect\eqref{global-power-fit-fct} and the parameters in
  \protect\tab{tab:global-power-fit-pars}. For $G_M^p$ the first value is
  for the data with $-t < 10 \gev^2$ and the second value for $-t > 10
  \gev^2$.}
\end{table*}

We find that a good representation of the data is possible if we represent
the form factor combinations \eqref{inverse-dec} as the product of two
fractional power laws:
\begin{align}
\label{global-power-fit-fct}
F_i^{q-s}(t) & =
  \frac{F_i^{q-s}(0)}{(1 - a_{iq}\, t/p_{iq})^{p_{iq}}_{\phantom{1}} \,
                      (1 - b_{iq}\, t/q_{iq})^{q_{iq}}_{\phantom{1}}}
\nonumber \\
\end{align}
with $q=u,d$ and $i=1,2$.  This ansatz makes no assumption about the size
of the strangeness contributions, but if they are neglected, then
\eqref{global-power-fit-fct} directly parameterizes the flavor form
factors $F_i^{u}$ and $F_i^{d}$.  The sum $a_{iq} + b_{iq}$ is equal to
the logarithmic derivative of $F_i^{q-s}$ at $t=0$.  We note in passing
that one has $(1-at/p)^{-p} \to \exp(at)$ for $p\to\infty$.

At variance with other approaches, we do not impose the asymptotic
behavior $F_i^q \sim 1/t^2$ or $F_2^q \sim 1/t^3$ that is predicted by
dimensional counting.  This is in line with the physical assumption behind
our GPD fit discussed later, namely that the hard-scattering mechanism
that gives rise to dimensional counting behavior is not relevant for the
electromagnetic form factors in the $t$ region where there is data.  In
this spirit, our global fit aims at describing the existing data and at
extrapolating them over a limited range, but it has no ambition to
describe the form factors in an asymptotic regime of large $t$.

A fit to our default set of form factor data, including the squared
neutron charge radius, provides the parameters compiled in
\tab{tab:global-power-fit-pars}.  The corresponding values of
$\chi^2_{\text{min}}$ are listed in \tab{tab:global-power-fit-chi}.  With
the exception of $q_{1u}-p_{1u}$ the differences $q_{iq}-p_{iq}$ of powers
were kept fixed in order to obtain a stable $\chi^2$ fit, their values
have been optimized by varying them in steps of $0.5$ and monitoring the
change of $\chi^2_{\text{min}}$ for the individual form factors, so as to
achieve a uniformly good description of all observables as much as
possible.

In terms of $\chi^2$ the fit is very good as
\tab{tab:global-power-fit-chi} reveals, with the minimal $\chi^2$ being
always smaller than the number of data points.  Plots comparing the fit
with data are shown in \sect{sec:def-fit}.  With a total of 16 parameters
and the ansatz \eqref{global-power-fit-fct} we can thus obtain an
excellent description of all the form factor data we have selected.  The
low $\chi^2$ of our fit does not imply that we have over-parameterized the
data: as explained earlier, systematic uncertainties are included in the
errors on the form factor data, so that their statistical point-to-point
fluctuations are not as large as suggested by the errors.
We note that a simpler fit with all $q_{iq}-p_{iq}$ set to zero give still
a rather good description of the data, with a global $\chi^2_{\text{min}}
= 182.9$.  It does, however, systematically overshoot the very precise
$G_M^p$ data for $-t$ below $1 \gev^2$ and correspondingly has a large
partial $\chi^2_{\text{min}} = 73.4$ for the 48 data points of $G_M^p$
with $-t < 10 \gev^2$.

Since we have a complete set of Sachs form factor data only up to $-t =
3.41 \gev^2$, the individual flavor form factors of our global power law
fit can only be considered reliable up to this $t$ value.  On the other
extreme, we have data only for $G_M^p$ in the range $8.5 \gev^2 < -t <
31.2 \gev^2$.  In this range $F_1^{u-s}$ as given by our fit contributes
more than $70\%$ to $G_M^{\,p}$ compared with the other flavor form
factors.  The same is true for the default GPD fit to be discussed in
\sect{sec:def-fit}.  The results of these two fits differ by at most
$17\%$ for all $-t < 31.2 \gev^2$.  In this sense, we may regard our fit
for $F_1^u$ as reasonably reliable over this $t$ range.

%% file: strange.tex
\section{Strangeness}
\label{sec:strange}

Although virtual $s\bar{s}$ pairs may not be rare in the proton, the
strangeness form factors $F_1^s(t)$ and $F_2^s(t)$ are expected to be
small, because they describe the \emph{difference} between the
distributions of strange quarks and antiquarks.  This is evident from the
corresponding sum rules \eqref{eq:SumRule} and should not be surprising
because the electromagnetic current probes the local excess of quarks over
antiquarks (or of antiquarks over quarks).

Since the proton has no net strangeness, the strange PDFs satisfy
\begin{equation}
\int_0^1\dd x \bigl[ s(x) - \bar{s}(x)\bigr] = 0
\label{eq:strange-sum-rule}
\end{equation}
and the strange Dirac form factor is normalized as
\begin{equation}
F_1^{s}(0) = 0 \,.
\end{equation}
By contrast, the strange Pauli form factor at $t=0$ is equal to the
strangeness magnetic moment, $F_2^s(0) = \mu_s$, and can be nonzero.

\begin{figure}[ht]
\begin{center}
\includegraphics[width=0.45\textwidth,%
viewport=70 50 395 295]{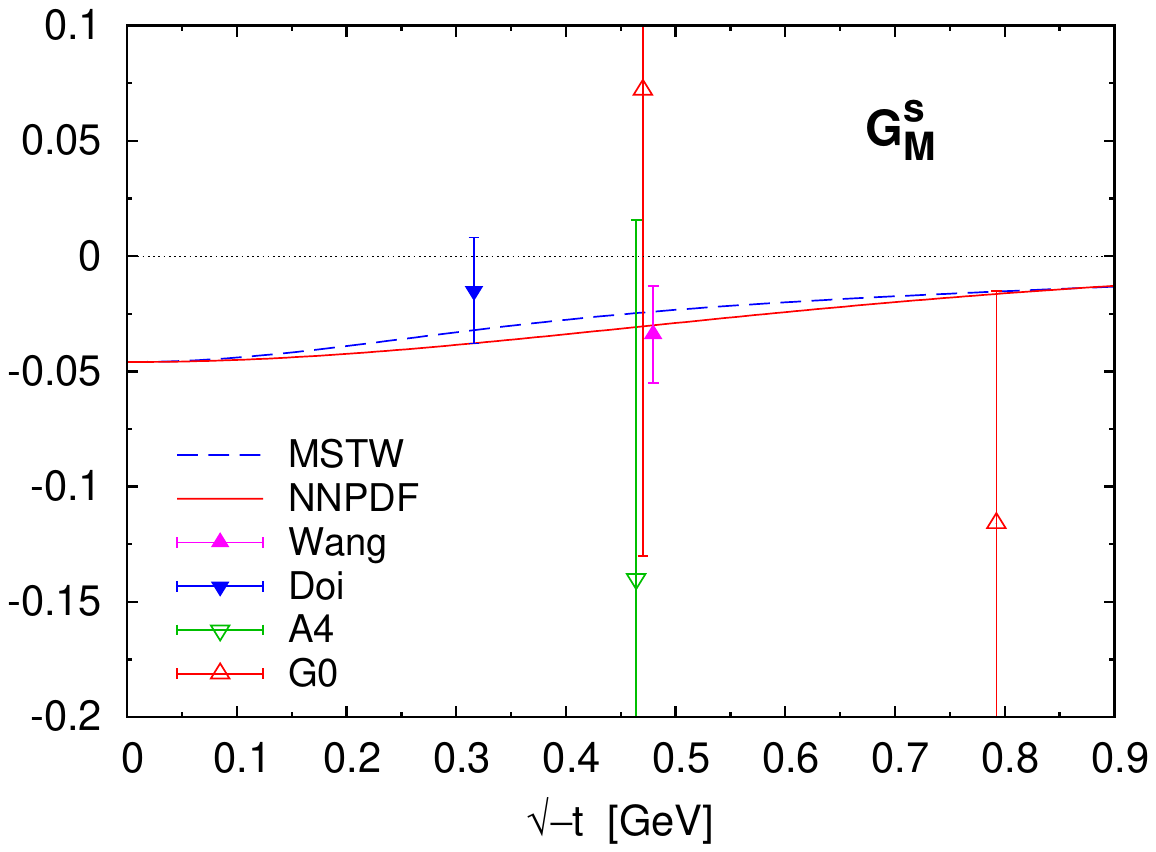} \\[1em]
\includegraphics[width=0.45\textwidth,%
viewport=70 50 395 295]{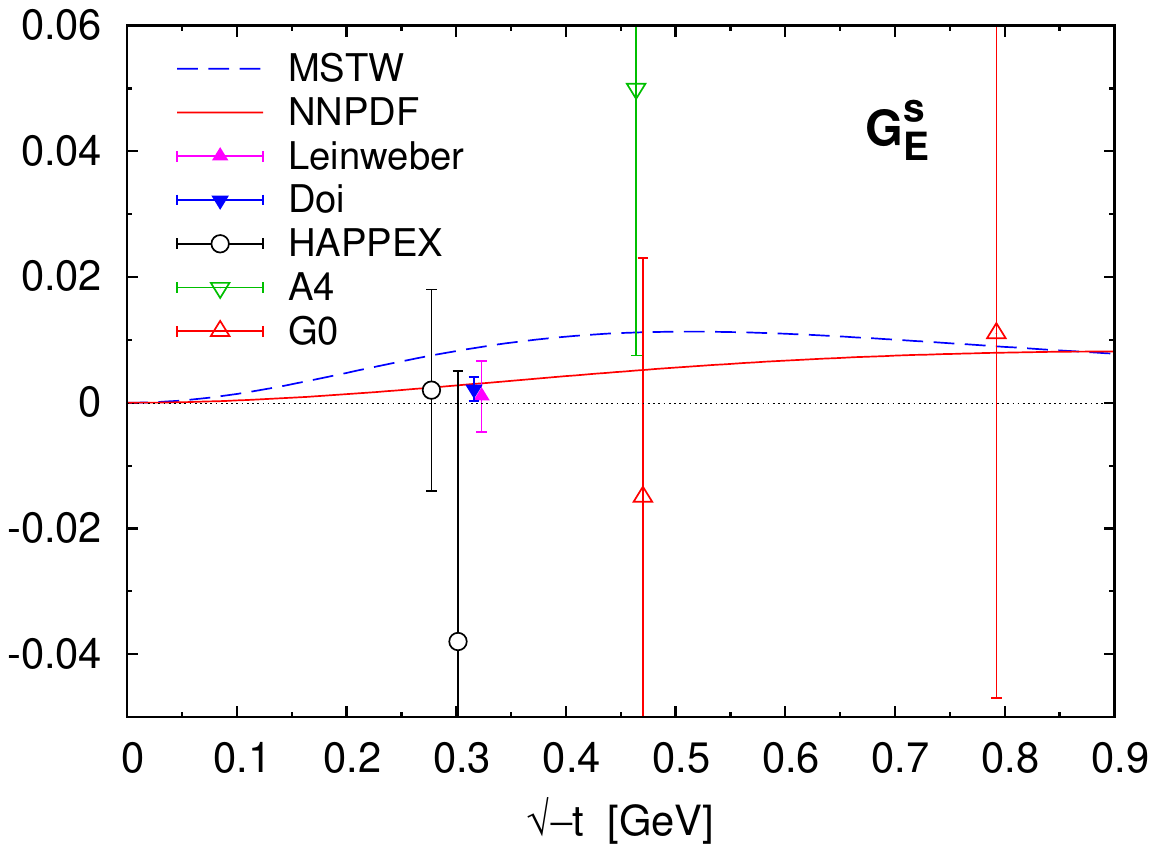}
\end{center}
\caption{\label{fig:strangeSachs} Data for the strangeness form factors by
  HAPPEX \protect\cite{Aniol:2005zf,Acha:2006my}, A4
  \protect\cite{Baunack:2009gy} and G0 \protect\cite{Androic:2009aa},
  together with lattice QCD results by Leinweber
  \protect\cite{Leinweber:2006ug}, Wang \protect\cite{Wang:1900ta} and Doi
  \protect\cite{Doi:2009sq}.  Our model predictions are shown as solid and
  dashed lines and labeled by the PDF set used in the model.}
\end{figure}

In recent years there has been an enormous activity to determine the
strangeness form factors $G_M^{s}$ and $G_E^{s}$ from parity violating
elastic scattering, see e.g.\
\cite{Maas:2004ta,Aniol:2004hp,Maas:2004dh,Aniol:2005zf,Aniol:2005zg,%
  Acha:2006my,Baunack:2009gy,Androic:2009aa}.  Despite this effort it is
not yet clear whether or not the strange form factors are significantly
nonzero, as can be seen in \fig{fig:strangeSachs}.  Likewise, recent
global fits of parton densities \cite{Martin:2009iq,Ball:2011gg} do not
provide unambiguous evidence for a nonzero difference
$s(x)-\bar{s}(x)$.  While the MSTW 2008 result \cite{Martin:2009iq} for
this quantity is compatible with zero within errors, this is not the case
for the analysis of NNPDF 2.2 \cite{Ball:2011gg}.  Lattice QCD results
\cite{Leinweber:2004tc,Leinweber:2006ug,Lin:2007gv,Wang:1900ta,Doi:2009sq}
have much smaller errors than the form factor measurements and the PDF
analyses.  Several lattice determinations obtain a nonzero value of
$\mu_s$ (see below), but $G_M^{s}$ and $G_E^{s}$ at finite $t$ are
typically compatible with zero within uncertainties, as shown in
\fig{fig:strangeSachs}.

For an estimate of the strangeness contributions we therefore take
recourse to a model.  Following our previous work \cite{Diehl:2007uc} we
parameterize the GPD $H_v^s$ in similar way as the $u$ and $d$ quark GPDs
in \cite{DFJK4} and in the present paper:
\begin{multline}
  \label{strange-ansatz}
H_v^s(x,t) = \bigl[ s(x)-\bar{s}(x) \bigr]
\\
\times \exp{\bigl[ t\ms \alpha_s' (1-x) \log{(1/x)} \bigr]} \,.
\end{multline}
In the exponent we set $\alpha'_s = 0.95 \gev^{-2}$.  According to the
study in \cite{Diehl:2007uc}, a variation of this value in the range $0.85
\gev^{-2} < \alpha'_s < 1.15 \gev^{-2}$ would not significantly change our
results for $F_1^s$.  The PDFs in the ansatz \eqref{strange-ansatz} are
taken either from MSTW 08 \cite{Martin:2009iq} or from NNPDF 2.2
\cite{Ball:2011gg} and are evaluated at the scale $\mu = 2 \gev$.  With
the sum rule \eqref{eq:SumRule} one can then compute the Dirac form factor
$F_1^s$.  In the region $-t < 36 \gev^2$, the result can be parameterized
as
\begin{equation}
F_1^s(t) = \frac{-c t}{(1 - a t /p)^p \, (1 - b t /q)^q}
\end{equation}
with an accuracy better than $2.5\%$, where $p=2$ and
\begin{align}
a &= 4.54 \gev^{-2} \,,
&
b &= 1.79 \gev^{-2} \,,
\nonumber \\
c &= 0.136 \gev^{-2} \,,
&
q &= 4.8
\end{align}
for MSTW, whereas $p=1$ and
\begin{align}
a &= 2.22 \gev^{-2} \,,
&
b &= 0.722 \gev^{-2} \,,
\nonumber \\
c &= 0.0244 \gev^{-2} \,,
&
q &= 1.92
\end{align}
for NNPDF.

For the strange Pauli form factor a similar parameterization cannot be
exploited, because the forward limit of the relevant GPD is completely
unknown.  We must therefore pursue a different strategy in this case.
Lattice simulations quote the following values for the strangeness
magnetic moment:
\begin{align}
  \label{lattice-mu-s}
\mu_s &= - 0.046 \pm 0.019 & \protect\cite{Leinweber:2004tc}  \,,
\nonumber \\
      &= - 0.066 \pm 0.026 & \protect\cite{Lin:2007gv}  \,,
\nonumber \\
      &= - 0.017 \pm 0.025 \pm 0.007 & \protect\cite{Doi:2009sq}  \,,
\end{align}
and we adopt the value $\mu_s = - 0.046$ from \cite{Leinweber:2004tc},
which is consistent with the other two determinations.
There are also estimates of $\mu_s$ from different variants of the
constituent quark model.  Results obtained before 2000 are typically an
order of magnitude larger than the lattice results \eqref{lattice-mu-s},
whereas more recent calculations, e.g.\ in
\cite{dahiya06,kiswandhi11,riska}, provide results in fair agreement with
them.

\begin{table*}[t]
\begin{center}
\renewcommand{\arraystretch}{1.2}
\begin{tabular}{lcccccc} \hline
\multicolumn{2}{c}{reference} & $-t~$
  & quantity & experimental & \multicolumn{2}{c}{model result} \\
  & & $[\gev^2]$ & & value & MSTW & NNPDF \\
\hline
HAPPEX & \protect\cite{Aniol:2004hp} & 0.477
 & $G_E^s+0.392\, G_M^s$ & $0.014 \pm 0.022$ & 0.0032 & $-0.00056$ \\
A4     & \protect\cite{Maas:2004dh} & 0.108
 & $G_E^s+0.106\, G_M^s$ & $0.071 \pm 0.036$ & 0.0057 & $-0.00078$ \\
HAPPEX & \protect\cite{Aniol:2005zg} & 0.099
 & $G_E^s+0.080\, G_M^s$ & $0.030 \pm 0.028$ & 0.0061 & $-7 \times 10^{-5}$ \\
HAPPEX & \protect\cite{Acha:2006my} & 0.109
 & $G_E^s+0.090\, G_M^s$ & $0.007 \pm 0.013$ & 0.0063 & $-0.00016$ \\
\hline
\end{tabular}
\end{center}
\caption{\label{tab:strange-combinations} Experimental values and model
  results for linear combinations of strangeness form factors.}
\end{table*}

For the $t$ dependence of $F_2^s$, we adopt a vector meson dominance
ansatz.  With three poles corresponding to the $\phi(1020)$ meson and its
excited states $\phi(1680)$ and $\phi(2170)$ \cite{Beringer:1900zz} we
have \footnote{%
  We emphasize that $\phi(1020)$ \emph{and} its excited states are
  necessary to obtain a proper large $t$ behavior of the strangeness form
  factors.  This sheds doubt on analyses that obtain the dipole behavior
  of isosinglet form factors ($G_M^p + G_M^{n\phantom{p}}$, $G_E^p +
  G_E^{n\phantom{p}}$ or $F_1^p + F_1^{n\phantom{p\!\!}\!}$) by a
  conspiracy of $\omega$ and $\phi$ exchange without excited states.  See
  also our discussion in section~3.1 of \protect\cite{Diehl:2007uc}.}
\begin{equation}
  F_2^{s}(t) = \mu_s \sum_{i=1}^3 \frac{a_i}{m_i^2 -t} \,.
\end{equation}
For lack of better knowledge, we require $F_2^s$ to decrease
asymptotically like $1/t^3$ as suggested by dimensional counting.
Together with the normalization condition at $t=0$, this gives the
constraints
\begin{align}
& \sum_i a_i /m_i^2 = 1 \,,
  \qquad \sum_i a_i = 0 \,, 
\nonumber \\
& \sum_i a_i \sum_{j\neq i} m_j^2 = 0 \,,
\end{align}
which imply
\begin{equation}
a_i = \frac{m_1^2\ms m_2^2\ms m_3^2}{%
            \prod\limits_{j\neq i}(m_i^2 -m_j^2)}
\end{equation}
for the residues.  With the mass values from \cite{Beringer:1900zz} we
obtain $a_1=2.115\,\gev^2$, $a_2=-4.113\,\gev^2$ and $a_3=1.998\,\gev^2$.
The first parameter is related to the tensor coupling between the
$\phi(1020)$ and the nucleon as
\begin{equation}
g^T_{\phi NN} = \mu_s\ms f_1^{} a_1^{} /m_1^2 \,.
\end{equation}
With $f_1 =13.4$ from the electronic decay width of the $\phi(1020)$
meson, we obtain
\begin{equation}
(g^T_{\phi NN})^2/(4\pi) = 0.13 \,,
\end{equation}
which is consistent with a dispersion analysis of nucleon-nucleon
scattering \cite{grein}.
We note in passing that for $-t < 5 \gev^2$ our parameterization of
$F_2^s$ can be approximated by a dipole form \eqref{general-dipole} with
$M_{\text{dip}}^2 = 1.13 \gev^2$ with 11\% accuracy.

Combining our models for $F_1^{s}$ and $F^{s}_2$ we obtain the Sachs form
factors shown in \fig{fig:strangeSachs}.  Both models are consistent with
experiment and with lattice results, except for a discrepancy between the
MSTW model and the lattice points of Leinweber and Doi at $-t = 0.1
\gev^2$.
Furthermore, our model results are in good agreement with the data listed
in \tab{tab:strange-combinations}, except for the A4 measurement, where
our values are about 2 standard deviations away from the experimental
value.  We note that the combination of the lattice results
\cite{Doi:2009sq} for $G_M^s$ and $G_E^s$ at $-t = 0.1 \gev^2$ gives
$G_E^s + 0.106\, G_M^s = 0.0006 \pm 0.0031$ (if we add errors in
quadrature), which is in better agreement with our model results and in
tension with the A4 value.

Comparing our strangeness form factors with the power-law fit of
$F_i^{u-s}$ and $F_i^{d-s}$ described in \sect{sec:power-law-fit}, we find
that the ratios $|F_i^s / F_i^u|$ and $|F_i^s / F_i^d|$ are below $6\%$
for $\sqrt{-t} < 1 \gev^2$ and below $12\%$ for $\sqrt{-t} < 2 \gev^2$.
With the interpolated set of flavor form factors described in the next
section, we find that our strangeness form factors are at most of the size
of the errors on $F_i^u$ and $F_i^d$.  In this respect the strangeness
contribution may be neglected when discussing the flavor decomposition of
the form factors, at the present level of accuracy.  We will return to the
strangeness form factors in \sect{sec:var-fits}.

%% file: interpol.tex
\section{Interpolated data}
\label{sec:interpol}

\subsection{Determination of flavor form factors}
\label{sec:interpol-procedure}

The valence quark GPDs $H_v^q$ and $E_v^q$ are constrained by the flavor
form factors through the sum rules \eqref{eq:SumRule}.  In order to
exploit all available information we will directly fit the GPDs to the
data on the Sachs form factors.  Nevertheless, we find it useful to
extract also the experimental values of the flavor form factors, which may
be regarded as the form factor set that is most suitable for an
interpretation in terms of quark and antiquark densities.  This extraction
will allow us to verify whether the different GPDs obtained in a global
fit satisfy the sum rules \eqref{eq:SumRule} with uniform quality.  It can
also be used to test simple functional forms for the flavor form factors.
This may be of use for lattice QCD studies, which typically require a
parameterization of simulation results for the extrapolation to the
physical values of parameters.  Finally, form factor fits are not
guaranteed to reproduce local structures in the data, because the
flexibility of the assumed parameterization might be insufficient for the
structure in question.  By contrast, the directly extracted flavor form
factors retain local structures present in the data.

\begin{figure*}
\begin{center}
\includegraphics[width=0.45\textwidth,%
viewport=70 50 395 295]{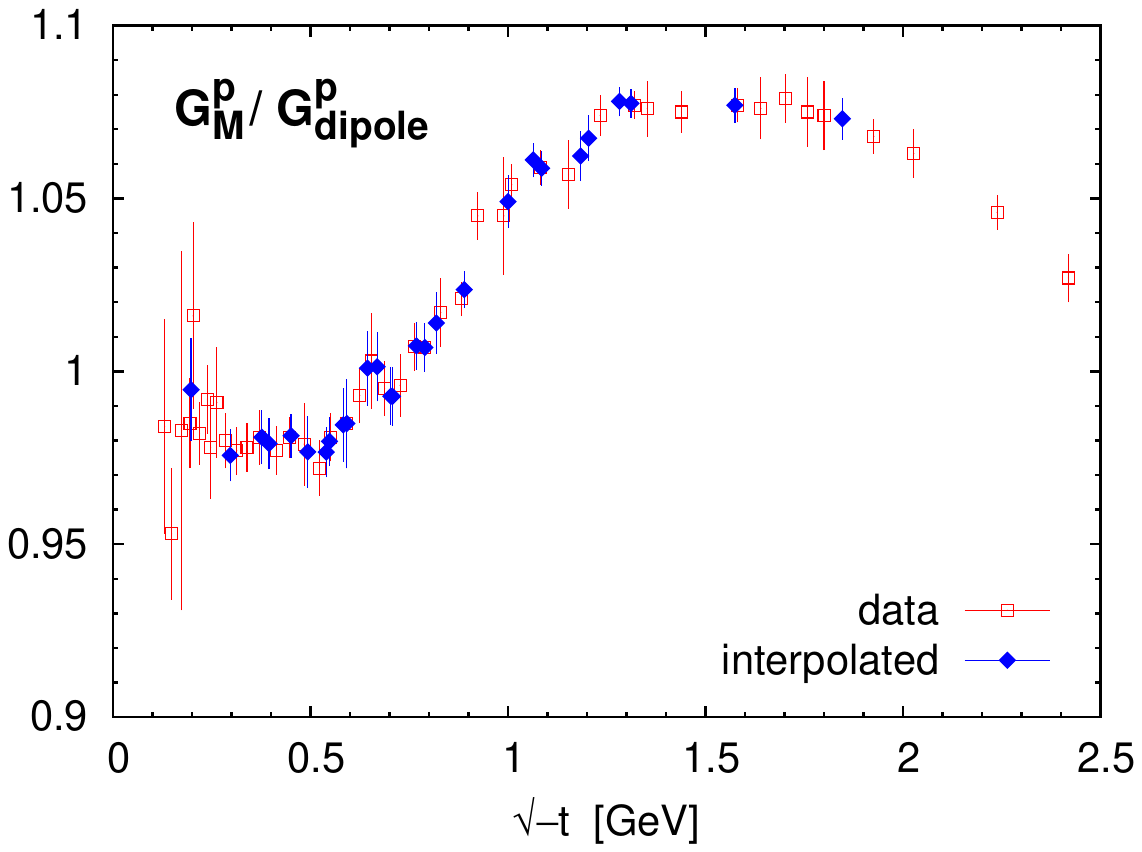}
\hspace{1.5em}
\includegraphics[width=0.45\textwidth,%
viewport=70 50 395 295]{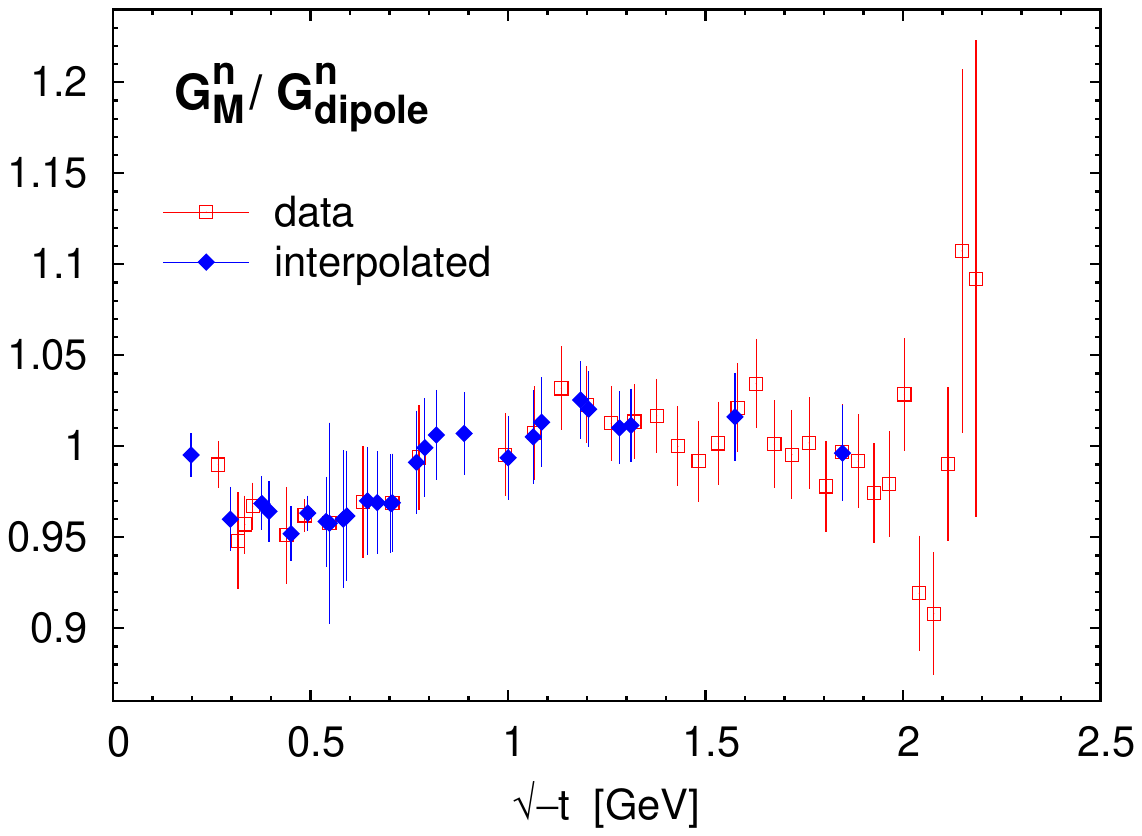} \\[1em]

\includegraphics[width=0.45\textwidth,%
viewport=70 50 395 295]{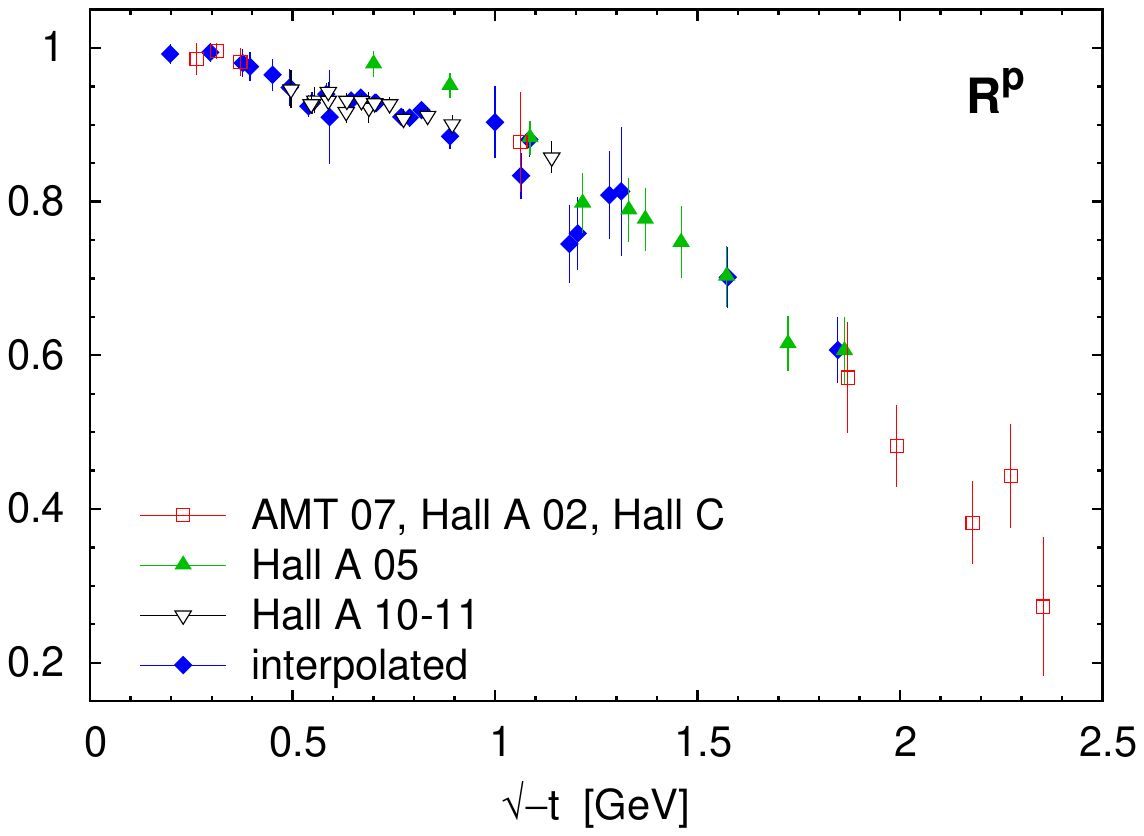}
\hspace{1.5em}
\includegraphics[width=0.45\textwidth,%
viewport=70 50 395 295]{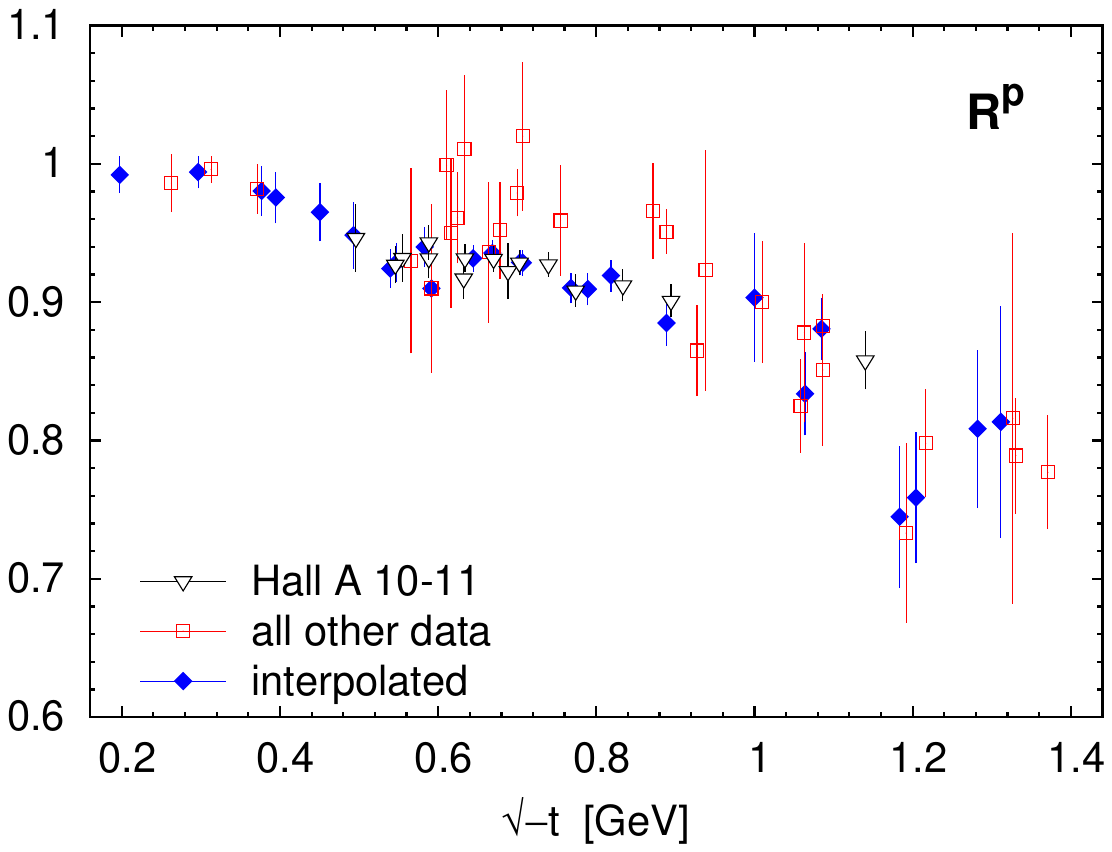} \\[1em]

\includegraphics[width=0.45\textwidth,%
viewport=70 50 395 295]{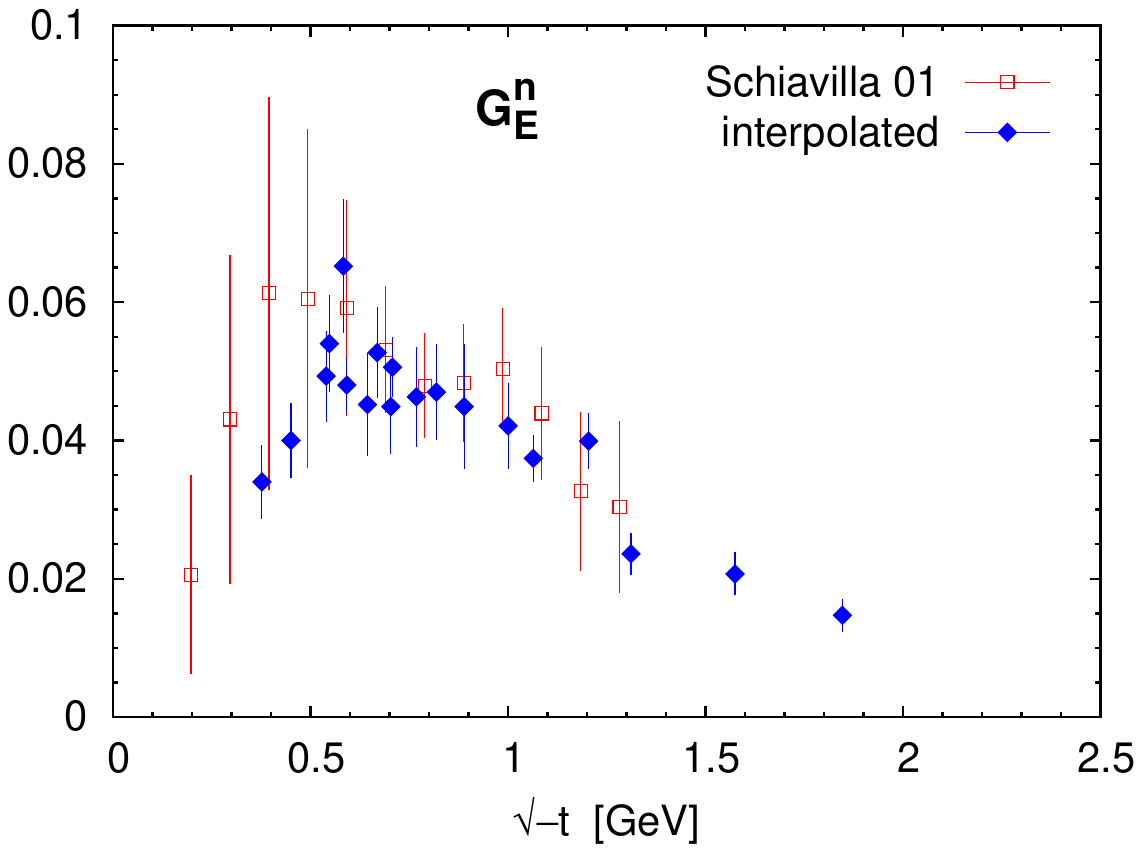}
\hspace{1.5em}
\includegraphics[width=0.45\textwidth,%
viewport=70 50 395 295]{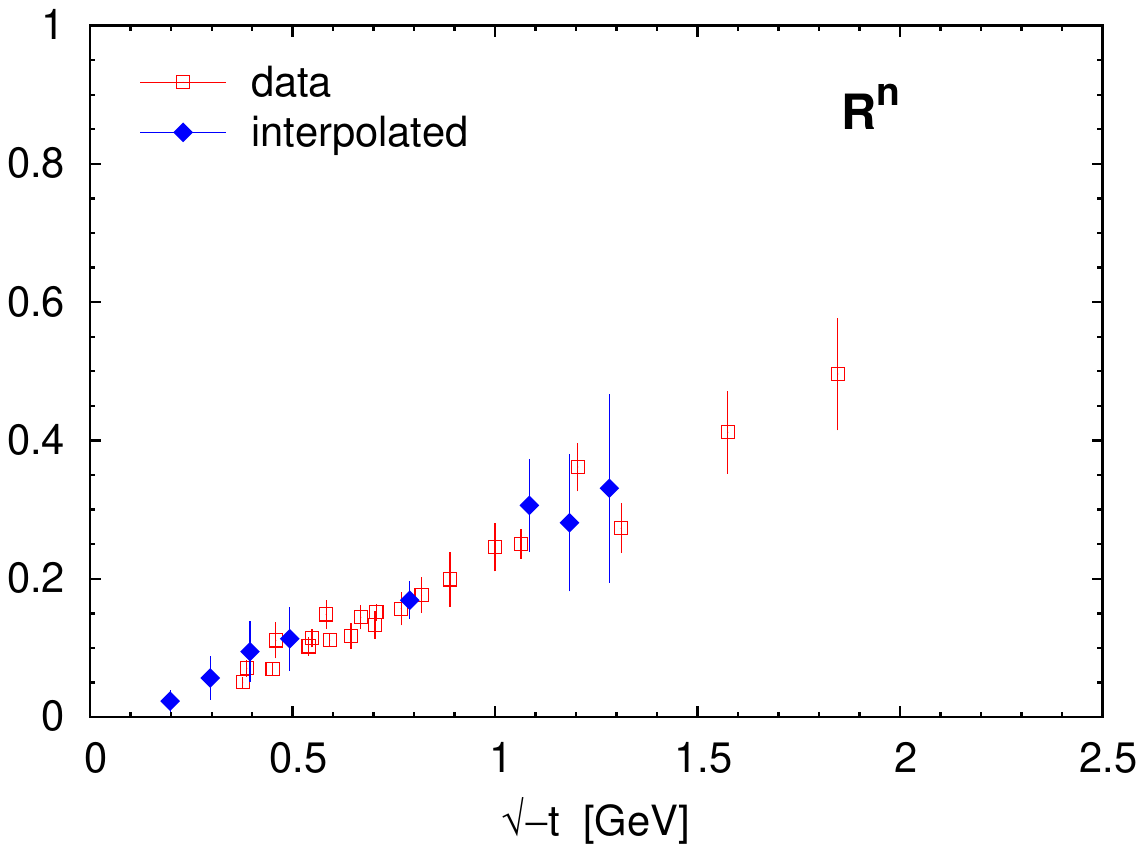} \\[1em]
\end{center}
\caption{\label{fig:data-vs-interpol} Comparison of interpolated (solid
  diamonds) and measured (open symbols) Sachs form factors and form factor
  ratios.  For $R^p$ the Hall A data of 2005 \cite{Punjabi:2005wq} and of
  2010--11 \cite{Paolone:2010qc,Ron:2011rd,Zhan:2011ji} have been used in
  the interpolation.  References for the other data are given in
  \protect\sect{sec:data-sel}.}
\end{figure*}

For the determination of the flavor form factors, the four Sachs form
factors are needed at the same set of $t$ values. This is in general not
the case for the available data, since each measurement has its own
criteria for a suitable choice of bins in $t$.  We therefore need an
interpolation procedure.  As basic set of $t$ values we choose those of
the electric neutron form factor $G_E^n$ and of the associated form factor
ratio $R^n$.  Compared to the other form factors or ratios, these
observables are measured at the smallest number of $t$ values.  In this
way we avoid having more interpolated data points for the other
observables than are actually measured.  We omit a few data points for
$G_E^n$, because there are $R^n$ measurements at exactly or approximately
the same value of $t$.  We thus obtain a basic set of 27 values of $-t$
between $0.039$ and $3.41 \gev^2$.  For these values we interpolate
$G_M^p$, $G_M^{n\phantom{p}}$ and $R^p$ using cubic splines.

In \fig{fig:data-vs-interpol} the resulting interpolated data points are
compared to the measured ones.  The data on the magnetic from factors of
proton and neutron are well represented by the interpolated set.  For the
ratio $R^p$ of electric and magnetic proton form factors, we again see the
tensions between different measurements for $\sqrt{-t}$ between $0.6$ and
$1.0 \gev$.  Following our discussion in \sect{sec:data-sel}, we use the
precise recent Hall A data \cite{Paolone:2010qc,Ron:2011rd,Zhan:2011ji}
for interpolating $R^p$, as well as the older Hall A measurement
\cite{Punjabi:2005wq} for the high-$t$ region.  We finally check the
compatibility between the original data on $G_E^{n}$ and on $R^n$, which
are both part of the interpolated form factor set.  To this end we take
our interpolated values of $G_M^n$ and compute $R^n$ when $G_E^n$ is
measured and vice versa.  As can be seen in \fig{fig:data-vs-interpol},
there is good agreement within the uncertainties.

\begin{figure}

\vspace{0.8em}

\begin{center}
\includegraphics[width=0.49\textwidth,%
viewport=70 50 395 295]{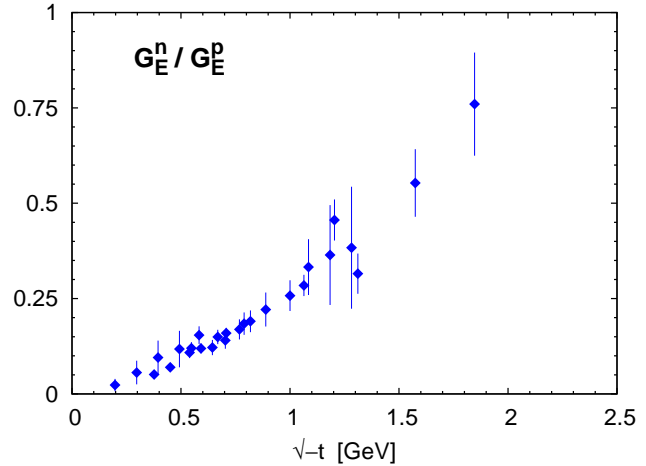}
\end{center}
\caption{\label{fig:genogep} The ratio $G_E^n / G_E^p$, evaluated with our
  set of interpolated data.}
\end{figure}

An interesting observation can be made from the set of interpolated data.
We see in \fig{fig:genogep} that $G_E^{n\phantom{p}} \ll G_E^{p}$ at low
$t$, as one may expect, whereas with increasing $\sqrt{-t}$ the ratio
$G_E^{n\phantom{p}} / G_E^p$ increases and becomes of order 1 as
$\sqrt{-t}$ approaches $2 \gev$.  This finding may be taken as a hint at a
zero crossing in the isovector combination $G_E^p - G_E^{n\phantom{p}}$ of
electric form factors.

\begin{figure*}[t]
\begin{center}
\includegraphics[width=0.45\textwidth,%
viewport=128 284 584 634]{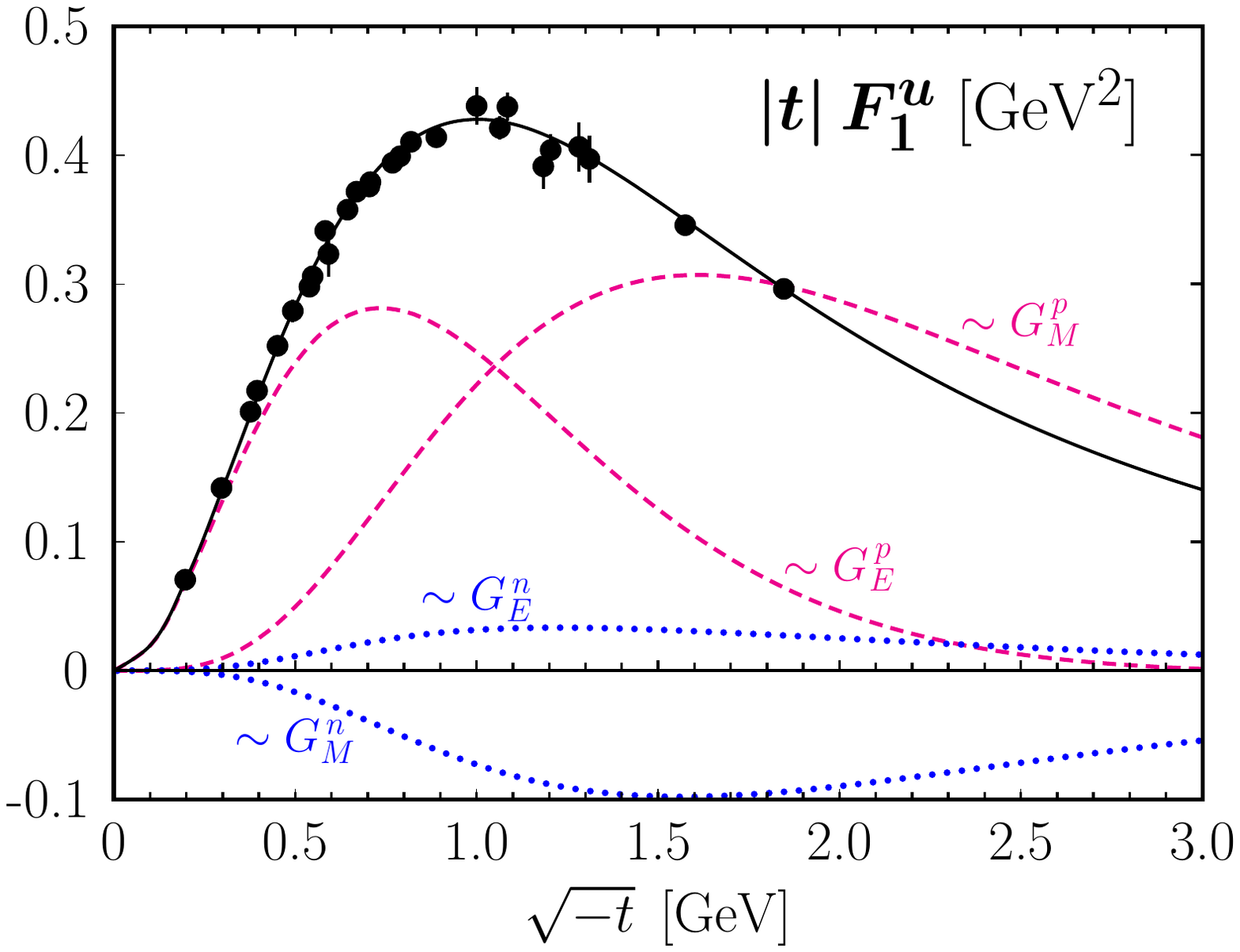}
\hspace{1.5em}
\includegraphics[width=0.46\textwidth,%
viewport=120 298 584 643]{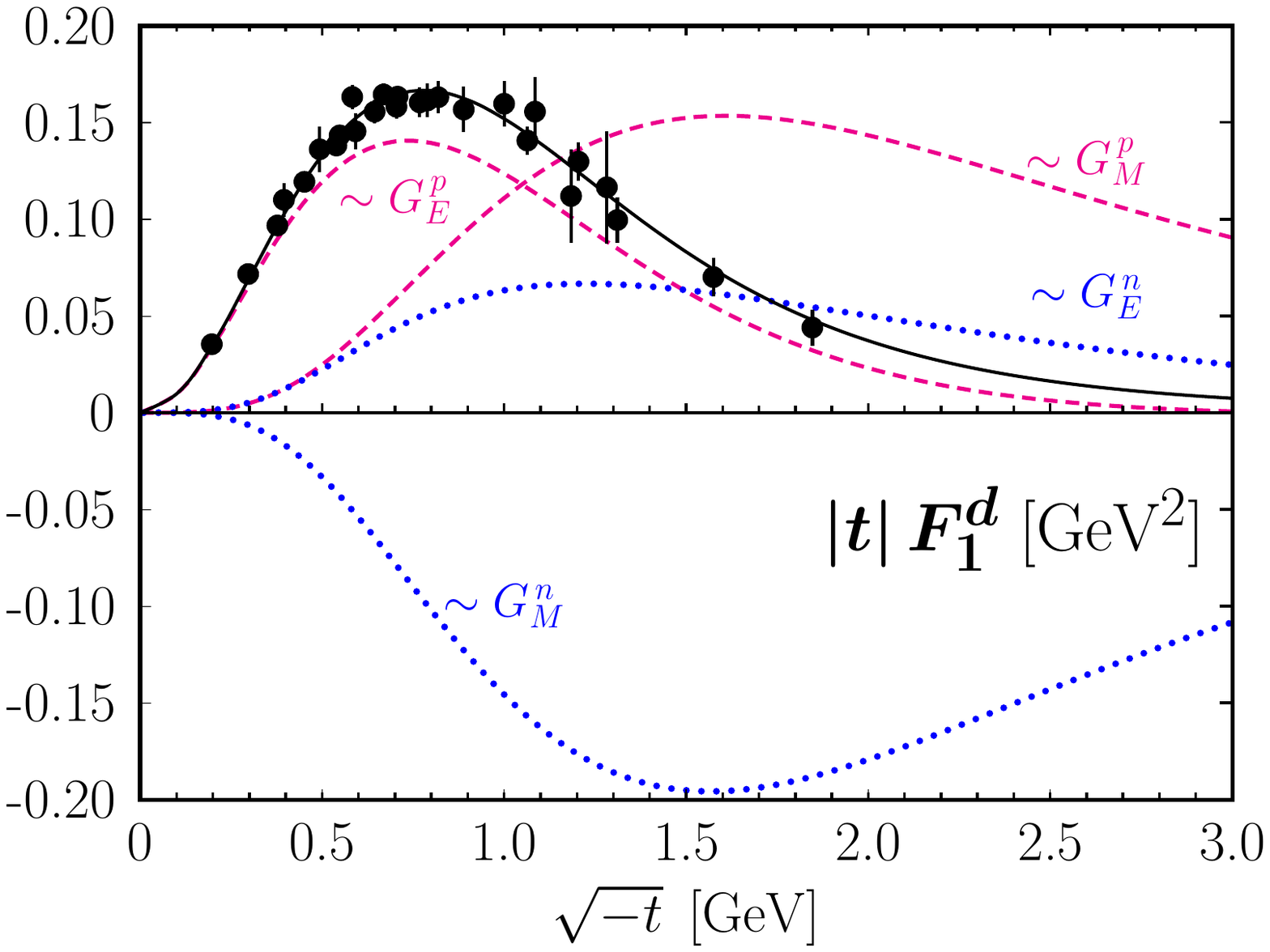}\\[2em]

\includegraphics[width=0.45\textwidth,%
viewport=128 284 584 634]{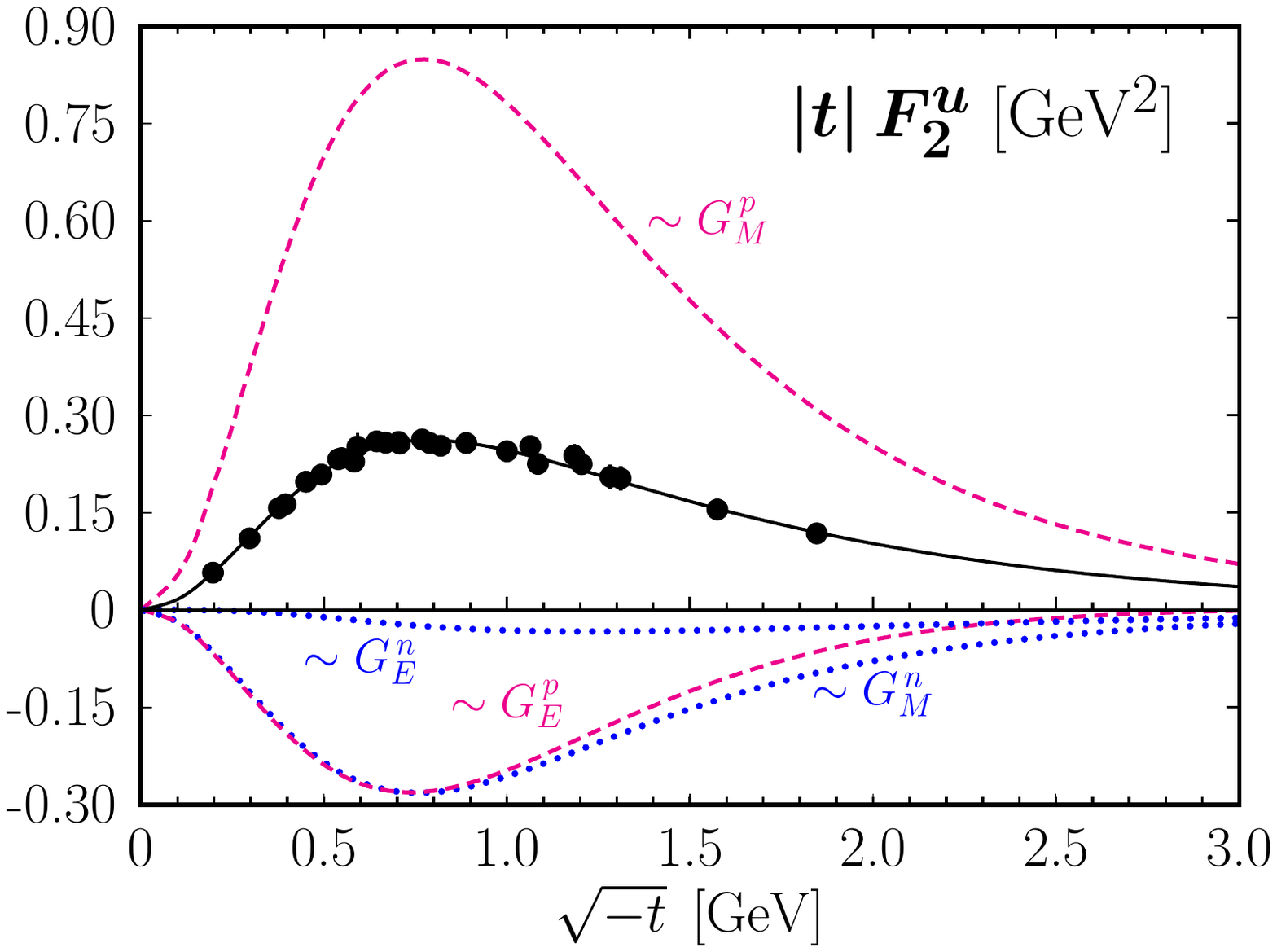}
\hspace{1.5em}
\includegraphics[width=0.45\textwidth,%
viewport=128 284 584 634]{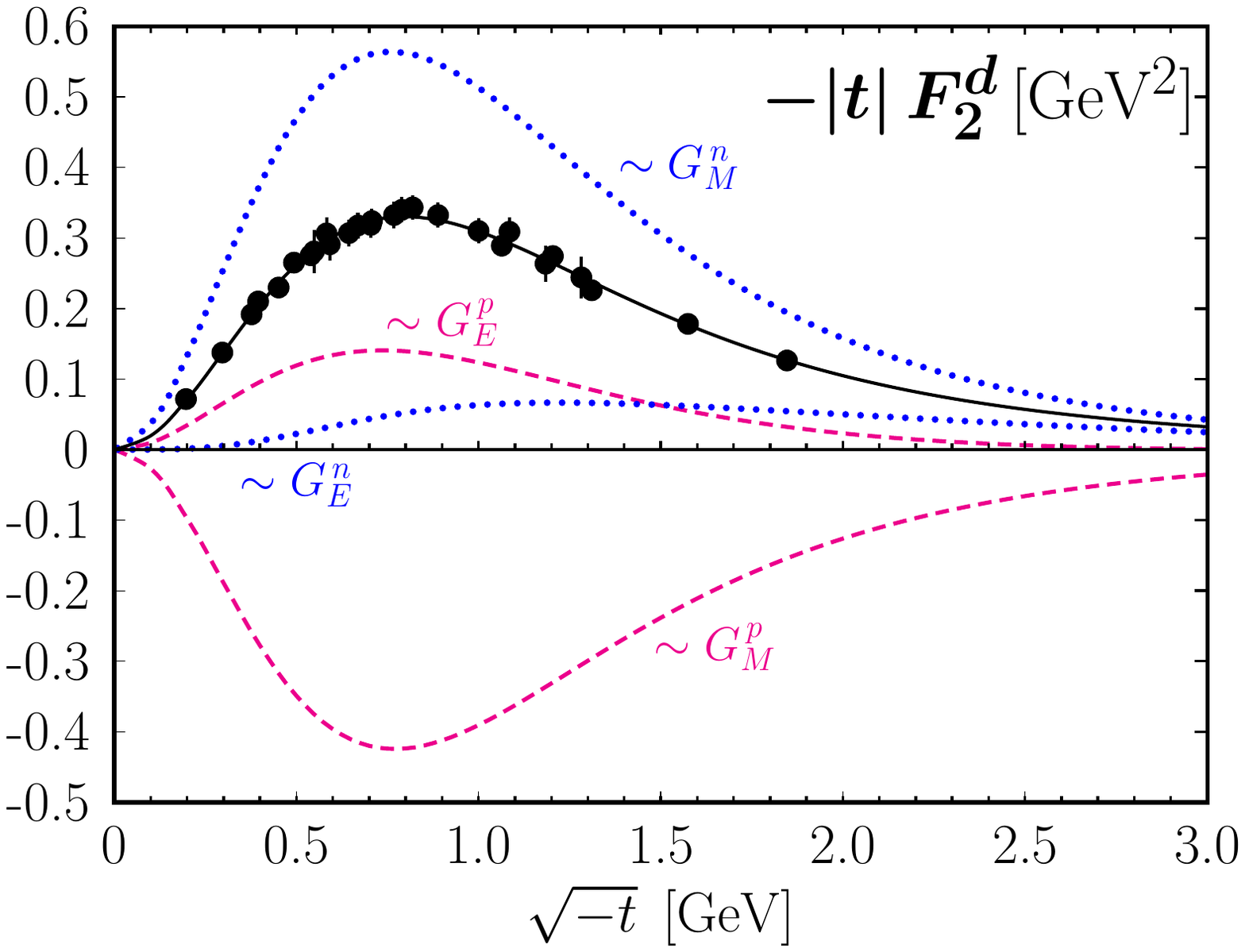}

\caption{\label{fig:flavorFF} The flavor combinations of the Dirac form
  factors, scaled by $|t|$ and plotted versus $\sqrt{-t}$. The individual
  contributions from the proton (neutron) Sachs form factors, computed
  from the power-law fit of \protect\sect{sec:power-law-fit}, are shown as
  magenta dashed (blue dotted) lines. The solid lines represent the sum of
  the individual contributions.}
\end{center}
\end{figure*}

From the interpolated Sachs form factors we compute the flavor form
factors using \eqref{Sachs-Dirac} and \eqref{inverse-dec}, where the
strangeness contributions are neglected.\footnote{%
  If one does not wish to neglect these contributions, one can simply
  re-interpret the form factors $F_i^{u}$ and $F_i^{d}$ of this section as
  $F_i^{u-s}$ and $F_i^{d-s}$, as discussed in \sect{sec:basics}.}
The errors of the flavor form factors are evaluated from those of the
uncorrelated Sachs form factors or form factor ratios with the help of
Gaussian error propagation.\footnote{%
  Strictly speaking, the values of $R^p$ are not independent of the
  results for $G_M^p$ in the global analysis \cite{Arrington:2007ux},
  because that analysis used several of the $R^p$ measurements contained
  in our data set.  Since we have no possibility to take this correlation
  into account, we treat the data for $R^p$ and $G_M^p$ as uncorrelated.}
In the same manner we can compute any combination of the flavor form
factors, including its errors.  Our results for $F_i^u$ and $F_i^d$ are
compiled in \app{app:tables} and shown in \fig{fig:flavorFF}.  One can see
that $F_2^{d}$ is negative while the other three form factors are
positive, and that
\begin{align}
F_1^{d} & < F_1^{u} \,,
& 
F_2^{d} & \simeq - F_2^{u} \,.
\end{align}
These properties reflect the normalization of the form factors at $t=0$,
\begin{align}
F_1^{u}(0) &= 2\,,       
&
F_1^{d}(0) &= 1\,,
\nonumber\\
F_2^{u}(0) &= \kappa_u\,,
&
F_2^{d}(0) &= \kappa_d\,,
\end{align}
where $\kappa_q$ is the contribution of quarks with flavor $q$ to the
anomalous magnetic moment of the proton ($\kappa_u=1.67$ and
$\kappa_d=-2.03$ if strangeness is neglected).  One may also notice that
with increasing $-t$ the ratio $F_1^{d}/F_1^{u}$ decreases, while
$F_2^{d}/F_2^{u}$ stays rather flat (see \fig{fig:ratios} below).  The
decrease of $F_1^{d}/F_1^{u}$ was already visible in the flavor form
factors extracted from our earlier work, see
\cite{Kroll:2006hx,Kroll:2007wn}, and its relation to the large-$x$
behavior of the parton densities was pointed out in \cite{DFJK4}.  We will
take up the discussion of the flavor form factors in \sects{sec:def-fit}
and \ref{sec:xmin-xmax}.

In \fig{fig:flavorFF} we also show the flavor form factors obtained with
the global power-law fit described in \sect{sec:power-law-fit}, neglecting
again the strangeness contributions.  Evidently, this fit describes the
flavor form factors very well.  The results of the fit are also decomposed
into the contributions from the individual Sachs form factors to a given
flavor form factor.  We notice that in the cases of the $d$-quark form
factors strong cancellations among the various contributions occur,
whereas $F_1^{u}$ is dominated by the proton form factors, with the
neutron form factors providing only small contributions.  These
observations tell us that the interpolated $F_1^{u}$ data are quite stable
against modifications of the Sachs form factors.  The $d$-quark form
factors, on the other hand, are rather sensitive to modifications of the
data.  One may also notice that the contribution to $F_1^{u}$ from
$G_E^{n}$, which is only measured up to $-t=3.41 \gev^2$, is very small.
Therefore, the power-law fit as well as the GPD fits described in
\sect{sec:gpd-fits} are still reliable at $-t$ above $3.4 \gev^2$.  By
contrast, the contributions of $G_E^n$ to the other flavor form factors,
in particular those for $d$-quarks, are rather important and, hence, our
fits to these form factors are to be taken with due caution at large $t$.
  
Recently, two extractions of the flavor form factors have been published
\cite{Qattan:2012zf,Cates:2011pz}. The results for $F_1^{u}$ and $F_1^{d}$
obtained in \cite{Qattan:2012zf} are rather similar to ours.  The same
holds for $F_2^{u}$, with slightly larger differences.  For $-F_2^{d}$,
however, the values in \cite{Qattan:2012zf} are systematically larger than
ours by up to $10\%$ for $-t$ around $1 \gev^2$ and smaller by a similar
amount for $-t$ around $3.2 \gev^2$. To understand this discrepancy, we
note that \cite{Qattan:2012zf} uses a fit for $G_M^{n}$ that includes the
new Hall B data \cite{Lachniet:2008qf} but also the older data sets Anklin
98 and Kubon 02 \cite{Anklin:1998ae, Kubon:2001rj} that we partially
discard for the reasons discussed in \sect{sec:neutron-data}.  As can be
seen in \fig{fig:flavorFF}, the impact of $G_M^{n}$ is largest on the
$d$-quark form factors; for $F_1^{d}$ the difference between our values
and those of \cite{Qattan:2012zf} is less visible because the overall
errors on it are larger than for $F_2^{d}$.

Comparing our flavor form factors with those extracted in
\cite{Cates:2011pz}, we find again rather similar values for $F_1^{u}$ and
$F_1^{d}$.  For $F_2^{u}$, however, the points of \cite{Cates:2011pz} are
below ours by up to $15\%$ for $-t$ around $1 \gev^2$, whereas for
$-F_2^{d}$ they are above ours by a similar amount in the same $t$ range.
These discrepancies are significant at the scale of the quoted errors in
the two analyses.  Their main origin is that \cite{Cates:2011pz} uses the
Kelly parameterization \cite{Kelly:2004hm} for $G_M^{n}$ and for the
proton form factors.  Apart from being in conflict with the Hall B data on
$G_M^{n}$, this parameterization closely follows the older data on $R^p$
and thus lies significantly above the recent Hall A results
\cite{Paolone:2010qc,Ron:2011rd,Zhan:2011ji} for $-t$ below $1 \gev^2$.
In \cite{Cates:2011pz} only the errors of $G_E^{n}$ and $R^n$ are taken
into account while the uncertainties of the Kelly parameterization of the
other Sachs form factors are ignored. Therefore the errors of the flavor
form factors quoted in \cite{Cates:2011pz} are smaller (for $F_2^{u}$ and
$F_2^{d}$ even substantially smaller) than ours.
We finally stress that, in contrast with \cite{Qattan:2012zf} and
\cite{Cates:2011pz}, we use only data points and no parameterizations to
construct our interpolated data set.

%%%%%%%%%%%%%%%%%%%%%%%%%%%%%%%%%%%%

\subsection{Simple fits to form factors}

Lattice QCD studies often require simple parameterizations of form factors
for the purpose of interpolation and extrapolation.  A representation like
the one used in our global power-law fit \eqref{global-power-fit-fct} can
normally not be used for this purpose, because it involves $4$ parameters
per form factor.
We can use our interpolated data set to investigate which functional forms
are suitable to describe the electromagnetic form factors.  In addition to
the dipole form \eqref{general-dipole}, we will consider the general power
law
\begin{align}
\label{general-power}
F(t) = \frac{F(0)}{\bigl( 1 - t / M^2_{p} )^p}
\end{align}
and as a special case also a tripole form, i.e.\ \eqref{general-power}
with $p=3$.  As another extension of the dipole parameterization we
consider the product of two single poles,
\begin{align}
\label{two-poles-power}
F(t) = \frac{F(0)}{\bigl(1 - t / M^2_{a} )\, \bigl( 1 - t / M^2_{b})} \,.
\end{align}
A summary of our fits is given in \tab{tab:simple-fits}.  For each form
factor we fit 27 data points in the range $0.039 \gev^2 \le -t \le 3.41
\gev^2$.

\begin{table*}
\renewcommand{\arraystretch}{1.2}
\begin{center}
\begin{tabular}{ccccc} \hline
form factor & \multicolumn{2}{c}{------ dipole ------}
            & \multicolumn{2}{c}{------ two poles ------} \\
            & accuracy & $\chi^2_{\text{min}}$
            & accuracy & $\chi^2_{\text{min}}$ \\
\hline
$G_M^{u-d}$ & $-4.5\%~\text{to}~2.5\%$ & 197.9
            & $-4.5\%~\text{to}~2.0\%$ & 49.1 \\
$G_M^{u+d}$ & $-8.0\%~\text{to}~5.5\%$ & 15.3
            & $-4.5\%~\text{to}~4.0\%$ & 6.7 \\
$G_E^{u+d}$ & $-5\%~\text{to}~13\%$    & 20.8
            & $-9\%~\text{to}~9\%$     & 13.5 \\
$G_E^{u-d}$ & $-75\%~\text{to}~10\%$   & 63.1 & & \\		
\hline
\end{tabular}

\vspace{2em}

\begin{tabular}{ccccc} \hline
form factor & \multicolumn{2}{c}{------ dipole ------}
            & \multicolumn{2}{c}{------ two poles ------} \\
            & accuracy & $\chi^2_{\text{min}}$
            & accuracy & $\chi^2_{\text{min}}$ \\
\hline
$F_1^{u+d}$ & $-5.0\%~\text{to}~9.5\%$ & 36.4
            & $-5.0\%~\text{to}~8.0\%$ & 27.3 \\
$F_1^{u-d}$ & $-12\%~\text{to}~29\%$   & 946
            & $-5.5\%~\text{to}~6.5\%$ & 69.2 \\
$F_2^{u-d}$ & $-19\%~\text{to}~4\%$    & 433  & & \\
$F_2^{u+d}$ & $-7.5\%~\text{to}~5.5\%$ & 24.2 & & \\
\hline
\end{tabular}

\vspace{2em}

\begin{tabular}{ccccc} \hline
form factor & \multicolumn{2}{c}{------ power law ------}
            & \multicolumn{2}{c}{------ tripole ------} \\
            & accuracy & $\chi^2_{\text{min}}$
            & accuracy & $\chi^2_{\text{min}}$ \\
\hline
$F_1^{u+d}$ & $-4.5\%~\text{to}~9.0\%$ & 30.9 & & \\
$F_1^{u-d}$ & $-6\%~\text{to}~8\%$     & 105  & & \\
$F_2^{u-d}$ & $-6.5\%~\text{to}~5.5\%$ & 113
            & $-7\%~\text{to}~28\%$    & 1037 \\
$F_2^{u+d}$ & & & $-6.5\%~\text{to}~5.5\%$ & 17.2 \\
\hline
\end{tabular}

\vspace{2em}

\begin{tabular}{cccccc} \hline
form factor & \multicolumn{2}{c}{------ dipole ------}
            & \multicolumn{3}{c}{--------- power law ---------} \\
            & accuracy & $\chi^2_{\text{min}}$
            & accuracy & $\chi^2_{\text{min}}$ & $p$ \\
\hline
$F_1^u$	    & $-7\%~\text{to}~14\%$     & 138
            & $-4.0\%~\text{to}~5.5\%$  & 17.9 & $1.13 \pm 0.03$ \\
$F_1^d$	    & $-50\%~\text{to}~10\%$    & 54.0
            & $-19\%~\text{to}~16\%$    & 13.0 & $2.81 \pm 0.18$ \\
$F_2^u$	    & $-12.5\%~\text{to}~8.5\%$ & 15.0
            & $-5.0\%~\text{to}~10.0\%$ & 9.6  & $2.16 \pm 0.07$ \\
$F_2^d$	    & $-24\%~\text{to}~8\%$	& 30.0
            & $-8.5\%~\text{to}~7.5\%$  & 11.2 & $2.38 \pm 0.10$ \\
\hline
\end{tabular}
\end{center}
\caption{\label{tab:simple-fits} Quality of fits to our interpolated form
  factor data with different simple functional forms.  For each form
  factor, 27 data points are fitted.  As ``accuracy'' we define the
  smallest and largest value of $(1 - \text{fit}/\text{data})$.  
  If a field is left empty, no stable fit of acceptable quality could be
  found.  The column ``power law'' refers to the form
  \protect\eqref{general-power} and the column ``two poles'' to
  \protect\eqref{two-poles-power}.  The last column in the last table
  gives the value of $p$ in the power-law fit.}
\end{table*}

\begin{figure}
\begin{center}
\includegraphics[width=0.45\textwidth,%
viewport=70 50 395 295]{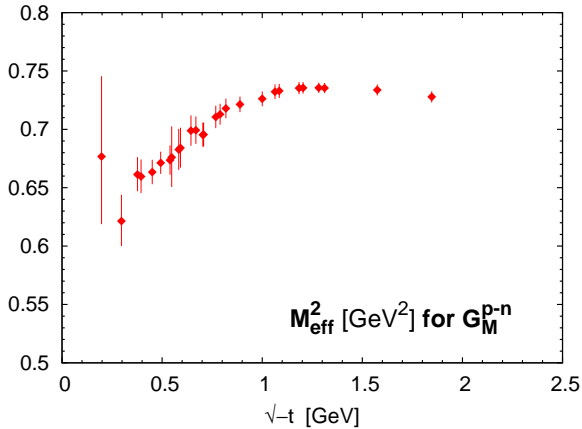}
\end{center}
\caption{\label{fig:GM3-dipmass} Effective dipole mass
  \protect\eqref{eff-dip-mass} for the isovector combination $G_M^{p} -
  G_M^{n\protect\phantom{p}}$ of magnetic form factors.}
\end{figure}

Let us first discuss the Sachs form factors in the isospin basis, i.e.\
the combinations $G_{M}^{u \pm d}$ and $G_{E}^{u \pm d}$.  They form the
most natural basis from the point of view of $t$-channel exchanges and of
analytic continuation to positive $t$, as is e.g.\ discussed in
section~4.2 of \cite{Diehl:2003ny}.  A dipole or a product of two poles
may hence appear as a natural candidate to describe these form factors.
We find that a dipole form describes our interpolated values for
$G_M^{u-d}$ with an accuracy better than 5\%.  On the scale of the errors
on $G_M^{u-d}$, this is however a poor description, and the associated
$\chi^2_{\text{min}}$ is very high as we see in \tab{tab:simple-fits}.
That $G_M^{u-d}$ cannot be described by a dipole within its uncertainties
is confirmed by a plot of the corresponding effective dipole mass (see
\sect{sec:dipole-par}) in \fig{fig:GM3-dipmass}.  A product of two poles
gives a better description, but still with a $\chi^2_{\text{min}}$ almost
twice as big as the number of data points.

The isosinglet combinations $G_{M}^{u+d}$ and $G_{E}^{u+d}$ have larger
error bars than $G_{M}^{u-d}$, and we find that they can be described
reasonably well by a dipole.  In both cases, a product of two poles, which
has one more free fit parameter, gives an even better description.  As for
$G_{E}^{u-d}$, we find that a dipole form gives a reasonable description
up to $\sqrt{-t} = 1.3 \gev$ but badly fails to reproduce the two data
points with $\sqrt{-t} = 1.57 \gev$ and $1.85 \gev$, as is reflected in
the accuracy and $\chi^2_{\text{min}}$ given in \tab{tab:simple-fits}.
This is not too surprising if we recall that this form factor may have a
zero around $\sqrt{-t} = 2 \gev$, as we observed when commenting on
\fig{fig:genogep}.  A two-parameter fit of $G_{E}^{u-d}$ to the product of
two poles is unstable.

Turning now to the Dirac and Pauli form factors in the isospin basis, we
find that $F_1^{u+d}$ is reasonably well described by a dipole, and even
better by the product of two poles.  By contrast, a dipole form is unable
to describe $F_1^{u-d}$.  The product of two poles permits a description
with 6\% accuracy, which has however still a large $\chi^2_{\text{min}}$.
A general power-law fit is slightly worse for this form factor, which has
very small errors.  The isotripolet Pauli form factor $F_2^{u-d}$ is also
known with high precision.  We find a very poor description by a dipole
fit, and an even worse one by a tripole form, whose asymptotic $t$
behavior corresponds to the dimensional counting prediction for
$F_2^{u-d}$ (and is obviously irrelevant for the $t$ range in question).
A general power law does better in comparison, with an accuracy of about
6\% but still a bad $\chi^2_{\text{min}}$.  The isoscalar combination
$F_2^{u+d}$ has rather large errors and is equally well described by a
dipole and a tripole fit.  A similar situation has often been found in
lattice studies, when the errors on the simulation did not permit to draw
strong conclusions on the $t$ dependence of certain form factors.

Turning finally to the flavor basis, we observe that a dipole fit gives a
rather poor description for $F_1^u$, $F_1^d$ and $F_2^d$.  The general
power law \eqref{general-power} works however very well in all three
cases.  It works also very well for $F_2^u$, where the fitted power is
close to 2, so that a dipole fit is adequate in this exceptional case.

In conclusion, we have not found a simple ``one fits all'' functional form
that would describe either the Sachs or the Dirac and Pauli form factors
in the isospin basis.  The only ansatz that gives a uniformly good
description of all form factor data for $-t$ up to $3.4 \gev^2$ is to fit
the flavor form factors to the general power law \eqref{general-power}
with its two free parameters.

%% file: fitgpd.tex
\section{GPD fit}
\label{sec:gpd-fits}

\subsection{Fit ansatz and positivity}
\label{sec:fit-ansatz}

Let us now briefly describe our ansatz for the GPDs, which will be used in
our fits to the form factors.  We largely follow the approach of our
earlier work \cite{DFJK4} and refer to it for a more detailed motivation
and discussion.  The main feature of our ansatz is an exponential $t$
behavior
\begin{align}
  \label{HE-ansatz}
H_v^q(x,t) &=   q_v(x)\, \exp\bigl[ t \ms f_q(x) \bigr] \,,
\nonumber \\
E_v^q(x,t) &= e_v^q(x)\, \exp\bigl[ t \ms g_q(x) \bigr] \,,
\end{align}
with an $x$ dependent width specified by the profile functions $f_q(x)$
and $g_q(x)$.  For polarized quarks we assume
\begin{align}
  \label{eq:htilde}
\widetilde{H}^q_v(x,t)
   &= \Delta q_v(x)\, \exp\bigl[ t \ms f_q(x) \bigr] \,,
\end{align}
where for lack of better knowledge we take the same $t$ dependence as for
$H_v^q$ (see \sect{sec:axialFF}).  The above forms refer to a definite
renormalization scale $\mu$, which we take equal to $2 \gev$ unless stated
otherwise.

An intuitive interpretation of GPDs at zero skewness can be given in
impact parameter space, where we define
\begin{align}
  \label{HE-impact}
q_v(x, \vbs^2)   &= \int \frac{\dd^2 \boldsymbol{\Delta}}{(2\pi)^2}\;
   e^{- i \vbs \boldsymbol{\Delta}}\,
   H_v^q(x,-\boldsymbol{\Delta}^2) \,,
\nonumber \\
e_v^q(x, \vbs^2) &= \int \frac{\dd^2 \boldsymbol{\Delta}}{(2\pi)^2}\;
   e^{- i \vbs \boldsymbol{\Delta}}\,
   E_v^q(x,-\boldsymbol{\Delta}^2) \,.
\end{align}
$q_v(x, \vbs^2)$ is difference of densities for quarks and antiquarks with
momentum fraction $x$ at a transverse distance $\vbs$ from the proton
center, with both the parton and the proton being unpolarized.  The
average impact parameter associated with this density difference is
\begin{align}
  \label{imp-par}
\langle \vbs^2 \rangle^q_x
  &= \frac{\int \dd^2\vbs\; \vbs^2\, q_v(x, \vbs^2)}{%
           \int \dd^2\vbs\; q_v(x, \vbs^2)}
   = 4 f_q(x) \,.
\end{align}
The corresponding density difference for longitudinally polarized partons
is
\begin{align}
 \label{Htilde-impact}
\Delta q_v(x, \vbs^2)
  &= \int \frac{\dd^2 \boldsymbol{\Delta}}{(2\pi)^2}\;
   e^{- i \vbs \boldsymbol{\Delta}}\,
   \widetilde{H}_v^q(x,-\boldsymbol{\Delta}^2) \,,
\end{align}
whereas for unpolarized partons in a proton polarized along the $x$-axis
one has \cite{Burkardt:2002hr}
\begin{align}
  \label{lateral-shift}
q^X_v(x, \vbs) &= q^{}_v(x, \vbs^2)
 - \frac{b^y}{m}\,
   \frac{\partial}{\partial\vbs^2}\, e_v^q(x, \vbs^2) \,.
\end{align}
Transverse polarization of the proton thus induces a sideways shift in the
distribution of partons.  The average amount of this shift is
\begin{align}
\langle b^y \rangle^q_x
  &= \frac{\int \dd^2\vbs\; b^y\, q^X_v(x, \vbs^2)}{%
           \int \dd^2\vbs\; q^X_v(x, \vbs^2)}
   = \frac{1}{2 m}\, \frac{e_v^q(x)}{q_v(x)} \,.
\end{align}

The interpretation as a density difference requires $q^X_v(x, \vbs) \ge 0$
when the antiquark contribution is negligible.  This implies a bound on
$\partial/(\partial \vbs^2)\, e_v^q(x, \vbs^2)$, which becomes even
stronger if we include information about polarized quarks.  In a region
where antiquarks can be neglected, we then have \cite{burkardt03}
\begin{multline}
  \label{bound-bspace}
\frac{\vbs^2}{m^2} \biggl[ \frac{\partial}{\partial \vbs^2}\,
  e_v^q(x, \vbs^2) \biggr]^2
\\[0.15em]
\le \Bigl\{ \bigl[ q(x, \vbs^2) \bigr]^2
         -  \bigl[ \Delta q(x, \vbs^2) \bigr]^2 \Bigr\} \,.
\end{multline}
With \eqref{HE-ansatz} and \eqref{eq:htilde}, the validity of
\eqref{bound-bspace} for all $\vbs$ at a given $x$ is equivalent to
\cite{DFJK4}
\begin{multline}
  \label{bound-final}
\frac{\bigl[ e_v^q(x) \bigr]{}^2}{8 m^2}  \le \exp(1)\, 
   \biggl[ \ms \frac{g_q(x)}{f_q(x)} \ms \biggr]^3\,
   \bigl[ f_q(x) - g_q(x) \bigr]
\\[0.15em]
\times \Bigl\{
  \bigl[ q_v(x) \bigr]^2 - \bigl[ \Delta q_v(x) \bigr]^2 \Bigr\} \,.
\end{multline}
Note that \eqref{bound-final} requires strict inequality $g_q(x) < f_q(x)$
of the profile functions, except for values of $x$ where $e_v^q(x) = 0$.
As we will see, the positivity bound on $e_v^q(x)$ severely constrains our
fits.

As a word of caution, we note that the density interpretation and the
associated positivity conditions do not strictly hold in QCD.  This is
because the ultraviolet renormalization that makes the GPDs well defined
and leads to their $\mu$ dependence involves subtractions that can in
principle invalidate positivity.  We nevertheless require the above
conditions to hold, so that a density interpretation is possible for the
results of our fits.  The technical implementation of the positivity
conditions is discussed in the next subsection.

We now specify our ansatz \eqref{HE-ansatz} and \eqref{eq:htilde}.  For
$q_v(x)$ we take a selection of up-to-date parton densities, which is
discussed in \sect{sec:pdfs}, and for the polarized densities $\Delta
q_v(x)$ we choose the recent determination \cite{DSSV}.  All these
distributions are defined with NLO evolution and evaluated at scale $\mu =
2 \gev$.  For $e_v^q(x)$, we make the ansatz
\begin{equation}
  \label{e-ansatz}
e_v^q(x) = \kappa_q\ms N_q\, x^{-\alpha_q} (1-x)^{\beta_q}\,
           \bigl(1 + \gamma_q \sqrt{x} \ms\bigr) \,,
\end{equation}
which has proven to work well for the parameterization of ordinary parton
densities.  The normalization factor $N_q$ ensures that
\begin{align}
\int_0^1 \dd x\, e_v^q(x) &= \kappa_q
\end{align}
as required by \eqref{eq:SumRule}.  The values of $\kappa_u$ and
$\kappa_d$ are computed from the measured magnetic moments of proton and
neutron and the assumed value for the strangeness contribution $\kappa_s =
\mu_s$.  The $\gamma_q$ dependent term in \eqref{e-ansatz} is new compared
with \cite{DFJK4} and significantly improves our fits, as we will discuss
in \sects{sec:def-fit} and \ref{sec:var-fits}.

For the profile functions $f_q(x)$ and $g_q(x)$ we assume the form
\cite{DFJK4}
\begin{align}
  \label{profile-ansatz}
f_q(x) &= \alpha_q'\ms (1-x)^3 \log(1/x) + B_q\ms (1-x)^3 
\nonumber \\
 & \quad + A_q\ms x (1-x)^2 \,,
\nonumber \\[0.1em]
g_q(x) &= \alpha_q'\ms (1-x)^3 \log(1/x) + D_q\ms (1-x)^3 
\nonumber \\
 & \quad + C_q\ms x (1-x)^2 \,.
\end{align}
The parameters $\alpha'_q$, $B_q$ and $D_q$ control the small-$x$ behavior
of these functions, whereas their behavior at large $x$ is controlled by
$A_q$ and $C_q$.  The factors of $(1-x)$ in $f_q(x)$ ensure that $\langle
\vbs^2 \rangle_x^q \sim (1-x)^2$ in the limit $x\to 1$, which follows from
requiring a finite transverse size of the proton in that limit (see
\sect{sec:distance}).

At small $x$, the $\log(1/x)$ term in $g_q(x)$ gives a $t$-dependent
contribution to the power behavior $E_v^q(x,t) \sim x^{- (\alpha_q^{} + t
  \alpha'_q)}$, in accordance with simple Regge phenomenology.  A
corresponding statement holds for $H_v^q(x,t)$ if the forward densities
$q_v(x)$ have a power behavior at small $x$.

%%%%%%%%%%%%%%%%%%%%%%%%%%%%%%%%%%%%%%%

\subsection{Selection of parton densities}
\label{sec:pdfs}

\begin{table*}[t]
\renewcommand{\arraystretch}{1.2}
\begin{center}
\begin{tabular}{lccccc} \hline
PDF & ref.     & $\alpha_u^{\text{eff}}$ & $\alpha_d^{\text{eff}}$
               & $\beta_u^{\text{eff}}$ & $\beta_d^{\text{eff}}$ \\
\hline
ABM 11 $n_f=4$ & \cite{Alekhin:2012ig} & $0.33$ & $0.34$ & $3.5$ & $5.0$ \\
CT 10          & \cite{Lai:2010vv} & $0.39$ & $0.42$ & $3.4$ & $3.6$~~$2.7$ \\
GJR 08 VF      & \cite{Gluck:2007ck}   & $0.54$ & $0.53$ & $3.8$ & $4.9$ \\
HERAPDF 1.5    & \cite{HERAPDF1.5}     & $0.33$ & $0.33$ & $4.2$ & $4.8$ \\
MSTW 2008      & \cite{Martin:2009iq}  & $0.54$ & $0.29$ & $3.5$ & $5.9$ \\
NNPDF 2.2      & \cite{Ball:2011gg}    & $0.43$ & $0.28$ & $3.5$ & $4.5$ \\
\hline
\end{tabular}
\end{center}
\caption{\label{tab:pdfs} The PDF sets used in our analysis and the
  effective powers for their behavior at small and large $x$ as defined in
  \protect\eqref{q-fits}.  All PDFs are evaluated at scale $\mu = 2\gev$.
  The two values of $\beta_d^{\text{eff}}$ for the CT 10 set correspond to
  separate fits in the ranges $0.65<x<0.75$ and $0.75<x<0.85$.}
\end{table*}

An important feature of our ansatz \eqref{HE-ansatz} is that for the
forward limit of $H_v^q(x,t)$ we can use the valence quark densities
obtained in global PDF analyses.  As we shall see, current PDF
determinations exhibit notable differences for the valence quark densities
$u_v(x)$ and $d_v(x)$, especially in the regions of small or large $x$.
Since these regions are of some importance in the sum rules
\eqref{eq:SumRule} (see \sect{sec:xmin-xmax}), we have explored several
recent PDF determinations in our fits.  They are all defined at NLO,
evaluated at $\mu = 2 \gev$, and listed in \tab{tab:pdfs}.  The numerical
values for all parton densities have been obtained with the routines of
the LHAPDF interface \cite{LHAPDF}, version 5.8.8.  From now on we will
denote the PDF sets only by the acronyms of their authors (ABM, CT, etc.)
since we only use one set from each group.

As already mentioned in the previous subsection, the power behavior at
small $x$, which is suggested by simple Regge phenomenology, is an
important ingredient of the physical motivation for our GPD ansatz.  We
shall see in \sect{sec:xmin-xmax} that the power behavior of the GPDs at
large $x$ is closely related to the large-$t$ behavior of the form
factors.  We have therefore taken a closer look at the behavior of the
parton densities at small and at large $x$.  To quantify this behavior, we
fit the PDFs to effective power laws
\begin{align}
  \label{q-fits}
q_v(x) &\sim\, x^{-\alpha_q^{\text{eff}}} 
 & \text{for}~ & 10^{-3} < x < 10^{-2}
\nonumber \\
         &\sim\, (1-x)^{\ms\beta_q^{\text{eff}}}
 & \text{for}~ & 0.65 < x < 0.85 \,.
\end{align}
The effective powers we obtain are given in \tab{tab:pdfs}.  In all cases
the accuracy of the fit is better than 5\%.  Our choice of $x$ intervals
in the fits comes from the requirements that they should be of importance
in the sum rules \eqref{eq:SumRule} and that the PDFs should indeed follow
an approximate power law behavior.  If we fit the small-$x$ behavior for
$10^{-4} < x < 10^{-3}$ then the effective powers decrease by $0.0$ to
$0.02$, with the following exceptions: for CT $\alpha_d^{\text{eff}}$
decreases by $0.04$, for NNPDF $\alpha_d^{\text{eff}}$ decreases by
$0.05$, and for MSTW $\alpha_d^{\text{eff}}$ decreases by $0.08$ and
$\alpha_u^{\text{eff}}$ by $0.10$

\begin{figure*}[t]
\begin{center}
\includegraphics[width=0.98\textwidth,%
viewport=0 0 610 190]{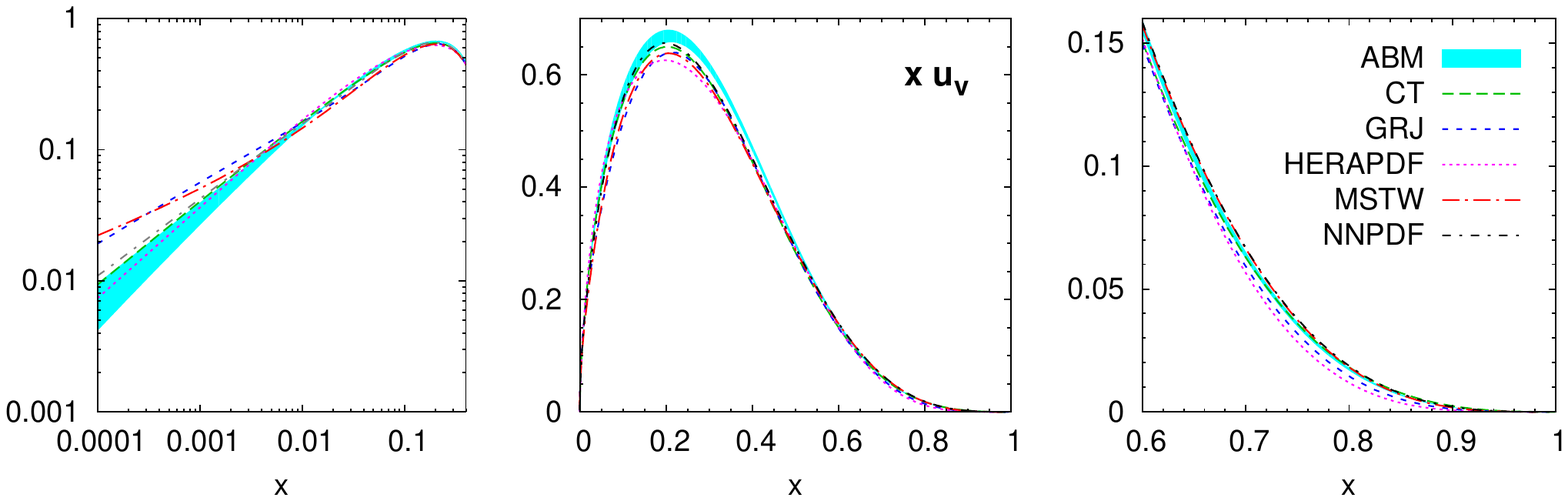} \\[0.5em]
\includegraphics[width=0.98\textwidth,%
viewport=0 0 610 190]{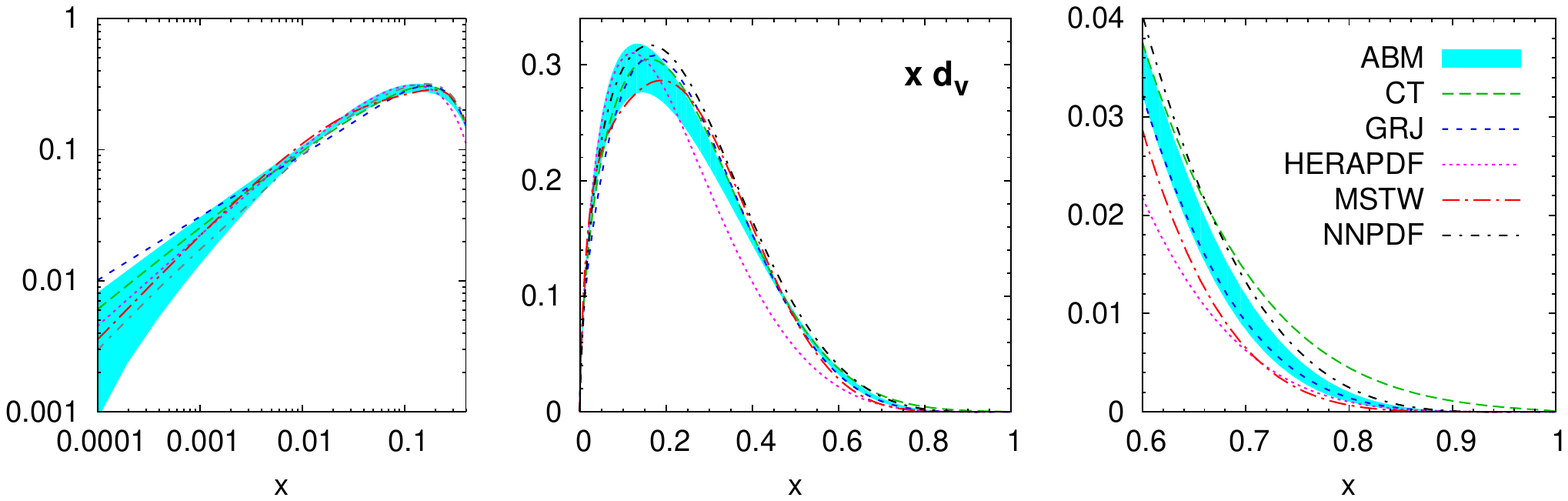}
\end{center}
\caption{\label{fig:pdfs} The valence densities of $u$ quarks (top) and
  $d$ quarks (bottom) for the PDF sets specified in \tab{tab:pdfs}.}
\end{figure*}

As it is evident from the effective powers in \tab{tab:pdfs}, there is a
significant variation between different PDF sets, and we must conclude
that neither the small-$x$ nor the large-$x$ behavior of the valence quark
distributions is presently known with certainty.  This is also seen in the
plots of the different PDFs in \fig{fig:pdfs}.  We further observe that
the spread between different PDF sets is larger than the error bands of
the individual PDFs, which is not surprising since the latter reflect
parametric errors of the PDF fits but not systematic uncertainties of the
fitting procedure.  Rather than the errors on a given PDF set, we will
hence use the variation from different sets in order to estimate the
uncertainty induced on our analysis of GPDs and form factors.

Returning to \tab{tab:pdfs}, we observe that in the MSTW and NNPDF sets
there is a significant difference in the effective powers for $u$ and $d$
quarks at small $x$, which we consider to be in tension with usual Regge
phenomenology.  At large $x$, we find that the $d$ quark distribution of
the CT analysis has a rather peculiar behavior: it does not follow an
approximate power behavior over a significant range in $x$, and for $x
\gsim 0.8$ it is significantly larger than in any other PDF set (including
the earlier set CTEQ 6M \cite{Pumplin:2002vw} of the same collaboration).

For our default GPD fit we have chosen the ABM set.  With the GJR set, our
fit gives a value of $\alpha$ in the parameterization of $e_v^q(x)$ that
we consider at the limit of what is plausible from Regge phenomenology,
and our fit with the HERAPDF set turns out to have a relatively large
$\chi^2$ compared with the other PDFs.

Let us note that for extremely large $x$, say above 0.9, some (although
not all) PDFs obtained with the LHAPDF interface behave unexpectedly,
either by not being monotonic in $x$ or by becoming negative.  Since the
PDFs are extremely small in that region, this may be due to numerical
instabilities, which would explain that the problems occur mostly for
$d_v(x)$.  We have however not investigated this issue further.  In any
case, the results of global PDF fits for such large $x$ must be regarded
as extrapolations, since there is no experimental data constraining them
in that region.  This is illustrated by the fact that already at $x=0.9$
the PDF sets we have chosen exhibit a spread of almost a factor 4 for
$u_v(x)$ and an even larger spread for $d_v(x)$.  Luckily, the uncertainty
in this $x$ region does not affect our analysis in a significant way,
since such values of $x$ do not dominate the integrals over GPDs that give
form factors in the $t$ range where there is data.  We will quantify this
in \sect{sec:xmin-xmax}.

Let us now specify how we implement the conditions from positivity in our
GPD fits.  We require the validity of \eqref{bound-final} for $x > 0.15$,
since for smaller $x$ antiquarks are found to become important in the
forward parton densities.
At large $x$, where it is plausible to neglect antiquarks, two types of
problem complicate using the positivity bound.
\begin{itemize}
\item As described in the previous paragraph, there are numerical
  instabilities in the parton densities.  To stay away from this region we
  do not enforce the bound \eqref{bound-final} for $x > 0.9$.  We do
  however require $g_q(x) < f_q(x)$ in that region, since this bound is
  independent of the PDFs.
\item With the polarized PDFs of \cite{DSSV} and some of the unpolarized
  PDF sets, the requirement $|\Delta q_v(x)| < q_v(x)$ is not satisfied
  for very large $x$.  Specifically, we find $\Delta u_v(x) > u_v(x)$ for
  the sets GJR ($x > 0.78$) and HERAPDF ($x > 0.72$), whereas $- \Delta
  d_v(x) >  d_v(x)$ for ABM ($x > 0.83$), GJR ($x > 0.82$), HERAPDF ($x >
  0.69$), MSTW ($x > 0.71$) and NNPDF ($x > 0.91$).  This is not too
  surprising, since positivity in conjunction with those unpolarized PDFs
  was not enforced in the DSSV extraction \cite{DSSV}, and it does not
  represent a physics problem given the overall uncertainties on the
  parton densities in the relevant $x$ region.

  To circumvent this problem at a technical level, we set the polarized
  PDFs to zero in the bound \eqref{bound-final} for those values of $x$
  where $|\Delta q_v(x)| > 0.9\ms q_v(x)$.
\end{itemize}

%%%%%%%%%%%%%%%%%%%%%%%%%%%%%%%%%%%%%

\subsection{The default fit}
\label{sec:def-fit}

We now have all elements needed for our fit of GPDs to the form factor
data.  In this section we discuss what we consider our best fit, where in
particular we set the strangeness form factors $F_1^s$ and $F_2^s$ to
zero.  Different variants of this fit are presented in the next section.
Compared with our analysis \cite{DFJK4} we have significantly extended the
form factor data used in the fit.  This allows for a larger number of free
parameters and for a simultaneous fit of $H_v^q$ and $E_v^q$ (with
$q=u,d$) to all data.  In \cite{DFJK4} we had instead first computed Dirac
and Pauli form factors from the experimental results and then performed
separate fits of $H_v^q$ and $E_v^q$.

Nevertheless, we cannot allow all parameters in \eqref{e-ansatz} and
\eqref{profile-ansatz} to vary independently.  Fits with too many free
parameters do not only give very large parameter uncertainties but also
tend to violate the positivity constraints.
To limit the number of free parameters, we appeal to Regge phenomenology.
Assuming that the small-$x$ behavior of both $E_v^q$ and $H_v^q$ is
dominated by the leading meson trajectories, namely those of the $\rho$
and the $\omega$, and that those trajectories are degenerate, we obtain
that to first approximation the small-$x$ powers $(\alpha_q^{} + t
\alpha'_q)$ in $E_v^q$ should be equal for $u$ and $d$ quarks.  They
should also be equal to the analogous powers in $H_v^q$, which is why we
have taken the same parameters $\alpha'_q$ in the profile functions $f_q$
and $g_q$, see \eqref{profile-ansatz}.  The $t$ dependent part of the
small-$x$ power for $H_v^q$ is not a fit parameter but a property of the
PDFs in our ansatz, and for the ABM parameterization we take in our
default fit the effective powers $\alpha^q_{\text{eff}}$ are nearly equal
for $u$ and $d$ quarks.

We emphasize that the equality of the small-$x$ powers in $E_v^u$,
$E_v^d$, $H_v^u$ and $H_v^d$ is not expected to be exact.  The meson
trajectories are not exactly degenerate and Regge phenomenology allows for
subleading trajectories and Regge cuts, which can all lead to different
effective powers when the GPDs are approximated by a single power law in a
certain $x$ range.  Moreover, one cannot expect to find the literal values
of meson trajectories in the small-$x$ behavior of GPDs or PDFs.  Indeed,
parton distributions are subject to scale evolution, which changes the
effective $x$ powers, although rather slowly as long as a single power law
gives a good description over a large interval in $x$.

Our approach is thus to take equal small-$x$ powers as long as our fit
does not require otherwise.  We thus set
\begin{equation}
\alpha_u = \alpha_d = \alpha
\end{equation}
for the small-$x$ powers in the forward functions $e_v^u$ and $e_v^d$,
given that we do not find significantly better fits if we allow $\alpha_u$
and $\alpha_d$ to differ.

On the other hand we find that, together with the abundant and precise low
$t$ data on several form factors, the very precise value
\eqref{r2nE-value} of the squared neutron charge radius $r^2_{nE}$
requires some deviation from full degeneracy of the small-$x$ powers.
With $\alpha'_u = \alpha'_d$ we obtain poor partial $\chi^2$ values for
$r^2_{nE}$ and for the $R^n$ data.  A good description can however be
obtained with a slight isospin breaking of the form $\alpha'_u >
\alpha'_{\smash{d}}$.  We therefore take
\begin{align}
  \label{def-fit-delta-alphap}
\alpha'_u - \alpha'_d &= 0.1 \gev^{-2}
\end{align}
in our fits.  A further increase of $\alpha'_u - \alpha'_d$ yields an even
better $\chi^2$, but we do not consider a large isospin splitting to be
physically motivated.

Let us now discuss the parameters $\gamma_u$ and $\gamma_d$ in the forward
functions $e_v^u(x)$ and $e_v^d(x)$, which are new compared with our study
\cite{DFJK4}.  We have varied these parameters independently in steps of
$1$ and selected the values
\begin{align}
  \label{def-fit-gammaq}
\gamma_u &= 4 \,,
&
\gamma_d &= 0
\end{align}
for our default fit.  Compared with setting both $\gamma_u$ and $\gamma_d$
to zero, this decreases the overall $\chi^2$ by about 30 units, with most
significant improvements for $G_M^p$ and $G_M^n$.  Taking $\gamma_u$ even
larger improves the $\chi^2$ only slightly and leads to larger fitted
values of $\alpha$.  We thus retain \eqref{def-fit-gammaq} as a compromise
between a good $\chi^2$ and parameters in line with Regge phenomenology.
If we take either $\gamma_u$ or $\gamma_d$ as free fit parameters, then
they have large errors of order $50\%$ while $\chi^2$ improves only
moderately compared with \eqref{def-fit-gammaq}.

\begin{table*}[t]
\renewcommand{\arraystretch}{1.15}
\begin{center}
\begin{tabular}{|cc|ccccccccc|}
\hline
 &      & \multicolumn{9}{c|}{$\beta_u$} \\
 &      & 4.65 & 4.70 & 4.75 & 4.80 & 4.85 & 4.90 & 4.95 & 5.00 & 5.05 \\
\hline
 & 5.20 &  ---  &  ---  &  ---  &  ---  &  ---  & 221.9 & 222.0 & \underline{\colorbox{yellow}{222.2}} & 222.3 \\ 
 & 5.25   & \underline{\colorbox{yellow}{221.2}} & 221.4 & 221.6 & 221.7 & 221.9 & 222.0 & \underline{\colorbox{yellow}{222.2}} & 222.3 & 222.4 \\
 & 5.30   & 221.3 & 221.5 & 221.7 & 221.9 & 222.0 & \underline{\colorbox{yellow}{222.2}} & 222.3 & 222.4 & 222.5 \\
 & 5.35   & 221.4 & 221.6 & 221.8 & 222.0 & 222.1 & 222.3 & 222.4 & 222.5 & 222.6 \\
 & 5.40   & 221.6 & 221.8 & 221.9 & 222.1 & 222.3 & 222.4 & 222.5 & 222.7 & 222.8 \\
$\beta_d$ & 5.45   & 221.7 & 221.9 & 222.1 & \underline{\colorbox{yellow}{222.2}} & 222.4 & 222.5 & 222.6 & 222.8 & 222.9 \\
 & 5.50   & 221.8 & 222.0 & \underline{\colorbox{yellow}{222.2}} & 222.3 & 222.5 & 222.6 & 222.8 & 222.9 & 223.0 \\
 & 5.55   & 221.9 & 222.1 & 222.3 & 222.4 & 222.6 & 222.7 & 222.9 & 223.0 & 223.1 \\
 & 5.60   & 222.0 & \underline{\colorbox{yellow}{222.2}} & 222.4 & 222.6 & 222.7 & 222.8 & 223.0 & 223.1 & 223.2 \\
 & 5.65   & 222.1 & 222.3 & 222.5 & 222.7 & 222.8 & 222.9 & 223.1 & 223.2 & 223.3 \\
 & 5.70   & \underline{\colorbox{yellow}{222.2}} & 222.4 & 222.6 & 222.8 & 222.9 & 223.1 & 223.2 & 223.3 & 223.4 \\
 & 5.75   & 222.3 & 222.5 & 222.7 & 222.9 & 223.0 & 223.1 & 223.3 & 223.4 & 223.5 \\
\hline
\end{tabular}
\end{center}
\caption{\label{tab:chi2-matrix} Values of $\chi^2$ for 
  fits with the same setting as our default fit (ABM 1).  Underlined
  values indicate the overall minimum and the one-sigma contour.  
  Positivity is violated in the fits with $\beta_d = 5.2$ and $\beta_u
  \le 4.85$, as well as for all fits with $\beta_d \le 5.15$ or $\beta_u
  \le 4.6$.}
\end{table*}

We observe that all our fits are very significantly influenced by the
positivity constraints, in accord with our previous analysis \cite{DFJK4}.
In particular, we find that if we leave $\beta_u$ and $\beta_d$ (or one of
them) as free parameters, then their fitted values are very low ($\beta_u
= 3.5$ and $\beta_d = 1.6$ if both are left free).  This badly violates
the positivity bound \eqref{bound-final}.
To circumvent this problem, we perform fits for fixed values of $\beta_u$
and $\beta_d$ on a grid with step size $0.05$.  The resulting values of
$\chi^2$ are given in \tab{tab:chi2-matrix}.  We see that the minimum of
$\chi^2$ in the $(\beta_u, \beta_d)$ plane occurs at the boundary of the
region allowed by our positivity conditions, with values
\begin{align}
  \label{def-fit-beta}
\beta_u &= 4.65 \,,
&
\beta_d = 5.25 \,.
\end{align}
We label this fit as ABM 1 and refer to it as our ``default fit'' in the
remainder of this work.  It yields the parameters
\begin{align}
  \label{eq:def-fit-pars}
\alpha'_d  &= (0.861 \pm 0.026) \gev^{-2} \,,
\nonumber \\[0.1em]
\alpha     &= 0.603 \pm 0.020
\end{align}
and the values in \tab{tab:def-fit-pars}.  We consider
\eqref{eq:def-fit-pars} to be consistent with expectations from Regge
phenomenology.  The value of $\alpha$ is somewhat large compared with the
intercepts of the leading meson trajectories, but we deem it still
acceptable.  We shall further discuss the parameters $\alpha$ and
$\beta_u$, $\beta_d$ in \sect{sec:var-fits}.

The parametric uncertainties in the fit are of reasonable size, with the
most precisely determined parameters being $\alpha$, $\alpha'_d$ (and
hence $\alpha'_u$), $A_u$ and $A_d$.  In general there are strong
correlations between all parameters.  We can obtain an uncertainty
estimate on $\beta_u$ and $\beta_d$ using the criterion $\Delta \chi^2 =
1$.  This gives an asymmetric contour in the $(\beta_u, \beta_d)$ plane,
which is marked in \tab{tab:chi2-values}.  We see that the large-$x$
powers of $e_v^q$ are determined with reasonable although not very high
precision.

\begin{table}[t]
\renewcommand{\arraystretch}{1.2}
\begin{center}
\begin{tabular}{crr} \hline
$q$   & \multicolumn{1}{c}{$u$} & \multicolumn{1}{c}{$d$} \\
\hline
$A_q$ &   $1.264 \pm 0.050$
      &   $4.198 \pm 0.231$ \\
$B_q$ &   $0.545 \pm 0.062$
      &   $0.206 \pm 0.073$ \\
$C_q$ &   $1.187 \pm 0.087$
      &   $3.106 \pm 0.249$ \\
$D_q$ &   $0.333 \pm 0.065$
      & $- 0.635 \pm 0.076$ \\
\hline
\end{tabular}
\end{center}
\caption{\label{tab:def-fit-pars} Parameters of the profile functions
  $f_q$ and $g_q$ (see \protect\eqref{profile-ansatz}) in our default fit.
  All quantities have the unit $\gev^{-2}$.}
\end{table}

When computing GPDs we use standard linear error propagation for the free
parameters in the fit; the necessary matrix is given in
\app{app:matrices}.  We use a simplified procedure to propagate the errors
on $\beta_u$ and $\beta_d$ into GPDs and observables that are derived from
them.  Namely, we compute the quantity in question for each of the 7 fits
that have $\Delta \chi^2 = 1$ w.r.t.\ the default fit in
\tab{tab:chi2-matrix} and compare the result with the value obtained with
the default fit.  If the difference is larger than what is obtained with
standard error propagation for the free parameters in the default fit, we
retain it for the error estimate.  A more elaborate procedure would also
scan the fits with $\Delta \chi^2 < 1$ in the $(\beta_u, \beta_d)$ plane,
but we refrain from doing so for the sake of simplicity.  We find that the
uncertainties due to the variation of $\beta_q$ are not important for the
electromagnetic form factors, whose error is therefore given by standard
error propagation for the free fit parameters.  The variation of $\beta_q$
is however relevant for the second and third $x$-moments of GPDs
(including the angular momentum sum rule), for the shift $s_q(x)$ to be
discussed in \sect{sec:distance} and for the model estimate of the Sivers
distributions in \sect{sec:lensing}.  In those cases, changes in
$e_v^q(x)$ due to the variation of $\beta_q$ are not compensated by
changes in the profile functions $g_q(x)$, in constrast to what happens
for the electromagnetic form factors to which the GPDs are fitted.

We note that our default fit corresponds to a local but not the global
minimum of $\chi^2$ in the $(\beta_u, \beta_d)$ plane.  The global minimum
is also assumed at the boundary of the $\beta_q$ values allowed by
positivity, namely at the largest possible $\beta_u$.  We find it at
$\beta_u = 18.85$ and $\beta_d = 5.25$.  It has $\chi^2 = 209.8$, which is
about 11 units smaller than in our default fit.  At the global minimum we
find that $g_u(x)$ is nearly zero at $x \sim 0.3$.  Both because of this
and because of the very high value of $\beta_u$, we do not consider this
fit to be physically plausible and retain the local minimum of $\chi^2$ at
the low end of the allowed $\beta_q$ values instead.

Let us now see how well our default fit describes the data.  Its overall
$\chi^2$ is $221.2$ for 178 data points.  Partial $\chi^2$ values are
given in \tab{tab:chi2-values} below, and plots in \fig{fig:Sachs}.  For
the sake of discussion we split the $G_M^p$ data into a low-$t$ and a
high-$t$ sample, with their boundary being at $-t = 10 \gev^2$.  We find
that the fit provides a very good description of the neutron form factors,
i.e.\ of $G_M^n$, $G_E^n$ and $R^n$, and also of $G_M^p$ at large $t$.
The description is still fair but less optimal for $G_M^p$ at low $t$ and
for $R^p$, with partial $\chi^2$ values of about 1.7 and 1.5 per data
point, respectively.  As shown in \fig{fig:Sachs} the fit slightly
overshoots the very precise $G_M^p$ data for $\sqrt{-t} < 0.7 \gev$, and
for $R^p$ it fails to reproduce the fine details of the data with
$\sqrt{-t} < 1 \gev$.  We find the same two shortcomings in all of our
alternative GPD fits to the same data set, as we will see in
\sect{sec:var-fits}.  It is not impossible to describe these data more
precisely, as is demonstrated by the power-law fit of
\sect{sec:power-law-fit}, which achieves partial $\chi^2$ values of $0.6$
and $1$ per data point for $G_M^p$ at low $t$ and for $R^p$, respectively.
We conclude that our GPD fits, with the significant constraints in
parameter space imposed by positivity, reach their limits of precision
here, and we must postpone a resolution of this shortcoming to the future.
We note that the low-$t$ data for $R^p$ are still subject to experimental
debate, as discussed in \sect{sec:proton-data}.

\begin{figure*}
\begin{center}
\includegraphics[width=0.47\textwidth,%
viewport=110 255 580 600]{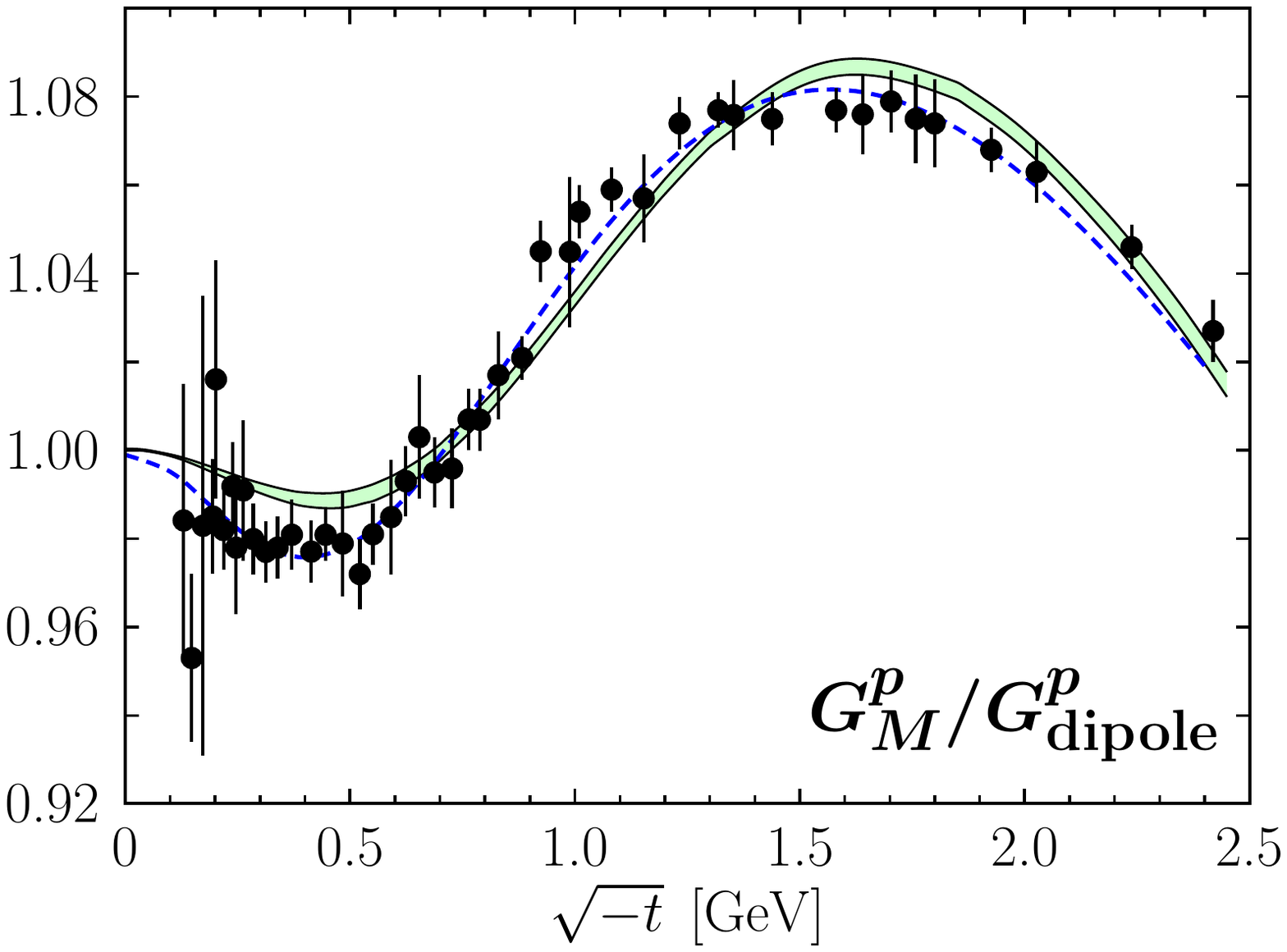} 
\hspace{0.8em}
\includegraphics[width=0.46\textwidth,%
viewport=128 288 584 683]{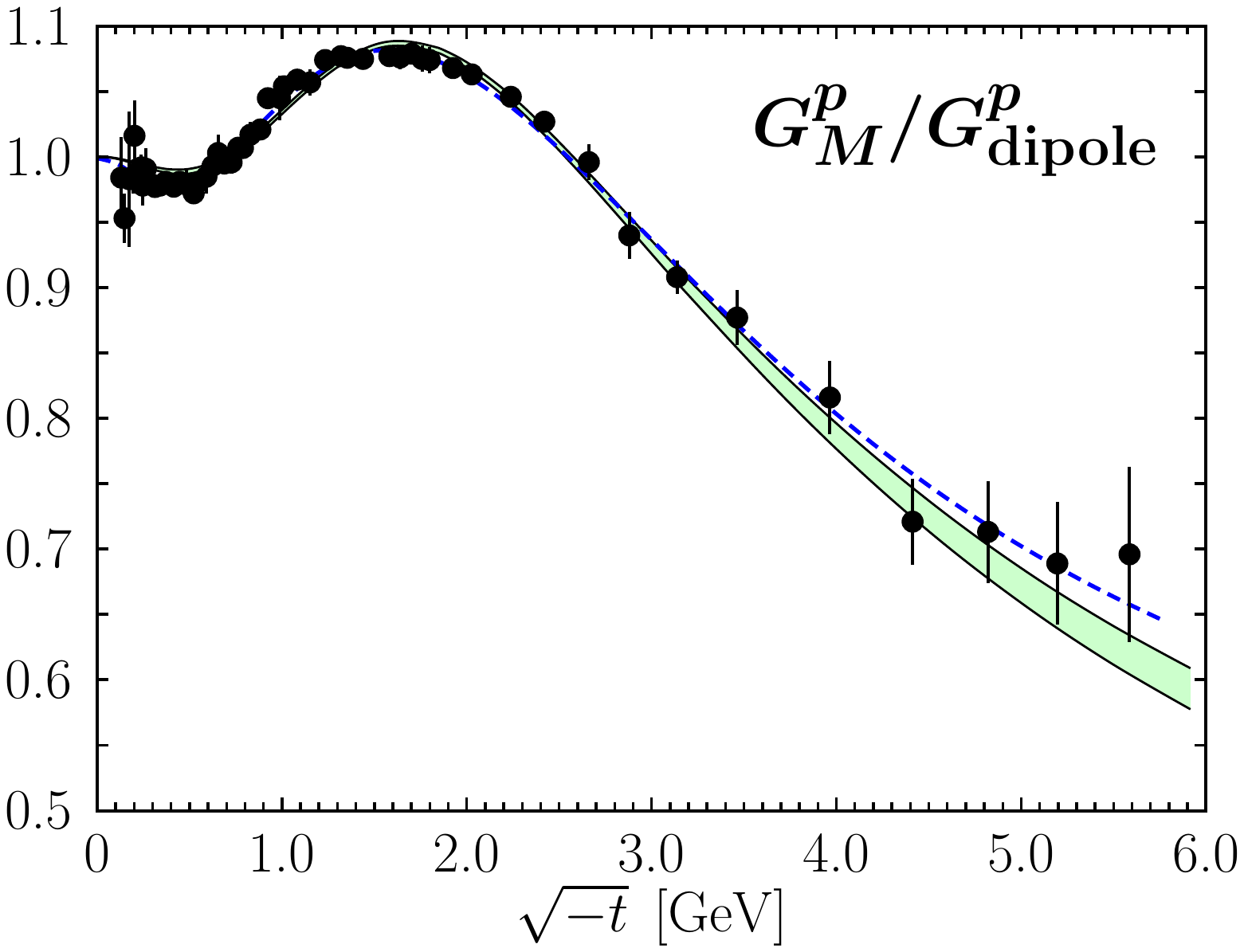} \\[1em]
\includegraphics[width=0.47\textwidth,%
viewport=110 290 565 635]{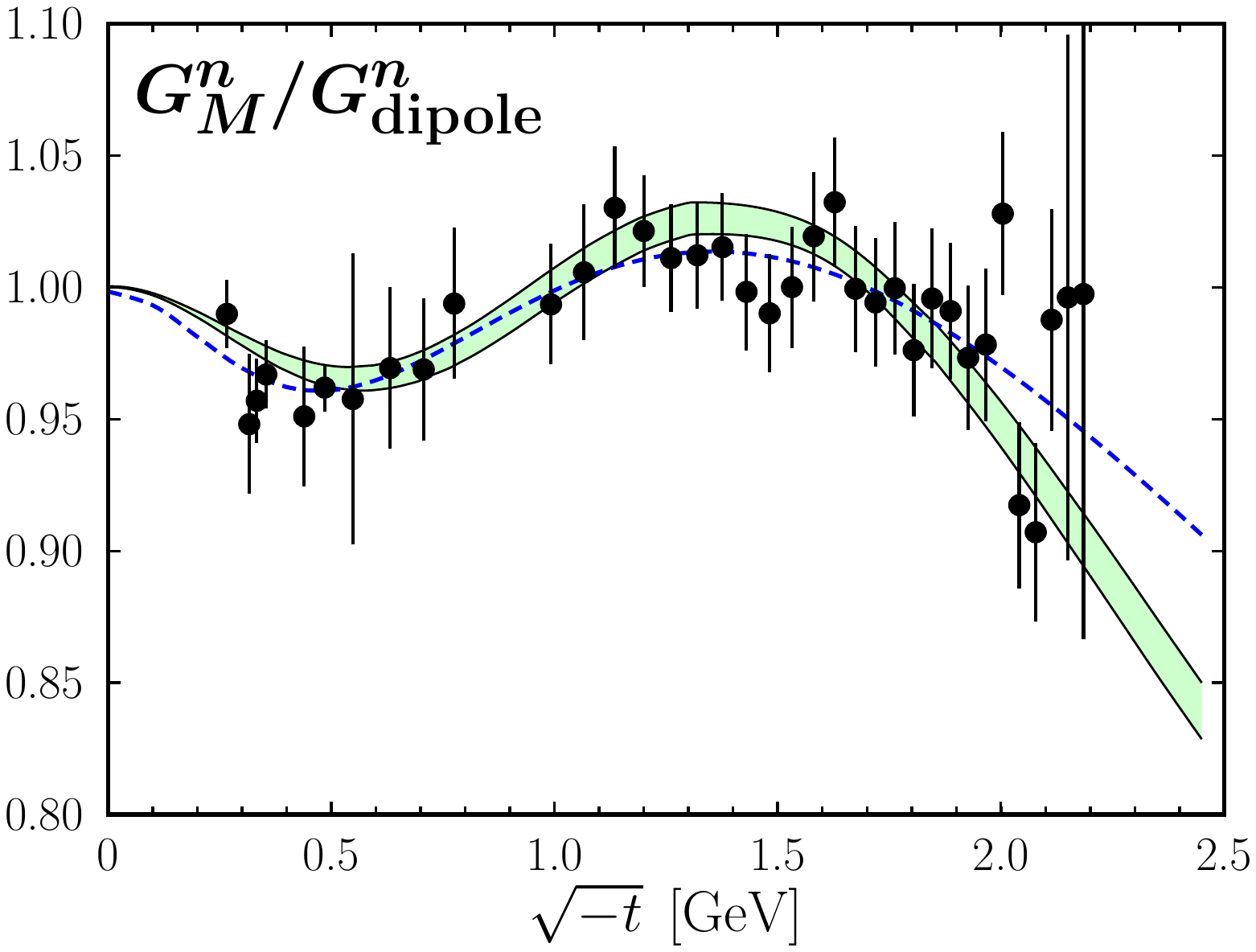}
\hspace{0.8em}
\includegraphics[width=0.468\textwidth,%
viewport=110 305 565 645]{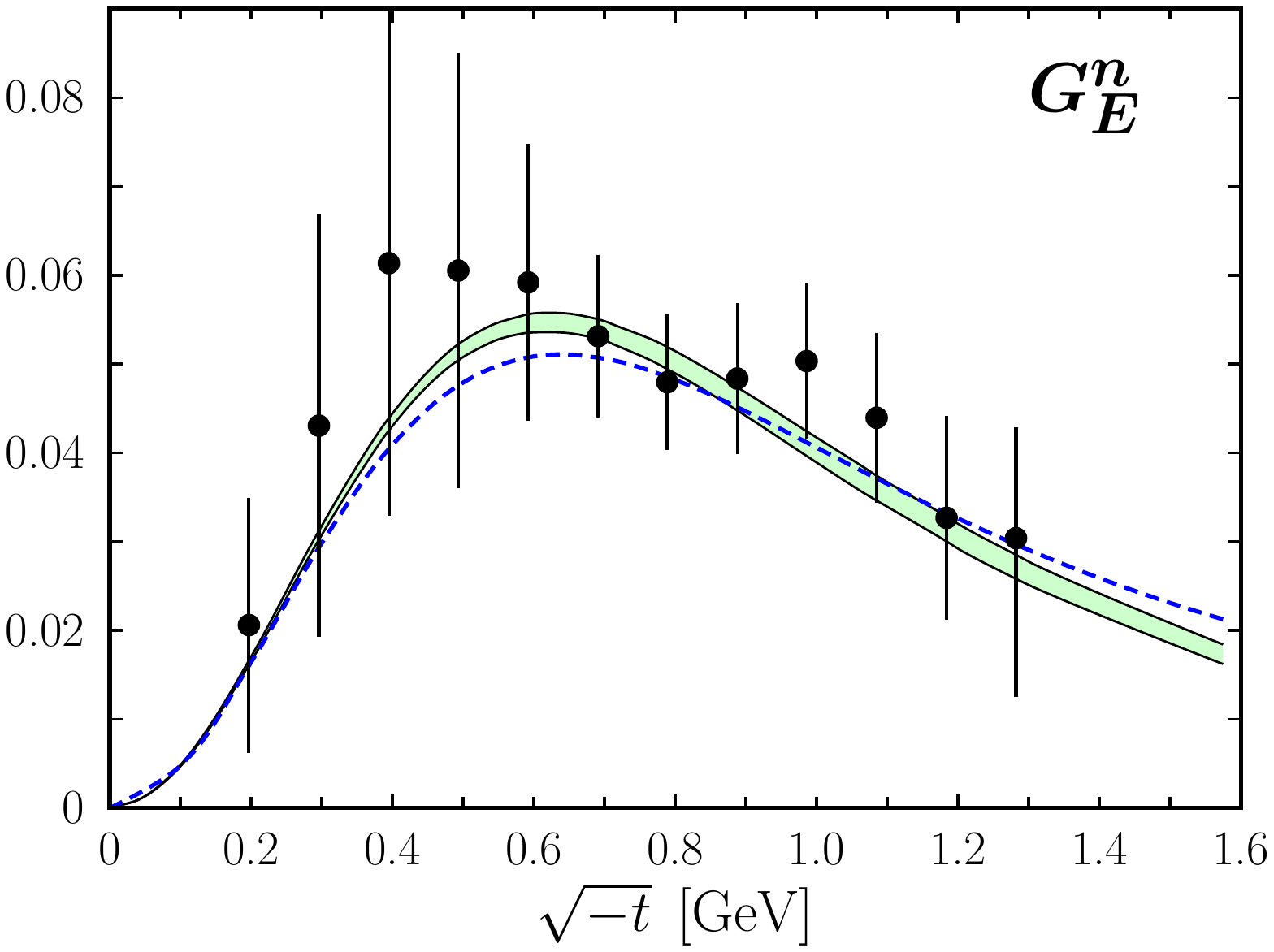} \\[1em]
\includegraphics[width=0.47\textwidth,%
viewport=115 320 570 685]{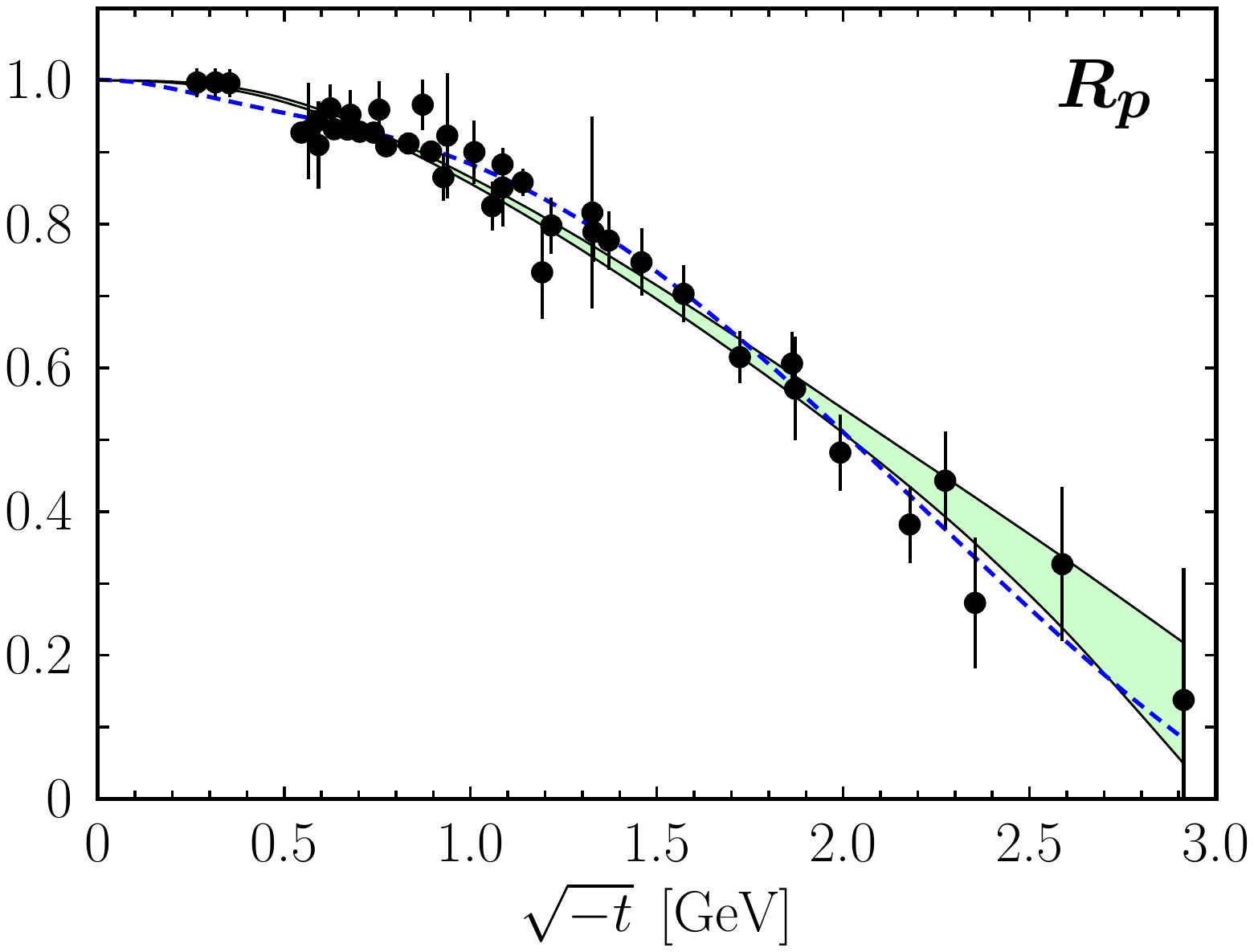}
\hspace{0.8em}
\includegraphics[width=0.47\textwidth,%
viewport=115 310 570 665]{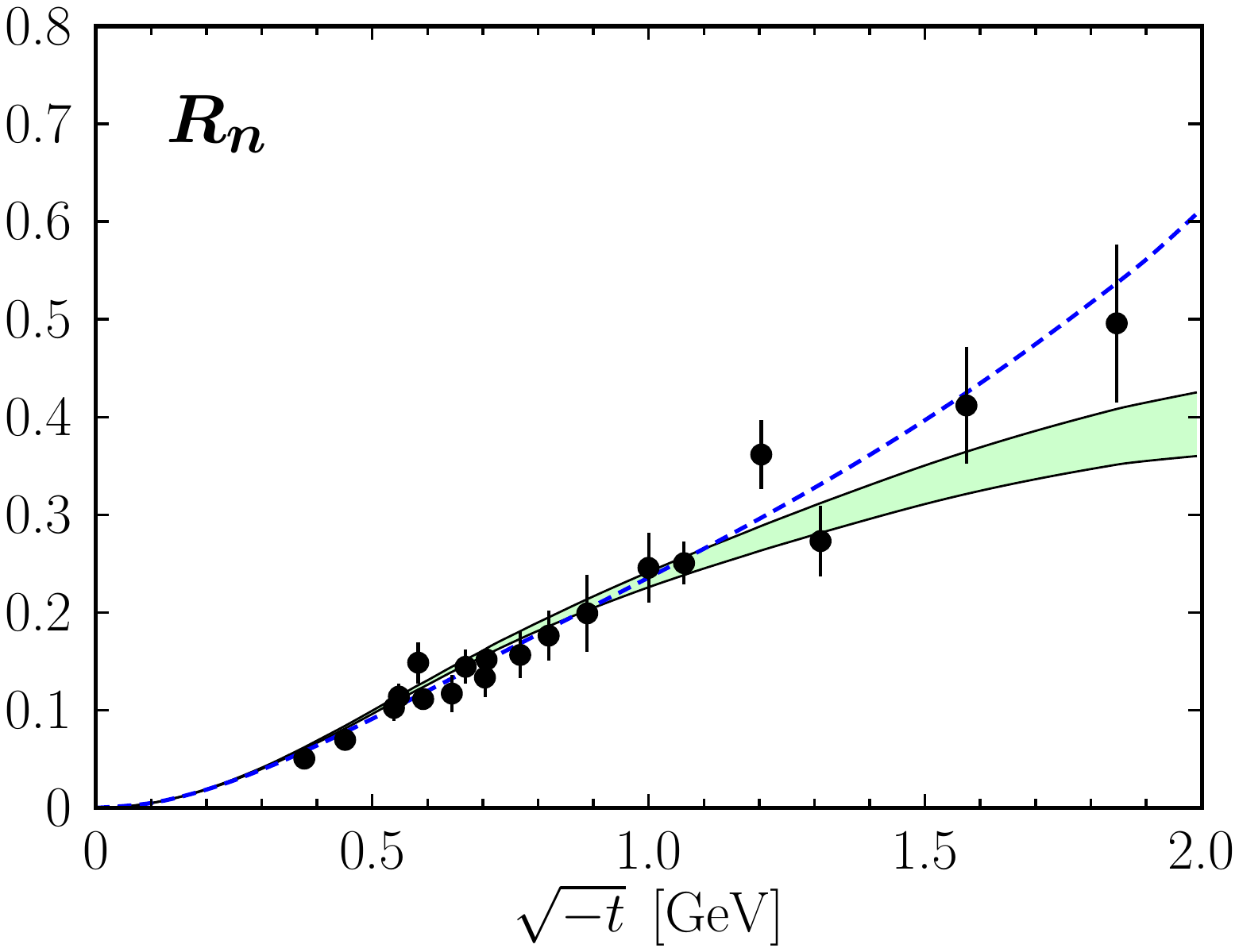}
\end{center}
\caption{\label{fig:Sachs} Results of our fits for the Sachs form factors
  or their ratios.  The dashed lines show the power-law fit described in
  \protect\sect{sec:power-law-fit} and the bands show the default GPD fit
  described in the present section.  The data points correspond to the
  default data set specified in \protect\tab{tab:data-overview}.}
\end{figure*}

\begin{figure*}
\begin{center}
\includegraphics[width=0.451\textwidth,%
viewport=130 275 590 630]{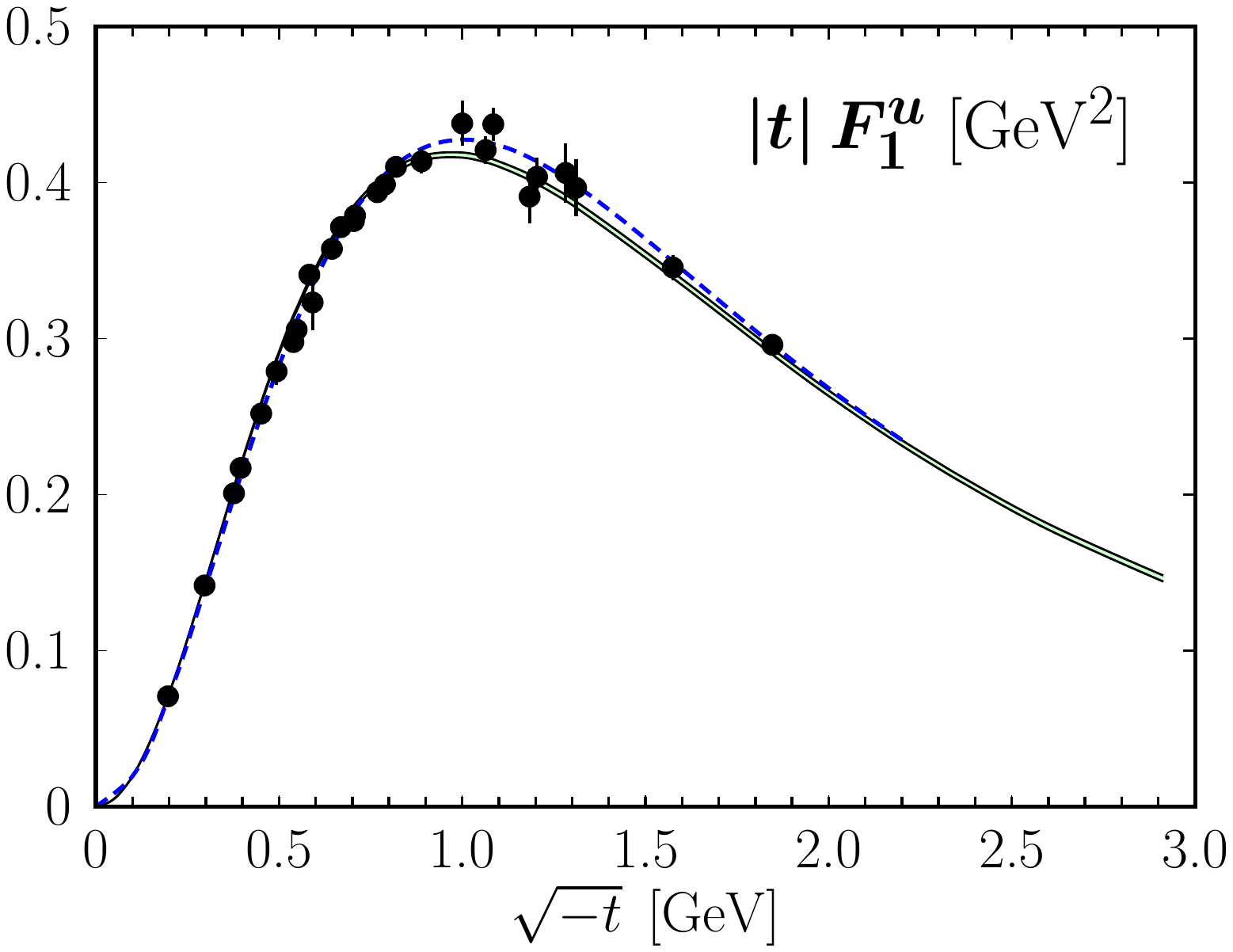}
\hspace{0.8em}
\includegraphics[width=0.447\textwidth,%
viewport=125 260 580 615]{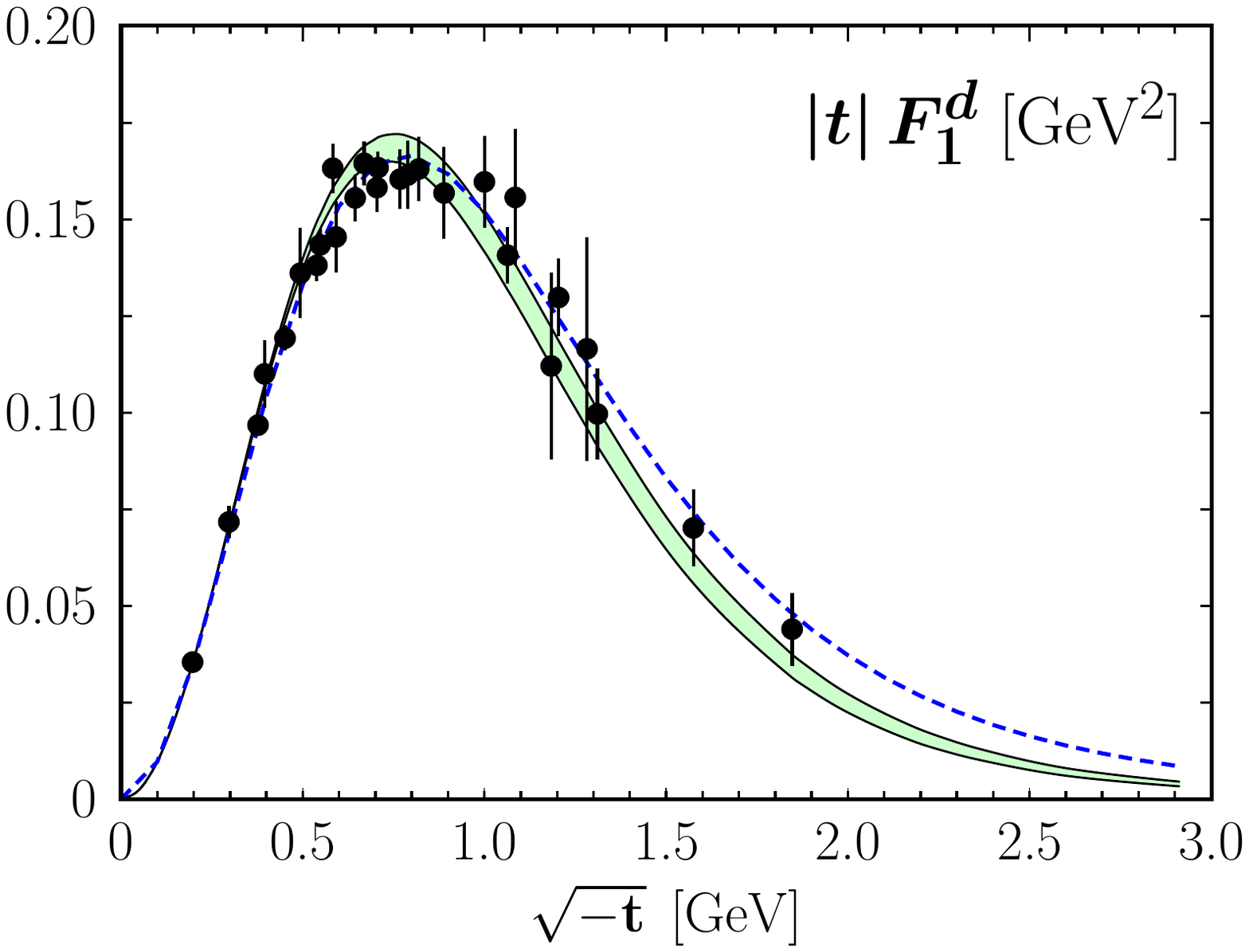} \\[1em]
\includegraphics[width=0.451\textwidth,%
viewport=129 305 588 660]{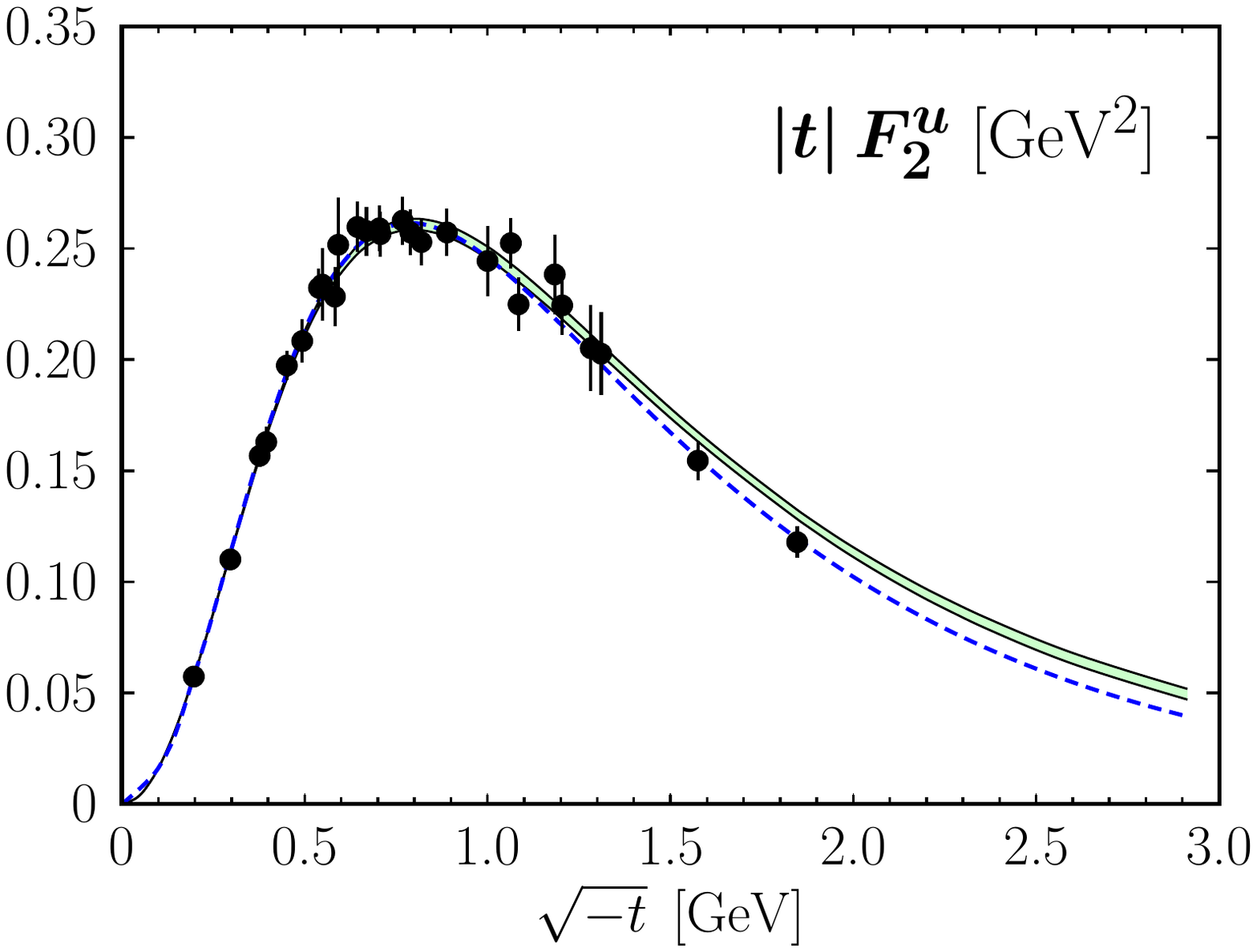}
\hspace{0.8em}
\includegraphics[width=0.444\textwidth,%
viewport=130 310 580 660]{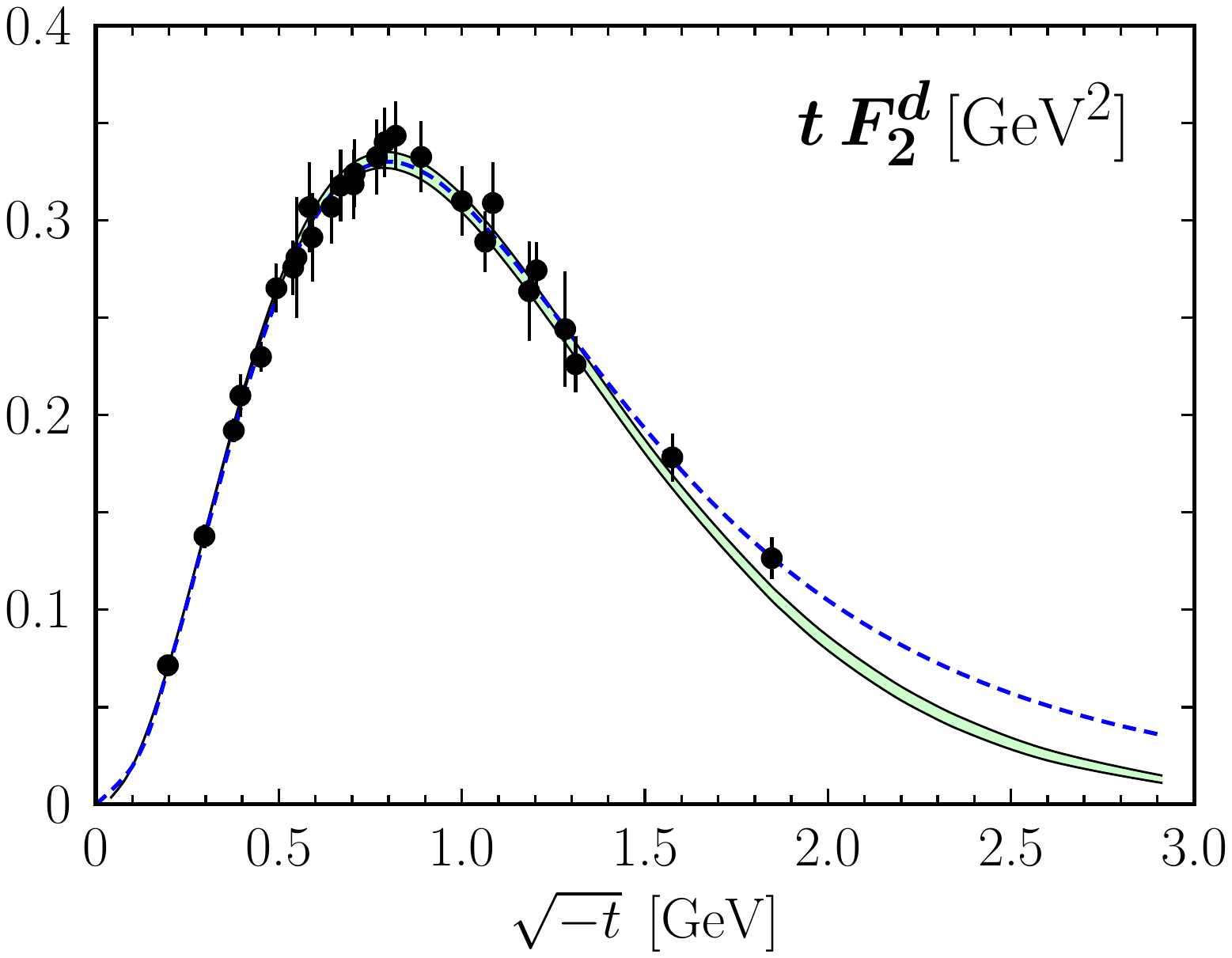} \\[1em]
\includegraphics[width=0.452\textwidth,%
viewport=100 315 560 666]{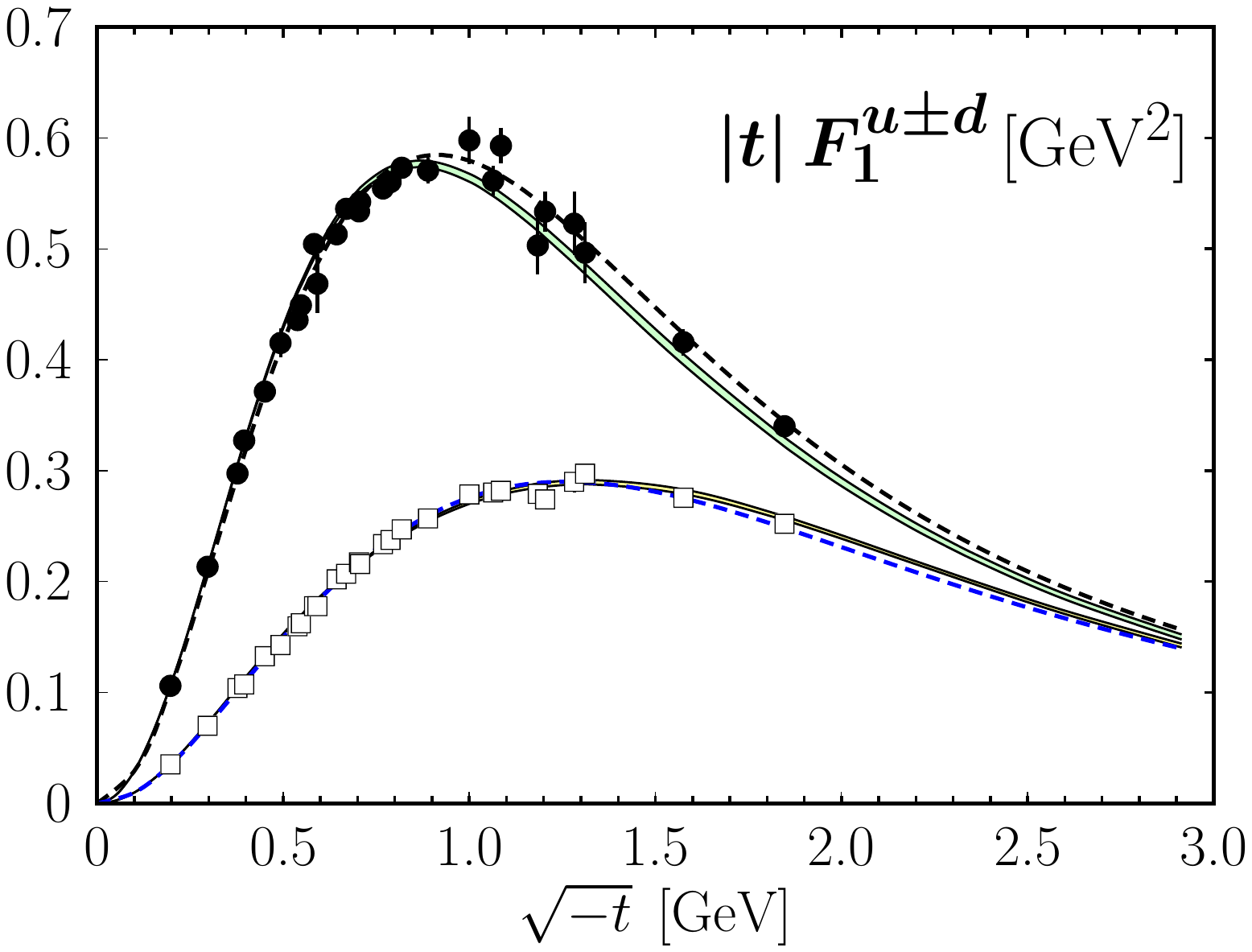}
\hspace{0.8em}
\includegraphics[width=0.447\textwidth,%
viewport=100 315 560 666]{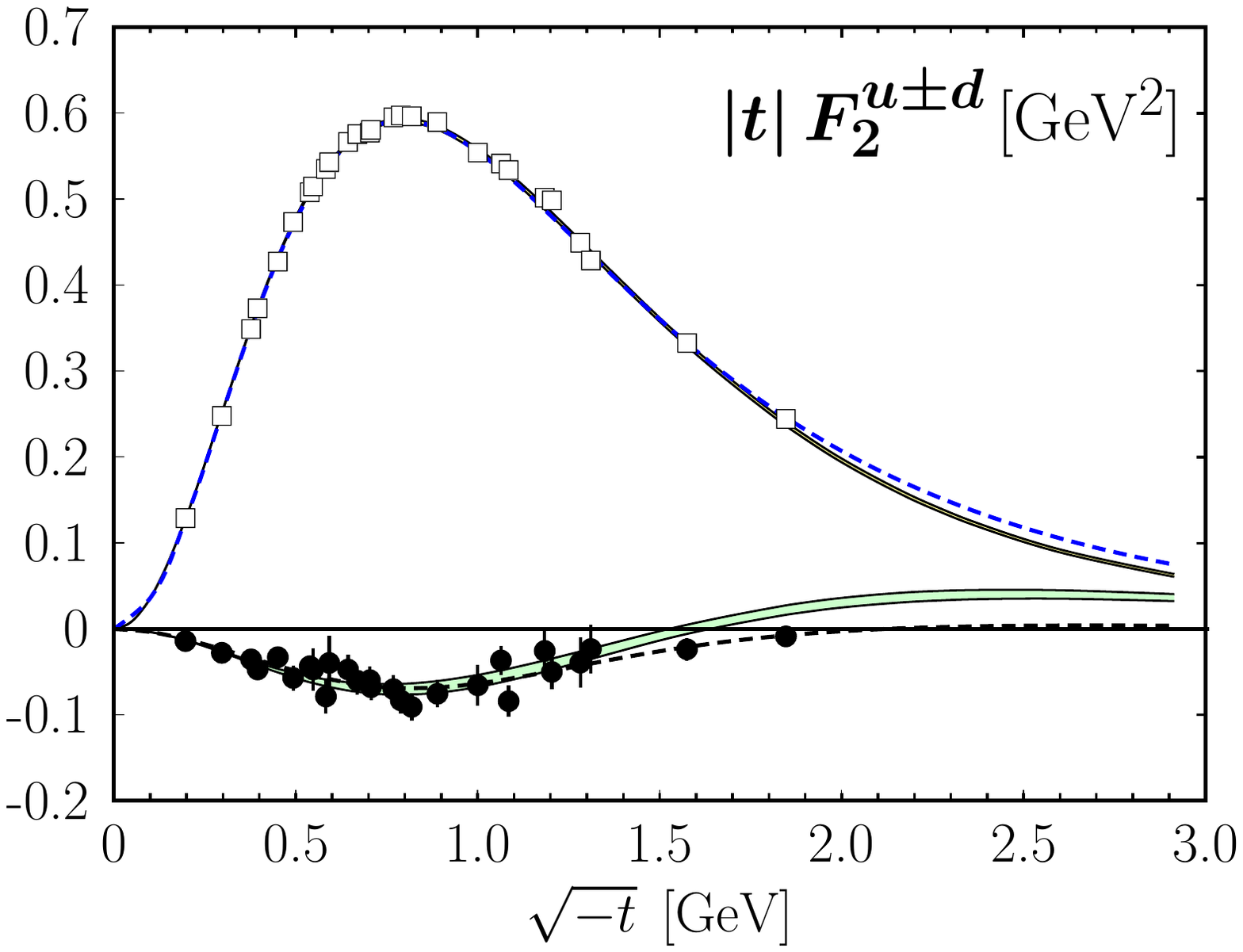}
\end{center}
\caption{\label{fig:flavor-gpd} Results of our fits for the flavor form
  factors, which are obtained by interpolation as described in
  \protect\sect{sec:interpol-procedure}.  All form factors are scaled with
  $|t|$.  The dashed lines show the power-law fit described in
  \protect\sect{sec:power-law-fit} and the bands show for the default GPD
  fit.  The data for the isotriplet combinations $F_1^{u-d}$ and
  $F_2^{u-d}$ are shown as open squares.}
\end{figure*}

In \fig{fig:flavor-gpd} we compare our default GPD fit (as well as our
power-law fit) with our interpolated data set for the Dirac and Pauli form
factors, both in the quark flavor basis and in the isospin basis.  Good
agreement can be observed in all cases.  We note that the isosinglet Pauli
form factor $F_2^{u+d}$ is very small due to a strong cancellation between
$u$ and $d$ quarks and therefore has large relative errors.  Our
interpolated data set suggests that it may have a zero crossing at
$\sqrt{-t}$ around $2 \gev$, but more precise data is needed for a
definite conclusion.

\begin{figure}[t]
\begin{center}
\includegraphics[width=0.45\textwidth,%
viewport=125 333 580 680]{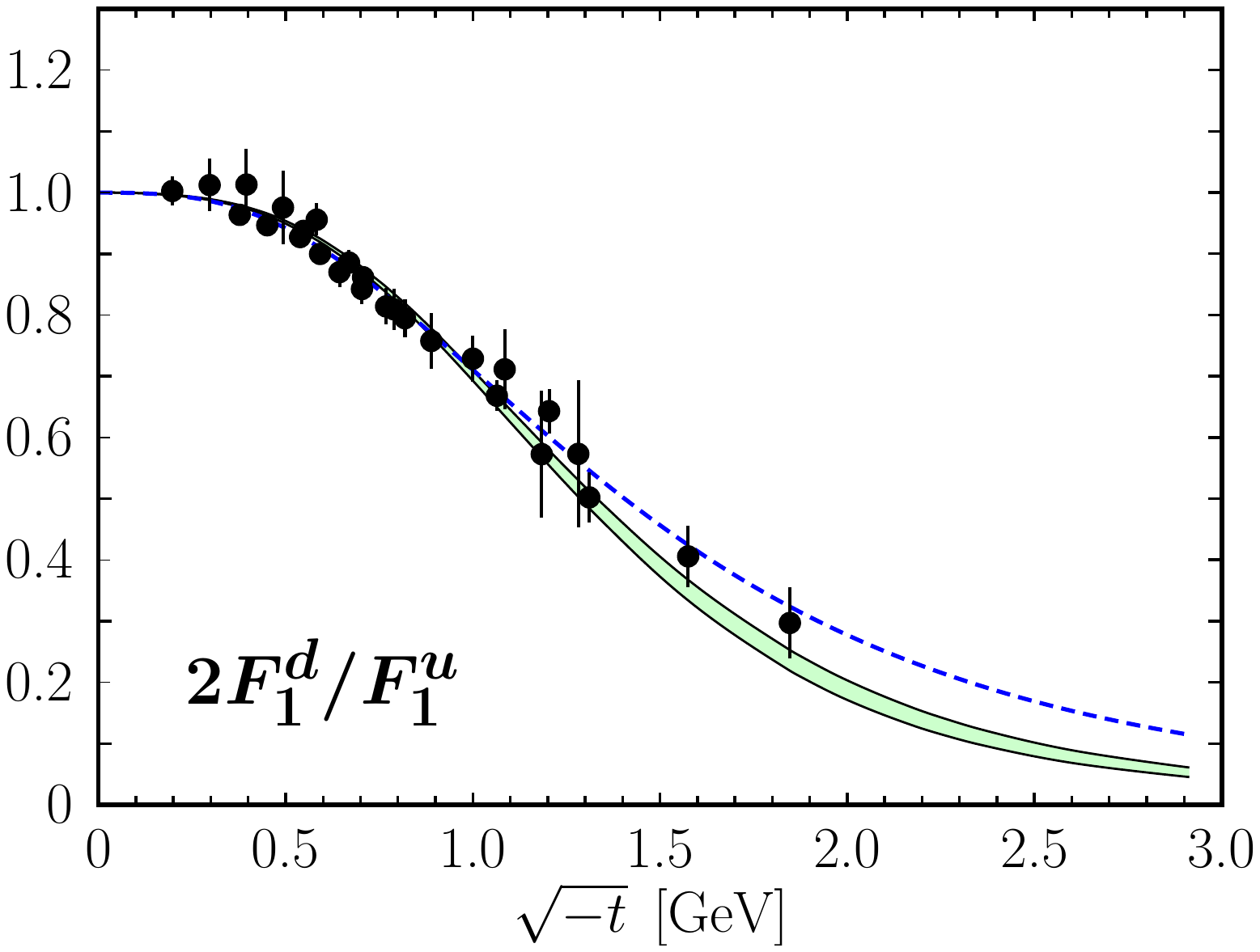} \\[1em]
\includegraphics[width=0.45\textwidth,%
viewport=125 333 580 680]{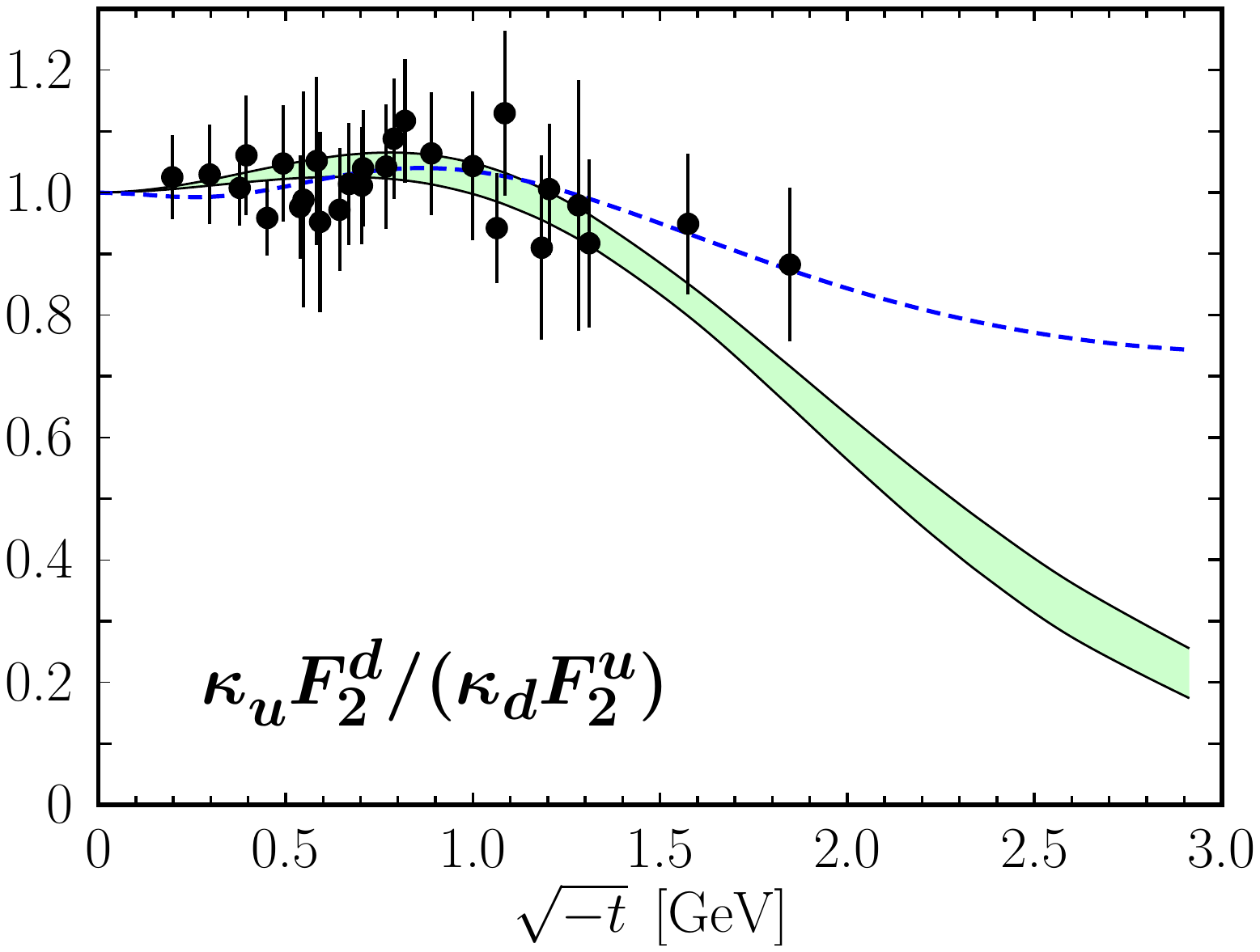}
\end{center}
\caption{\label{fig:ratios} Results of our fits for the ratios of the
  Dirac and Pauli form factors for $d$ and $u$ quarks, normalized to unity
  at $t=0$.  Data and curves are as in \protect\figs{fig:Sachs} and
  \ref{fig:flavor-gpd}.}
\end{figure}

In \fig{fig:ratios} we finally show the ratios of Dirac and Pauli form
factors for $u$ and $d$ quarks.  As already mentioned in
\sect{sec:interpol-procedure}, we have a strong decrease of this ratio for
the Dirac form factors, whereas for the Pauli form factors it stays nearly
flat.  In our default GPD fit (but not in the power-law fit) we obtain a
decrease of $|F_2^d/F_2^u|$ for $\sqrt{-t} > 1.5 \gev$, which is not
suggested by the data but consistent within their errors.  It will be
interesting to see the behavior of this ratio with data for larger $-t$.

%%%%%%%%%%%%%%%%%%%%%%%%%%%%%%%%%%%%%

\subsection{Variations of the default fit}
\label{sec:var-fits}

We have performed a number of alternative fits to explore the dependence
of our results on several choices we have made in the fitting procedure.
A brief description of the different alternative fits is given in
\tab{tab:fit-basics}, and the corresponding values of $\chi^2$ can be
found in \tab{tab:chi2-values}.  For all fits described in the following,
we find again that a local minimum of $\chi^2$ in the $(\beta_u, \beta_d)$
plane is taken at the boundary of values allowed by the positivity
conditions, as was the case for our default fit.

Let us first discuss the alternative fits using the same PDFs as the
default fit.  We remark that a fit with $\gamma_u = \gamma_d = 0$ has
severe problems with positivity.  Specifically, it tends to have $A_u <
C_u$ and thus to violate the condition $g_q(x) < f_q(x)$, except for
rather large values $\beta_u \ge 6.20$ and $\beta_d \ge 5.95$.  Given this
and the significantly larger $\chi^2$ already mentioned in the previous
subsection, we conclude that the data clearly prefer a nonzero $\gamma_u$
and do not consider this parameter setting any further.

\begin{table*}
\renewcommand{\arraystretch}{1.1}
\begin{center}
\begin{tabular}{lccl}
\hline
fit     & $\beta_u$ & $\beta_d$ &  remarks \\
\hline
ABM 1   &   $4.65$  &  $5.25$   &  default fit (see
                                   \protect\sect{sec:def-fit}) \\
ABM 2   &   $5.05$  &  $5.80$   &  $\gamma_d = \gamma_u = 4$ \\
ABM 3   &   $4.55$  &  $5.20$   &  with strangeness (MSTW) \\
ABM 4   &   $4.60$  &  $5.25$   &  with strangeness (NNPDF) \\
ABM 5   &   $4.40$  &  $5.00$   &  alternative data for $R^p$ \\
ABM 6   &   $4.95$  &  $5.50$   &  alternative data for $G_M^p$ \\[0.1em]
\hline
ABM 0   &   $4.35$  &  $5.10$ 	&  as ABM 1 but at scale $\mu=1 \gev$ \\
\hline
CT      &   $4.60$  &  $4.15$   &  \\
GJR     &   $4.65$  &  $5.10$   &  \\
HERAPDF &   $4.70$  &  $5.35$   &  \\
MSTW    &   $4.65$  &  $5.90$   &  $\alpha'_u - \alpha'_d = 0$ \\
NNPDF   &   $4.70$  &  $5.15$   &  \\
\hline
\end{tabular}
\end{center}

\vspace{-0.5em}

\caption{\label{tab:fit-basics} Overview of our GPD fits.  All fits have
  $\gamma_u = 4$, $\gamma_d = 0$, $\alpha'_u - \alpha'_d = 0.1 \gev^{-2}$
  and PDFs evaluated at $\mu = 2 \gev$ unless stated otherwise.}
%\end{table*}

\vspace{1em}

%\begin{table*}
\renewcommand{\arraystretch}{1.1}
\begin{center}
\begin{tabular}{lccrccccc}
\hline
  & total & \multicolumn{2}{c}{--- $G_M^p$ ---}
  & $R^p$ & $G_M^n$ & $R^n$ & $G_E^n$ & $r^2_{nE}$ \\
data points & $178$ & $48$ & $6$ & $54$ & $36$ & $21$ & $12$ & $1$ \\
\hline
ABM 1  & $221.2$ & $79.7$ &  $3.8$ & $78.8$ & $29.5$ & $24.1$ & $3.2$ & $2.1$ \\
ABM 2  & $219.2$ & $71.1$ &  $4.4$ & $85.9$ & $27.7$ & $25.0$ & $3.2$ & $1.9$ \\
ABM 3  & $230.5$ & $81.5$ &  $2.9$ & $76.7$ & $29.7$ & $31.1$ & $3.2$ & $5.5$ \\
ABM 4  & $225.0$ & $79.2$ &  $3.9$ & $81.1$ & $29.0$ & $26.1$ & $3.0$ & $2.7$ \\
ABM $5^{\,a}$
       & $166.4$ & $64.5$ &  $1.9$ & $40.8$ & $30.5$ & $22.0$ & $2.9$ & $3.7$ \\
ABM $6^{\,b}$
       & $139.3$ & $14.9$ &  $4.0$ & $63.8$ & $27.2$ & $23.8$ & $3.1$ & $2.4$ \\
\hline
ABM 0  & $198.1$ & $77.1$ &  $2.3$ & $65.6$ & $27.1$ & $20.9$ & $3.1$ & $1.9$ \\
\hline
CT     & $212.6$ & $75.7$ &  $2.9$ & $75.6$ & $28.3$ & $24.5$ & $3.2$ & $1.4$ \\
GJR    & $189.7$ & $64.3$ &  $3.5$ & $70.9$ & $27.0$ & $19.6$ & $3.3$ & $1.1$ \\
HERAPDF& $254.5$ & $83.4$ & $10.6$ & $96.4$ & $33.3$ & $25.7$ & $3.3$ & $1.7$ \\
MSTW   & $167.9$ & $53.1$ &  $2.9$ & $63.1$ & $24.4$ & $17.9$ & $4.9$ & $1.6$ \\
NNPDF  & $196.6$ & $72.6$ &  $3.9$ & $73.2$ & $27.5$ & $15.4$ & $4.0$ & $0.0$ \\
\hline
power law & $122.3$ & $28.8$ & $1.8$ & $52.7$ & $20.4$ & $15.3$ & $3.4$ & $0.0$ \\
\hline
\multicolumn{9}{l}{${}^{a}$ $172$ data points, $48$ for $R^p$ \hfill
                   ${}^{b}$ $174$ data points, $44+6$ for $G_M^p$
                  \rule{0pt}{2.8ex}}
\end{tabular}
\end{center}

\vspace{-0.5em}

\caption{\label{tab:chi2-values} Partial and total values of $\chi^2$ for
  the GPD fits and for the power-law fit of
  \protect\sect{sec:power-law-fit}.  The first value for $G_M^p$ refers to
  the data with $-t < 10 \gev^2$ and the second value to the data with $-t
  > 10 \gev^2$.}
\end{table*}

We now discuss the other fits in turn, referring to their labels in
\tab{tab:fit-basics}.
\begin{description}
\item[ABM 2.] In this fit we set $\gamma_d = 4$, i.e.\ equal to
  $\gamma_u$.  The description of the data is globally as good as for the
  default fit ABM 1, with some improvement for $G_M^p$ and a somewhat
  worse $\chi^2$ for $R^p$.  The fitted value $\alpha = 0.654 \pm 0.017$
  is relatively high, which is why we prefer the fit ABM 1 (where $\alpha
  = 0.603 \pm 0.020$) as our default.
\item[ABM 3 and 4.] In these fits we include the models for the
  strangeness form factors described in \sect{sec:strange}, taking either
  the strangeness PDFs of MSTW or of NNPDF (all other parton density sets
  we consider have $s(x) = \bar{s}(x)$ and hence cannot be used to model
  $F_1^s$).  The global description of the data is just slightly worse
  than for fit ABM 1, where strangeness is neglected, with a slight
  preference for the variant using the NNPDF densities in the model for
  $F_1^s$.  An exception is the poor description of the neutron charge
  radius in the case of ABM 3.  We note that we get a more satisfactory
  result with $\alpha'_u - \alpha'_d = 0.2 \gev^{-2}$ in this case.

  Overall, we conclude that the influence of the small strangeness form
  factors in our fit is visible in the details but does not change the
  overall picture significantly.  This is in line with our findings
  discussed at the end of \sect{sec:strange}.
\item[ABM 5.] In this fit we omit the recent Hall A data for $R^p$, namely
  the sets Paolone 10, Ron 11 and Zhan 11, and instead include the
  polarized target data of Jones 06 and Crawford 07 (see
  \tab{tab:Rp-data}).  As can be seen in \fig{fig:Rp-data}, this removes
  the tension between the recent Hall A data and earlier JLab measurements
  that is present in our default data set.  The fit ABM 5 describes the
  alternative data set for $R^p$ very well, with a $\chi^2$ of less than 1
  per data point.  It also gives a significantly improved (although not
  perfect) description of $G_M^p$ at low $t$.  Only for the neutron charge
  radius does the description become notably worse compared with fit ABM
  1.  The data on $R^p$ thus has a clear impact on our fits, which
  confirms the urgency for experimental clarification of the tensions
  between the current data sets.
\item[ABM 6.] To investigate the impact of the $G_M^p$ data on our fit, we
  take as an alternative data set the results of Arrington 05
  \cite{Arrington:2004ae}, which covers the range $0.141 \gev^2 \le -t \le
  9.121 \gev^2$.  To have data at higher $-t$, we include the results of
  our default set, AMT 07 \cite{Arrington:2007ux}, for $-t \ge 9.848
  \gev^2$.  The Arrington 05 extraction has significantly larger errors
  than the one of AMT 07 for the reasons discussed in
  \sect{sec:proton-data}, and we correspondingly find a very small
  $\chi^2$ of $0.4$ per data point for the low-$t$ data on $G_M^p$.  The
  description of the other data sets is not much changed compared with fit
  ABM 1, except for a slight improvement for $R^p$.

  The dependence of $\chi^2$ on the choice of $(\beta_u, \beta_d)$ is much
  less pronounced in fit ABM 6 than in all other fits, so that those
  parameters are less well determined.  Clearly, the high precision of the
  data on $G_M^p$ we adopted in our default fit is of great importance for
  a precise determination of the GPDs.
\item[ABM 0.] In this fit we take the PDFs from the ABM set, but evaluate
  them at scale $\mu = 1 \gev$ instead of $2 \gev$.  The electromagnetic
  form factors are scale invariant and can be described by the sum rules
  \eqref{eq:SumRule} with GPDs at any scale, but of course it does matter
  at which scale we make the functional ansatz for GPDs specified in
  \sect{sec:fit-ansatz}.

  Compared with fit ABM 1, the description of the data in fit ABM 0 is
  improved significantly for $G_M^p$ at low $t$ and for $R^p$, with little
  change for the other observables.  Since the usual PDFs at $\mu = 1
  \gev$ are determined only indirectly (data corresponding to such a low
  scale is not included in the PDF analyses in order to limit the
  importance of higher-twist corrections), we prefer the fits with $\mu =
  2 \gev$ as our default.
\end{description}

We have also repeated the fit with the other PDF sets discussed in
\sect{sec:pdfs}, neglecting the strangeness form factors and fixing
$\gamma_u = 4$, $\gamma_d = 0$ as in our default fit.

For all PDFs we find that $\alpha'_u - \alpha'_d = 0.1 \gev^{-2}$ clearly
improves the description of $r^2_{nE}$ and of $R^n$ compared with
$\alpha'_u - \alpha'_d = 0$.  An exception is the fit with the PDFs of
MSTW, where this parameter setting makes the fit slightly worse, so that
we take $\alpha'_u = \alpha'_d$ in this case.  This exception is not
implausible, since the MSTW set already has a large isospin breaking
between $\alpha_u^{\text{eff}}$ and $\alpha_d^{\text{eff}}$ (see
\tab{tab:pdfs}).  A further isospin breaking in the profile functions of
$H_v^q$ and $E_v^q$ is then visibly not preferred by the fit.

Whereas the fit with the HERAPDF densities describes the form factor data
less well than the fit ABM 1, the variants using CT, GJR, MSTW or NNPDF
have lower values of $\chi^2$.  Like in the default fit, the description
is least optimal for $G_M^p$ at low $t$ and for $R^p$, whereas the other
observables are described very well (except for the fit using HERAPDF,
which undershoots the $G_M^p$ data at high $t$).  As spelled out in
\sect{sec:pdfs}, we choose the fit using the ABM set as our default, where
the behavior of both $q_v(x)$ and $e_v^q(x)$ at small and at large $x$
corresponds best to the physical picture underlying our parameterization
of the GPDs.

\begin{figure*}[t]
\begin{center}
\includegraphics[width=0.98\textwidth,%
viewport=0 0 600 210]{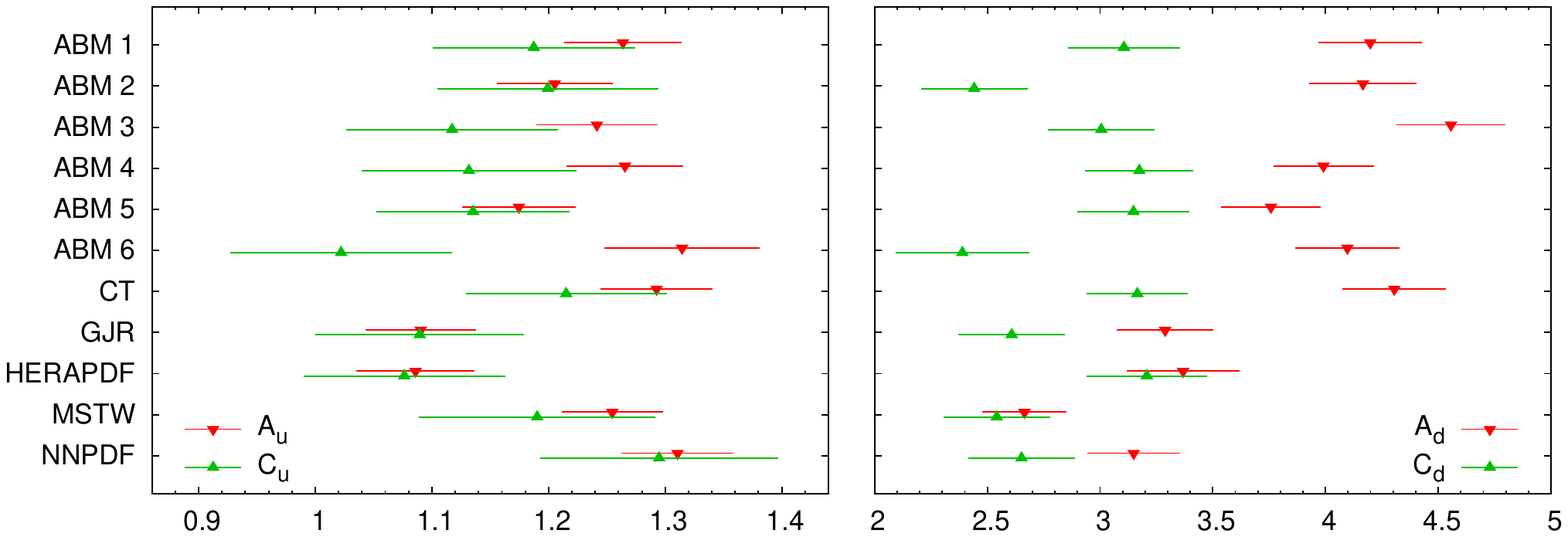} \\[0.5em]
\includegraphics[width=0.98\textwidth,%
viewport=0 0 600 210]{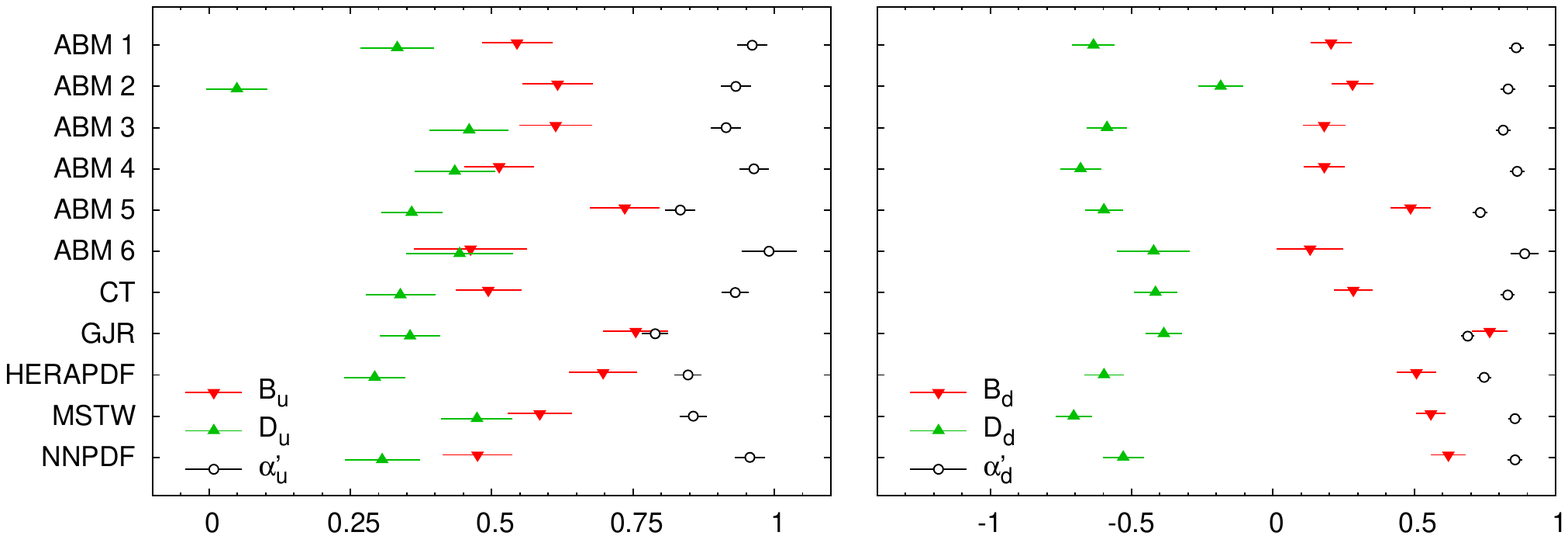}
\end{center}
\caption{\label{fig:par-values} Parameters of the profile functions and
  their errors obtained in the fits described in
  \protect\tab{tab:fit-basics}.  Not shown are the results of fit ABM 0,
  which refers to GPDs at a different scale $\mu$.  Only the error bars
  for fit ABM 1 include the uncertainty that results from varying
  $\beta_u$ and $\beta_d$, which is computed as described in
  \protect\sect{sec:def-fit}.}
\end{figure*}

The parameters of the profile functions obtained in our GPD fits are shown
in \fig{fig:par-values} (the results for $\alpha$ can be found in
\fig{fig:2nd-mom} below).  Clearly, the parametric uncertainties on
$f_q(x)$ and $g_q(x)$ in a given fit are smaller than their variation due
to choosing different PDFs and to other choices made in the fit.  This
also illustrates the strong correlation between the forward limit and the
$t$ dependence of $H_v^q(x,t)$ in a fit to the form factors using the sum
rule \eqref{eq:SumRule}.  Nevertheless, certain parameters turn out to be
relatively stable against variations in the fit, namely
\begin{align}
1.08 \gev^{-2} & < A_u < 1.32 \gev^{-2} \,,
\nonumber \\
1.02 \gev^{-2} & < C_u < 1.30 \gev^{-2} \,,
\nonumber \\
0.68 \gev^{-2} & < \alpha'_d \, < 0.90 \gev^{-2} \,,
\end{align}
with the range of $\alpha'_u$ being shifted to larger values by $0.1
\gev^{-2}$.  Also, a clear hierarchy between the profile functions for $u$
and $d$ quarks at high $x$ is seen throughout all fits, with $A_d > A_u$
and $C_d > C_u$.  We also find $D_d < 0$ and $D_u > 0$ in all cases.  By
contrast, the variation in the values of $A_d$ is appreciable.

Let us now investigate the variation of the forward limit $e_v^q(x)$ in
our fits.  To investigate its behavior at small and large $x$ we have
performed the same type of fits as for the PDFs in \eqref{q-fits}, with a
power behavior
\begin{align}
e_v^q(x) &\sim\, x^{-\alpha_q^{\text{eff}}} 
 & \text{for}~ & 10^{-3} < x < 10^{-2} \,,
\nonumber \\
         &\sim\, (1-x)^{\ms\beta_q^{\text{eff}}}
 & \text{for}~ & 0.65 < x < 0.85 \,.
\end{align}
For our default fit ABM 1 we find
\begin{align}
\alpha_u^{\text{eff}} &= 0.526 \,,
&
\alpha_d^{\text{eff}} &= 0.622 \,,
\nonumber \\
\beta_u^{\text{eff}} &= 4.72 \,,
&
\beta_d^{\text{eff}} &= 5.44 \,,
\end{align}
where all effective power laws describe $e_v^q(x)$ within 2\% for the
stated $x$ ranges.  We did the same exercise for the results of our
alternative GPD fits.

In all cases except for fit ABM 2 we find that $\alpha_d^{\text{eff}}$
differs from $\alpha$ by at most 0.02, whereas $\alpha_u^{\text{eff}}$ is
smaller than $\alpha_d^{\text{eff}}$ by about 0.1.  The slight difference
between $\alpha_d^{\text{eff}}$ and $\alpha$ is due to the factor
$(1-x)^{\beta_d}$ in the parameterization of $e_v^d(x)$, whereas for $u$
quarks the extra factor $\bigl( 1 + \gamma_u \sqrt{x} \bigr)$ with
$\gamma_u = 4$ is responsible for the large difference between
$\alpha_u^{\text{eff}}$ and $\alpha$.  For fit ABM 2, where $\gamma_u =
\gamma_d = 4$ we find $\alpha_u^{\text{eff}} \approx \alpha_d^{\text{eff}}
\approx 0.58$, which is smaller than the fitted value of $\alpha$ by about
0.07.  Since fits ABM 1 and ABM 2 describe the data nearly equally well,
we cannot conclude whether or not the data favor a slight isospin breaking
between the effective small-$x$ powers in $e_v^u(x)$ and $e_v^d(x)$.

The range of the effective powers in the different fits is
\begin{align}
0.52 & < \alpha_u^{\text{eff}} < 0.60 \,,
\nonumber \\
0.58 & < \alpha_d^{\text{eff}} < 0.70 \,.
\end{align}
A comparison with their analogs for the PDFs in \tab{tab:pdfs} reveals
that, for both $u$ and $d$ quarks, $\alpha_q^{\text{eff}}$ is clearly
larger in $e_v^q$ than in $q_v$ for all our fits (except for the fit with
MSTW PDFs, where the effective powers in $e_v^u$ and $u_v^{}$ nearly
equal).

Turning to the large-$x$ behavior, we observe that $\beta_q^{\text{eff}}$
is slightly smaller than $\beta_q$, by at most $0.1$ for $u$ quarks and at
most $0.2$ for $d$ quarks, so that the parameter $\beta_q$ gives a good
representation of the behavior of $e_v^q(x)$ at large but not extremely
large $x$.  The effective power is larger in $e_v^q$ than in $q_v$, with a
larger difference for $u$ quarks.  In both $q_v$ and $e_v^q$ the effective
power of $(1-x)$ is larger for $d$ quarks than for $u$ quarks (with the
exception of $e_v^q$ in the fit using the CT PDFs, whose large-$x$
behavior we find problematic as discussed in \sect{sec:pdfs}).

% the following table belongs to the next section (A note on nucleon
% radii).  It is placed here in order to end up on the correct page.
%
\begin{table*}[t]
\renewcommand{\arraystretch}{1.3}
\begin{center}
\begin{tabular}{lcccc} \hline
          & $r_{pM} \, [\fm]$ & $r_{nM} \, [\fm]$ & $r_{pE} \, [\fm]$
          & $r^2_{nE} \, [\fm^2]$ \\
\hline
ABM 1     & $0.832 \pm 0.002$ & $0.854 \pm 0.004$ & $0.838 \pm 0.003$
          & $-0.1129 \pm 0.0019$ \\
power law & $0.868 \pm 0.006$ & $0.887 \pm 0.012$ & $0.917 \pm 0.013$
          & $-0.1161 \pm 0.0022$ \\
\hline
PDG 2012  & $0.777 \pm 0.016$ & $0.862^{+0.009}_{-0.008}$\hspace{0.8em}
          & $0.8775 \pm 0.0051$ & $-0.1161 \pm 0.0022$ \\
\hline
\end{tabular}
\end{center}
\caption{\label{tab:radii-Sachs} Magnetic and charge radii of the
  proton and neutron obtained in our fits, compared with the values quoted
  in the Review of Particle Physics (PDG 2012)
  \protect\cite{Beringer:1900zz}.  We have added in quadrature the
  statistical and systematic uncertainties given for $r_{pM}$ in PDG
  2012.}
\end{table*} 

The effective large-$x$ power for $e_v^q$ is however not always larger
then the one for $q_v$ by $1$ unit or more, as is suggested by the
positivity bound \eqref{bound-final}, where the factor depending on the
profile functions $f_q$ and $g_q$ behaves like $(1-x)^2$ at large $x$.
This illustrates that considerations based on the mathematical limit $x\to
1$ cannot always be taken literally in the region $0.65 < x < 0.85$, which
is of relevance in the integrals that yield the electromagnetic form
factors.

%%%%%%%%%%%%%%%%%%%%%%%%%%%

\subsection{A note on nucleon radii}
\label{sec:charge-radii}

From our fits to the nucleon form factors we can evaluate the associated
radii.  The values we obtain are given in \tab{tab:radii-Sachs}.  The
magnetic radii of proton and neutron are defined as
\begin{equation}
r^2_{iM} = \frac{6}{G_M^i(0)}\, \frac{\dd G_M^i(t)}{\dd t} \bigg|_{t=0}
\end{equation}
with $i=p,n$.  They are compared with their values given in the Review of
Particle Physics 2012 (PDG 2012) \cite{Beringer:1900zz}, which are based
on the same method as we use here, i.e.\ on fits of the magnetic form
factors measured in elastic electron-nucleon scattering.  We observe
reasonable agreement for $r_{nM}$ but a clear discrepancy for $r_{pM}$.
We do not consider this discrepancy to be a serious shortcoming of our
fits.  A precise determination of the radii requires form factor
measurements at $t$ as small as possible in order to determine the local
derivative at $t=0$, together with excellent control over theoretical
uncertainties in the form factor extraction and in the fit from which the
derivative is extracted.  These requirements are clearly not satisfied in
our fits, which aim at a global description of all electromagnetic form
factors in the full $t$ region where data is available, rather than at a
precise description in the vicinity of $t=0$.  We do, however, note that
the value for $r_{pM}$ quoted by PDG 2012 and reproduced in our table is
based on a single experiment \cite{Bernauer:2010wm}, which uses a strongly
simplified treatment of two-photon exchange in the extraction, and which
obtains a significantly smaller result than previous determinations (see
the full listing in \cite{Beringer:1900zz}).

\begin{table*}[t]
\renewcommand{\arraystretch}{1.3}
\begin{center}
\begin{tabular}{lcccc} \hline
fit & $b_{u1} \, [\fm]$ & $b_{d1} \, [\fm]$
    & $b_{u2} \, [\fm]$ & $b_{d2} \, [\fm]$ \\
\hline
ABM 1     & $0.627 \pm 0.003$ & $0.638 \pm 0.003$
          & $0.717 \pm 0.004$ & $0.693 \pm 0.004$ \\
power law & $0.697 \pm 0.011$ & $0.704 \pm 0.011$
          & $0.705 \pm 0.007$ & $0.736 \pm 0.016$ \\
\hline
\end{tabular}
\end{center}
\caption{\label{tab:imp-par} Average impact parameters
  \protect\eqref{avg-imp-par} for the flavor form factors $F_i^{q}$
  computed from our fits.}
\end{table*}

The proton charge radius is defined as
\begin{equation}
r^2_{pE} = 6\, \frac{\dd G_E^p(t)}{\dd t} \bigg|_{t=0} \,.
\end{equation}
We observe a significant discrepancy for this radius between our two fits
in \tab{tab:radii-Sachs} and between our fits and the value given by PDG
2012.  We note that the PDG value is a combination of determinations from
the Lamb shift in electronic hydrogen and from measurements of $G_E^p$ in
$ep$ scattering.  It is fully consistent with the result
\begin{align}
r_{pE} &= (0.8779 \pm 0.0094) \fm
\intertext{obtained from the Lamb shift in electronic hydrogen alone (see
  figure~2 in \cite{Pohl:2013yb}), but in strong disagreement with the
  very precise value}
r_{pE} &= (0.84184 \pm 0.00067) \fm
\end{align}
that has been extracted from the Lamb shift in muonic hydrogen
\cite{Pohl:2010zza}.  The origin of this discrepancy is currently unclear,
see the discussion and references in \cite{Pohl:2013yb}.  Given this
unresolved problem, we have refrained from using a value of $r_{pE}$ as
independent input to our form factor fits.

Our results for the squared neutron charge radius, defined in
\eqref{neutron-radius}, are also given in \tab{tab:radii-Sachs}.  Here the
PDG 2012 value, which is obtained from neutron scattering off the shell
electrons in nuclear targets, is used as a data point in our fits.  As it
happens, the power law fit reproduces this value perfectly, whereas the
default GPD fit does not. As mentioned above, we do not consider this to
be a serious problem, given the global nature of our fits.

For the flavor form factors we can define average impact parameters by
\begin{equation}
  \label{avg-imp-par}
b^2_{qi} = \frac{4}{F_i^q(0)} \, \frac{\dd F_i^q(t)}{\dd t} \bigg|_{t=0}
\end{equation}
with $i=1,2$ and $q=u,d$.  Notice the difference between the factor $4$ in
these two-dimensional quantities and the factor $6$ in the
three-dimensional radii discussed above.  The definition
\eqref{avg-imp-par} is analogous to the one of the impact parameter
$\langle \vbs^2 \rangle^q_x$ in \eqref{imp-par} but includes an average
over the momentum fraction $x$.  The results of our fits are given in
\tab{tab:imp-par} for reference.  As discussed above, they should not be
taken as precision determinations of these quantities, and the
discrepancies between the two fits within their respective parametric
errors should not be regarded as a problem.

%% file: properties.tex
\section{Properties of the fit}
\label{sec:fit-properties}

In this section we take a closer look at the GPDs we have extracted from
the form factor data.  Unless stated otherwise, the results shown are
obtained with the default fit ABM 1.

%%%%%%%%%%%%%%%%%%%%%%%

\subsection{Sensitive $x$ range}
\label{sec:xmin-xmax}

\begin{figure*}[t]
\begin{center}
\includegraphics[width=0.48\textwidth,%
viewport=125 300 580 650]{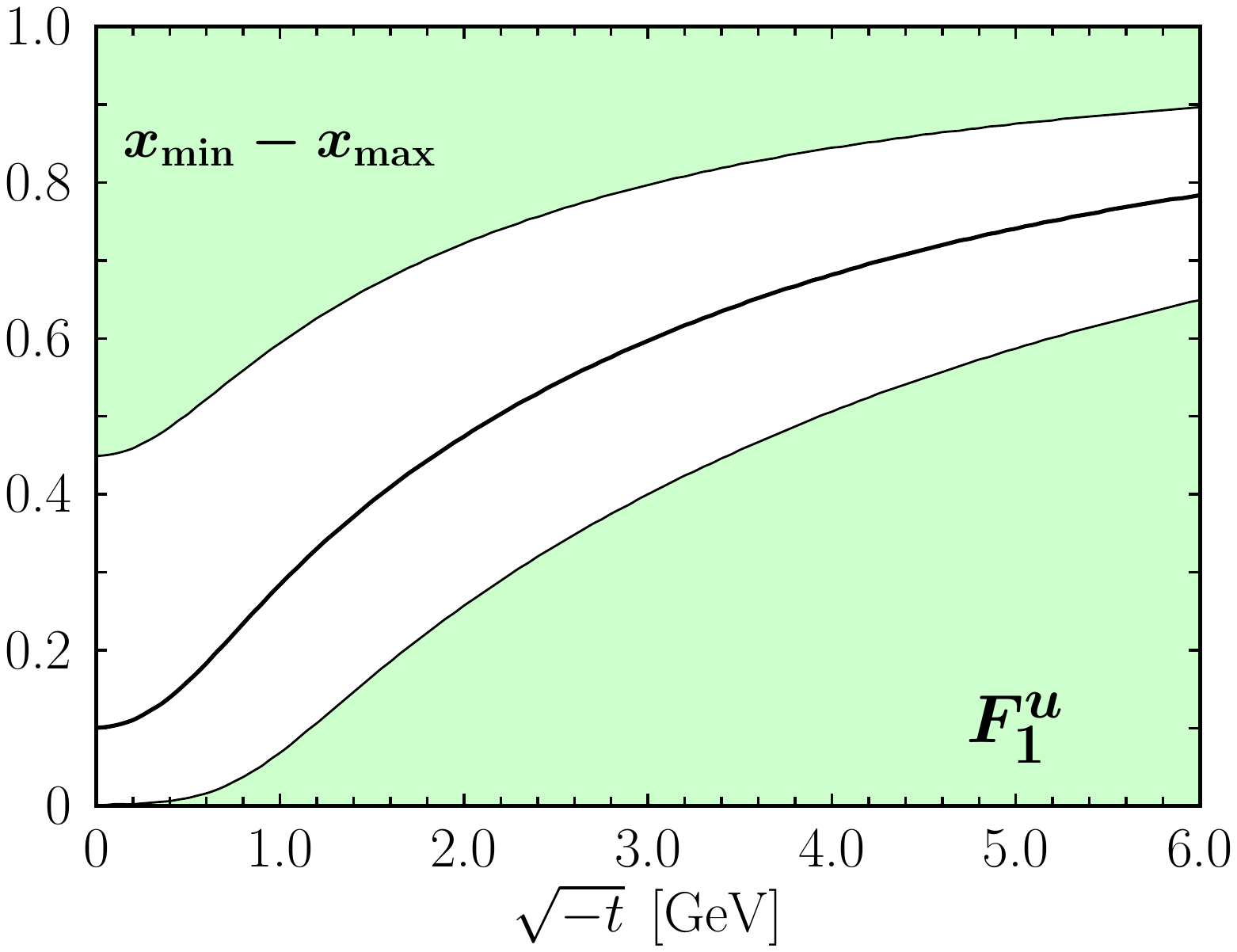}
\hspace{1em}
\includegraphics[width=0.48\textwidth,%
viewport=125 300 580 650]{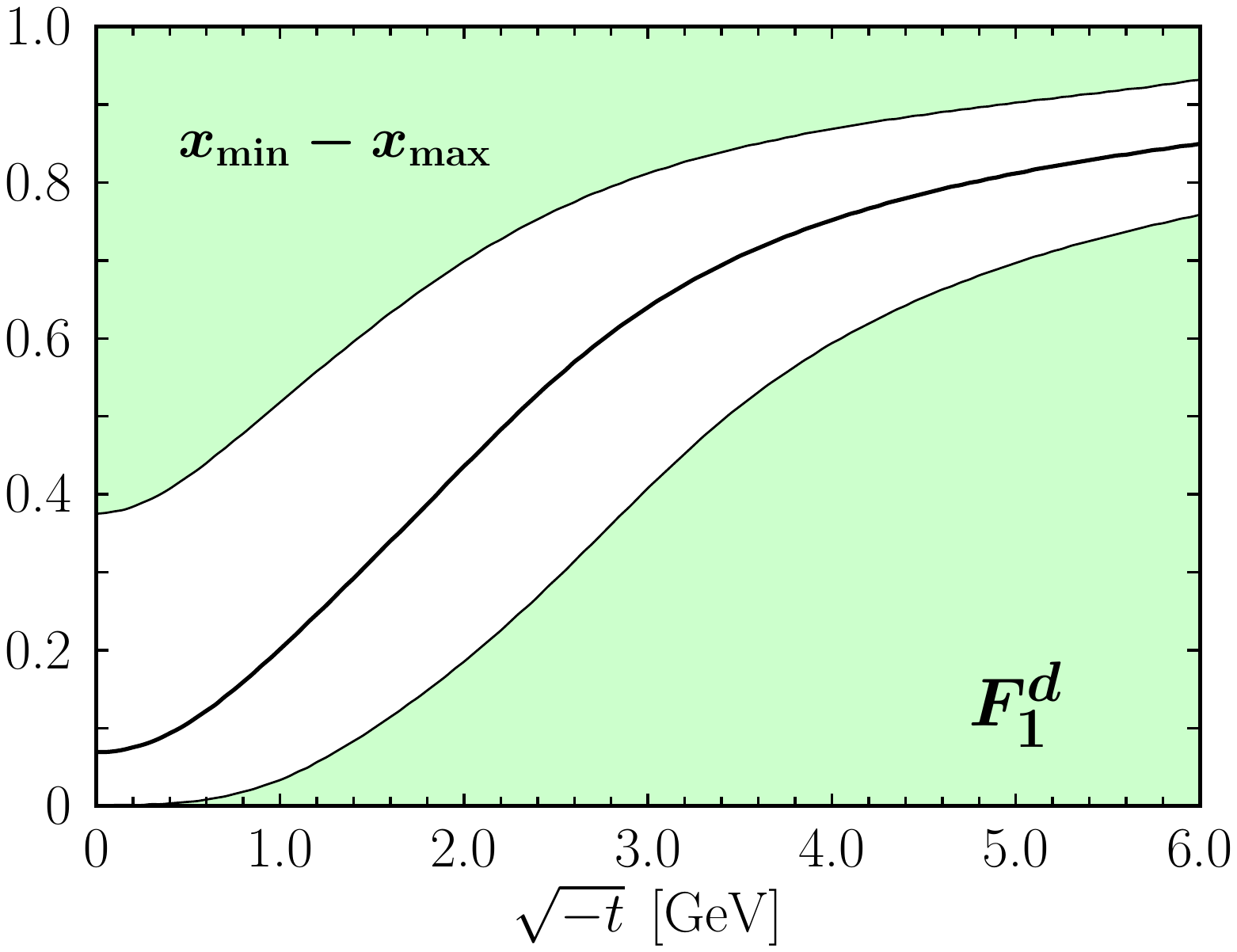} \\[1em]

\includegraphics[width=0.48\textwidth,%
viewport=105 300 560 650]{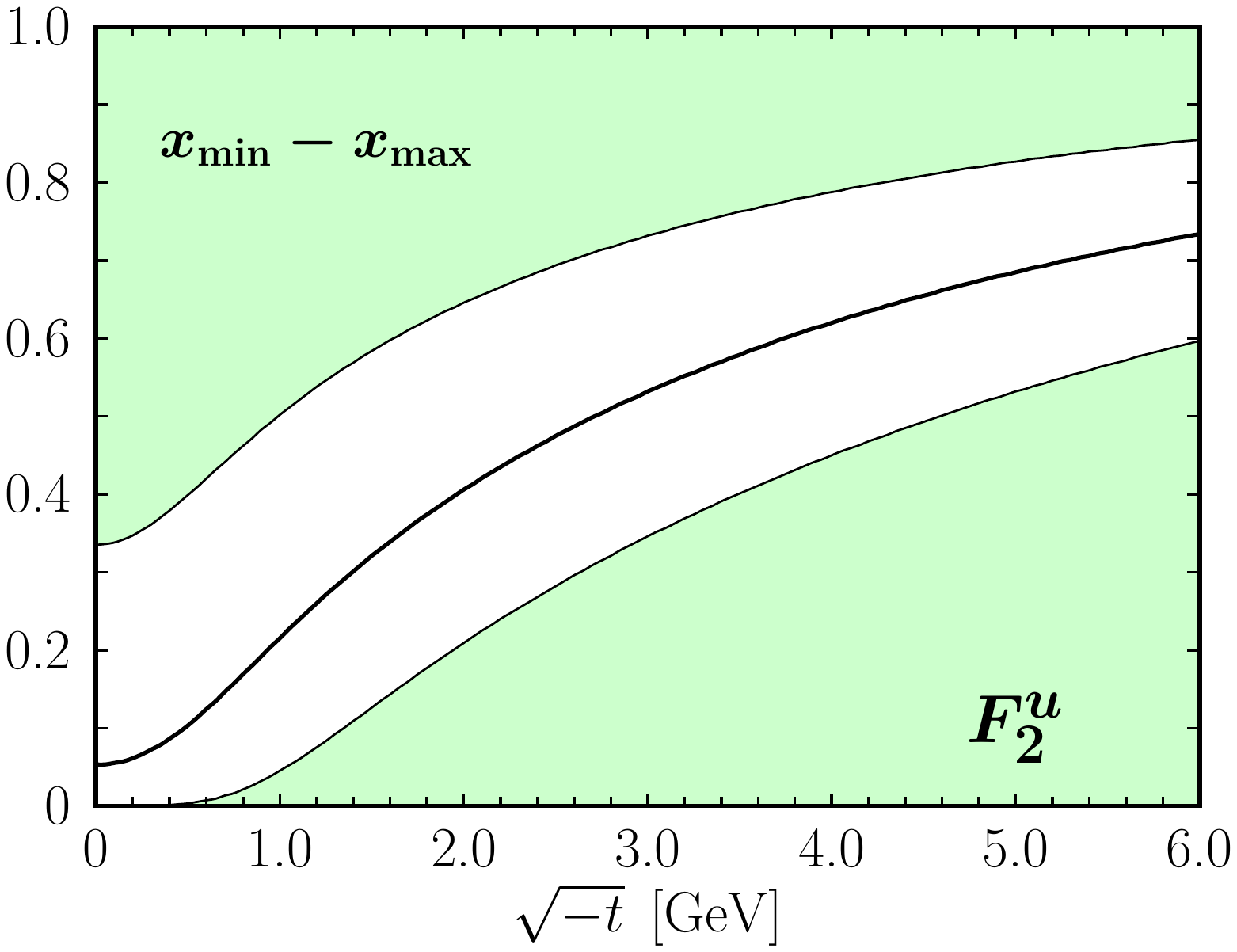}
\includegraphics[width=0.48\textwidth,%
viewport=105 300 560 650]{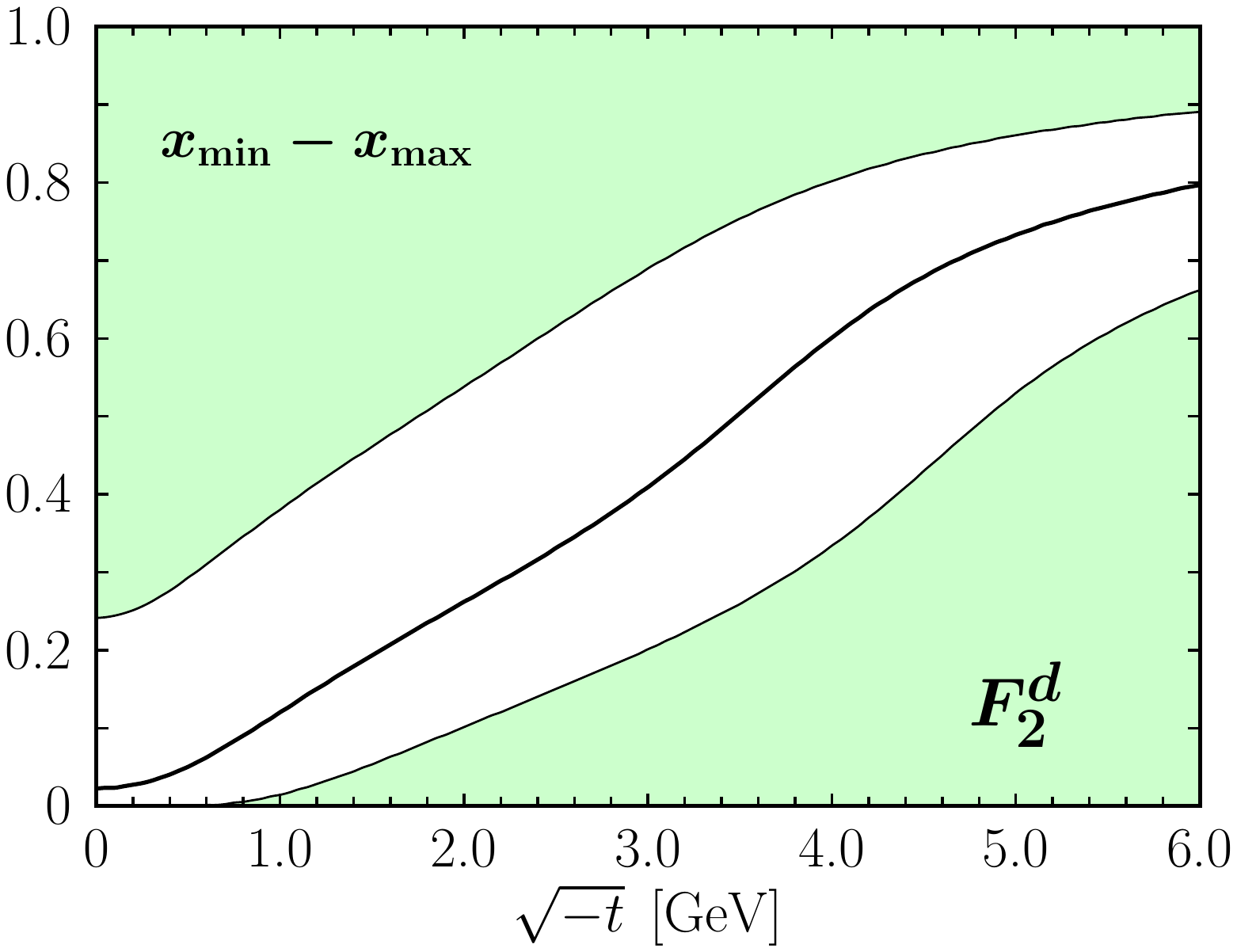}
\end{center}
\caption{\label{fig:xmin-xmax} Sensitive (white) and insensitive
  (shadowed) regions of $x$ for the flavor from factors in the default
  fit. The upper and lower shaded $x$-regions each account for $5\%$, the
  white regions for $90\%$ of the form factor integrals in
  \protect\eqref{eq:SumRule}.  The thick solid lines represent the median
  values of $x$.}
\end{figure*}

To begin with, we investigate which region of $x$ is most important in the
form factor integrals \eqref{eq:SumRule}.  We follow our previous work
\cite{DFJK4} and introduce a minimal and a maximal value of $x$ by
\begin{align}
    \int_0^{x_{\text{min}}(t)} {\dd} x\, K_v^q(x,t)
 &= \int^1_{x_{\text{max}}(t)} {\dd} x\, K_v^q(x,t)
\nonumber \\[0.3em]
 &= 0.05\, F^q_i(t)
\end{align}
for each of the GPDs $K_v^q = H_v^q, E_v^q$. The $x$ range from
$x_{\text{min}}$ to $x_{\text{max}}$ accounts for $90\%$ of the flavor
form factor $F^q_i$ in the sum rule. For $x < x_{\text{min}}$ and $x >
x_{\text{max}}$ the GPDs are obviously not well determined in our
analysis.  A typical value of $x$ in the integrals is given by the median
\begin{equation}
  \int_0^{x_{\text{med}}(t)} {\dd} x\, K_v^q(x,t) 
  = 0.5\, F^q_i(t) \,.
\end{equation}
In \fig{fig:xmin-xmax} we show these quantities for the four flavor form
factors.  We observe strong correlations between $x$ and $t$.  The form
factor data at small (large) $t$ determine the GPDs at small (large)
$x$. One also sees in \fig{fig:xmin-xmax} that the $x$ regions where the
GPDs are best determined are rather narrow and even shrink with increasing
$-t$. For $\sqrt{-t} = 6 \gev$ the sensitive $x$ values are between $0.6$
to $0.9$, while $x_{\text{max}}$ at $t=0$ is between $0.25$ and $0.45$
depending on the form factor.  The values of $x_{\text{min}}$ at $t=0$ are
very small and given in \tab{tab:xmin-xmax}.

\begin{table}[t]
\renewcommand{\arraystretch}{1.2} 
\begin{center}
\begin{tabular}{cc}
\hline     
form factor & $x_{\text{min}}(0)$ \\
\hline
$F_1^{u}$ & $1.5 \times 10^{-3}$ \\
$F_1^{d}$ & $9.2 \times 10^{-4}$ \\
$F_2^{u}$ & $2.5 \times 10^{-4}$ \\
$F_2^{d}$ & $3.6 \times 10^{-5}$ \\
\hline  
\end{tabular}
\end{center}
\caption{\label{tab:xmin-xmax} $x_{\text{min}}$ for the flavor form
  factors at $t=0$, evaluated from the default fit ABM 1.}
\end{table}

\begin{figure*}[t]
\begin{center}
\includegraphics[width=0.46\textwidth,%
viewport=85 185 420 640]{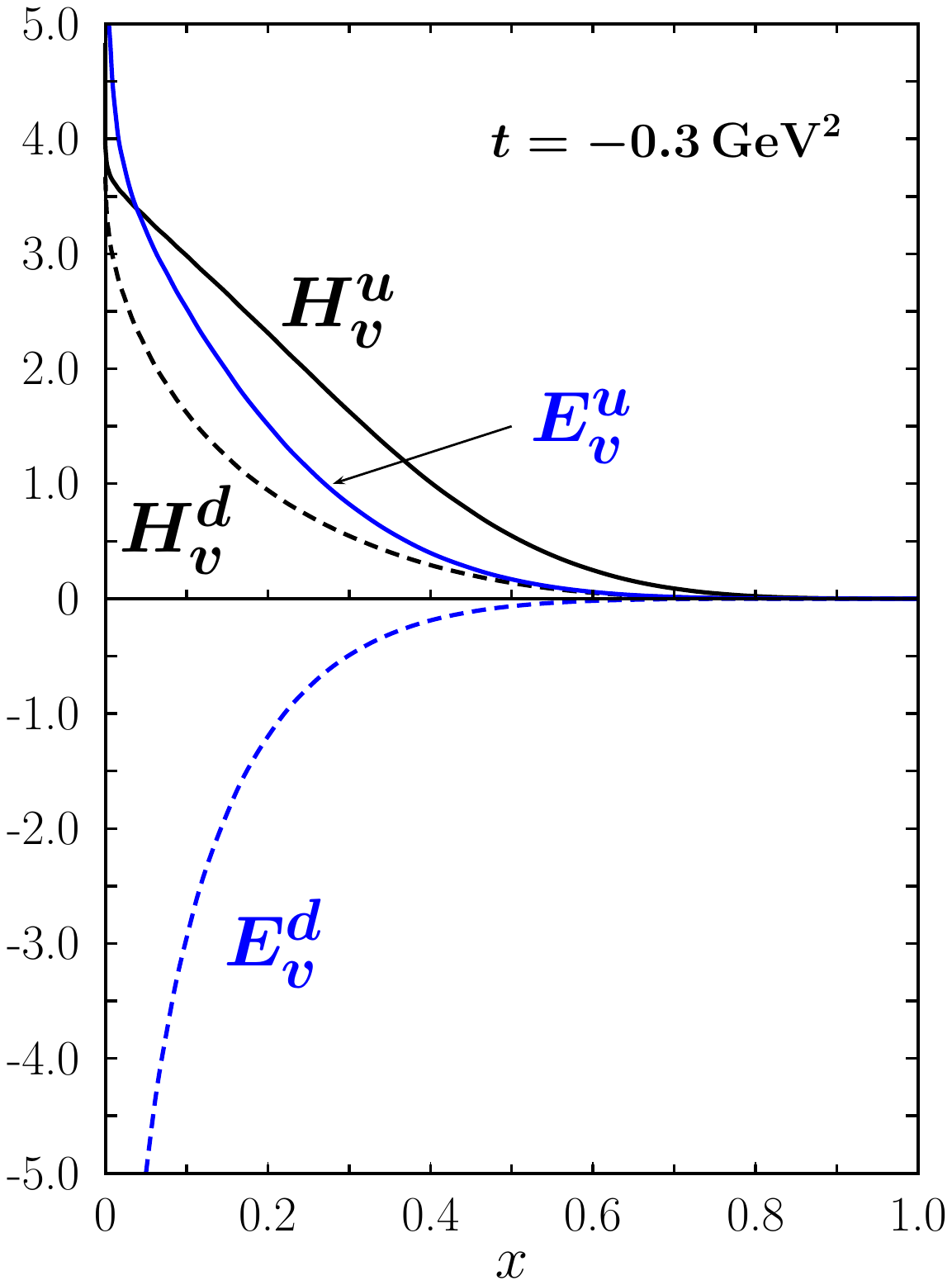}
\hspace{1em}
\includegraphics[width=0.49\textwidth,%
viewport=75 162 430 610]{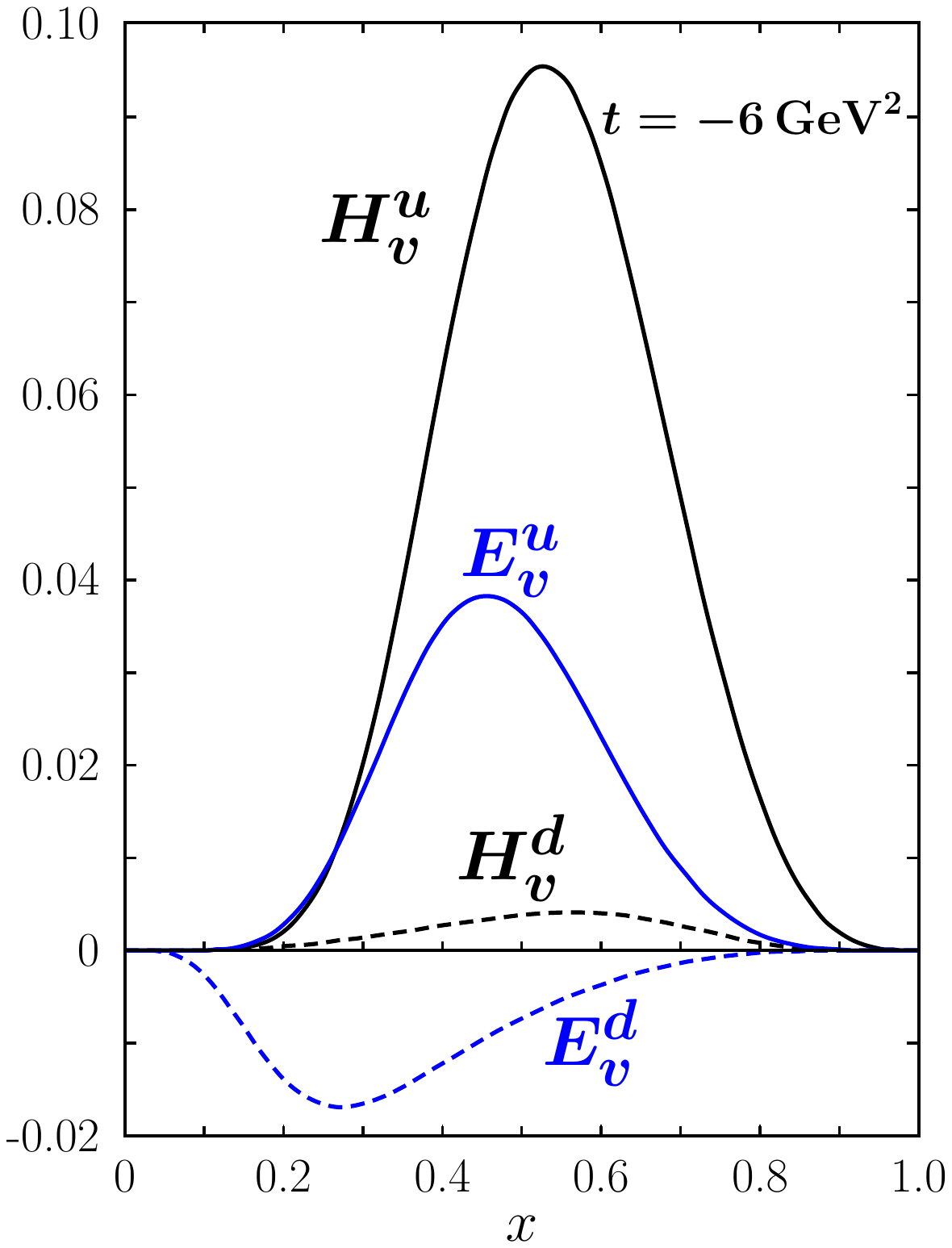}
\end{center}
\caption{\label{fig:gpd} The valence quark GPDs at zero skewness and $-t=
  0.3$ or $6 \gev^2$.  They have been obtained in the default fit and
  refer to the scale $\mu=2 \gev$.}
\end{figure*}

The strong $x$ -- $t$ correlation is a consequence of our parameterization
of the GPDs.  Their forward limits are singular as $x^{-\alpha}$ which is
explicit for $E_v^q$ and hidden in the PDFs for $H_v^q$.  The values of
the Regge intercept $\alpha$ range from $0.33$ to $0.62$ for our default
fit.  The profile functions, on the other hand, provide factors $x^{-t
  \alpha'}$.  Combining both $x$-factors one has a Regge behavior with a
linear trajectory in the exponent of $x$ which crosses zero at $t_0 =
-\alpha /\alpha'$.  This happens near $-t= 0.5 \gev^2$ and turns
the singular behavior at $x=0$ for small $t$ into a zero of the GPD at
$x=0$ for larger $t$.  For yet larger $t$, each of the GPDs develops a
pronounced maximum at an $x$ value that increases with $-t$.  This
behavior is responsible the $x$ -- $t$ correlation we observe.  In
\fig{fig:gpd} we show as an illustration the GPDs at $-t=0.3$ and $6.0
\gev^2$.

We are now in a position to investigate the large-$t$ behavior of the
flavor form factors.  Since it is controlled by the large-$x$ behavior of
the GPDs, we can approximate their forward limits by $(1-x)^\beta$ and the
profile function by the third term $A (1-x)^2$.  (For simplicity we use
here the same notation for $H_v^q$ and $E_v^q$ and drop sub- and
superscripts on $\beta$ and $A$.)  At sufficiently large $t$ the integral
$F(t) = \int \dd x K(x,t)$ can be evaluated in the saddle-point
approximation, obtained by minimizing the exponent in $K \sim \exp{\bigl[
  \beta \log{(1-x)} + t A (1-x)^2 \bigr]}$ with respect to $x$.  One finds
\cite{DFJK4}
\begin{align}
  \label{eq:power-law} 
F(t)  \sim (-t)^{-(1+\beta)/2}_{\phantom{1}}
\end{align}
with the saddle point being at
\begin{align}
x_s = 1 - \sqrt{\beta} \big/ \sqrt{-2tA} \,.
\end{align}
We see that the saddle point lies in the so-called soft region
\begin{equation}
1-x \sim \Lambda/\sqrt{-t}
\end{equation}
with $\Lambda$ of hadronic size, where for sufficiently large $t$ the
active parton carries most of the proton's momentum while all spectators
are soft.  The dominance of this soft region has been assumed by Drell and
Yan \cite{drell-yan} (see also \cite{burkardt03}) in order to derive the
famous relation between the large-$t$ behavior of the form factors and the
large-$x$ behavior of the deep inelastic structure functions.

\begin{figure}
\begin{center}
\includegraphics[width=0.49\textwidth,%
viewport=101 370 560 710]{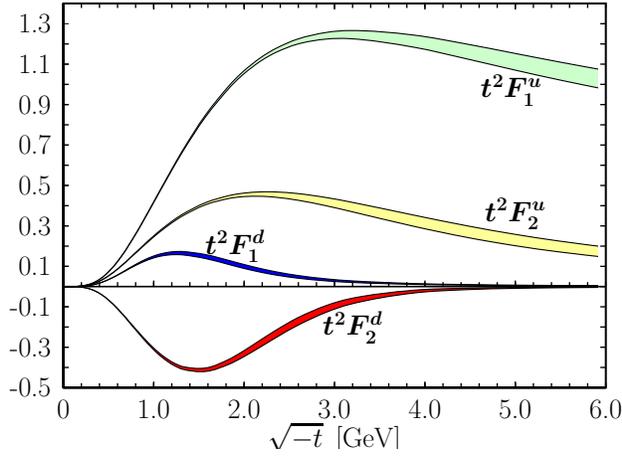}
\end{center}
\caption{\label{fig:moments-extrapolation} The flavor form factors scaled
  by $t^2$ and shown in units of $\gev^4$, calculated from our default
  fit ABM 1.}
\end{figure}

The derivation of \eqref{eq:power-law} requires the relevant GPD to behave
like $(1-x)^\beta\ms \exp{\bigl[\ms t A (1-x)^2 \bigr]}$ in the sensitive
$x$ region and to be sufficiently peaked around the saddle point $x_s$.
Let us see how well \eqref{eq:power-law} works quantitatively, restricting
ourselves to the region $\sqrt{-t} < 6 \gev$ where form factor data is
available.  From \tabs{tab:pdfs} and \ref{tab:fit-basics} we obtain the
following powers of $1/(-t) \ms$:
\begin{align}
2.25 & ~\text{for}~ F_1^u \,,
&
3.00 & ~\text{for}~ F_1^d \,,
\nonumber \\
2.83 & ~\text{for}~ F_2^u \,,
&
3.13 & ~\text{for}~ F_2^d \,.
\end{align}
Both $F_1^u$ and $F_2^u$ are well described with these powers when
$\sqrt{-t} > 4 \gev$, whereas for $F_1^d$ and $F_2^d$ the prediction
\eqref{eq:power-law} works only qualitatively.  In our fit result for
$F_1^d$ we find a power of $3.35$ when $\sqrt{-t} > 4 \gev$, whereas for
$F_2^d$ we see a clear power-law behavior only above $\sqrt{-t} = 4.5
\gev$, with the power being $4.2$.  Regarding their absolute size, we see
in \fig{fig:moments-extrapolation} that the $d$ quark form factors are
suppressed compared to their $u$ quark counterparts for $\sqrt{-t} > 2.5
\gev$ or so.  A consequence of this suppression is that
\begin{equation}
  F_i^n \approx \frac{e_d}{e_u}\, F_i^p
\end{equation}
at large $t$. Our results approach this behavior slowly; for $\sqrt{-t} >
4 \gev$ is holds within $10\%$.

%%%%%%%%%%%%%%%%%%%%%%%%%%%%%%%%%%%%

\subsection{Mellin moments and Ji's sum rule}
\label{sec:second-moments}

\begin{figure*}[t]
\begin{center}
\includegraphics[width=0.45\textwidth,%
viewport=100 370 570 720]{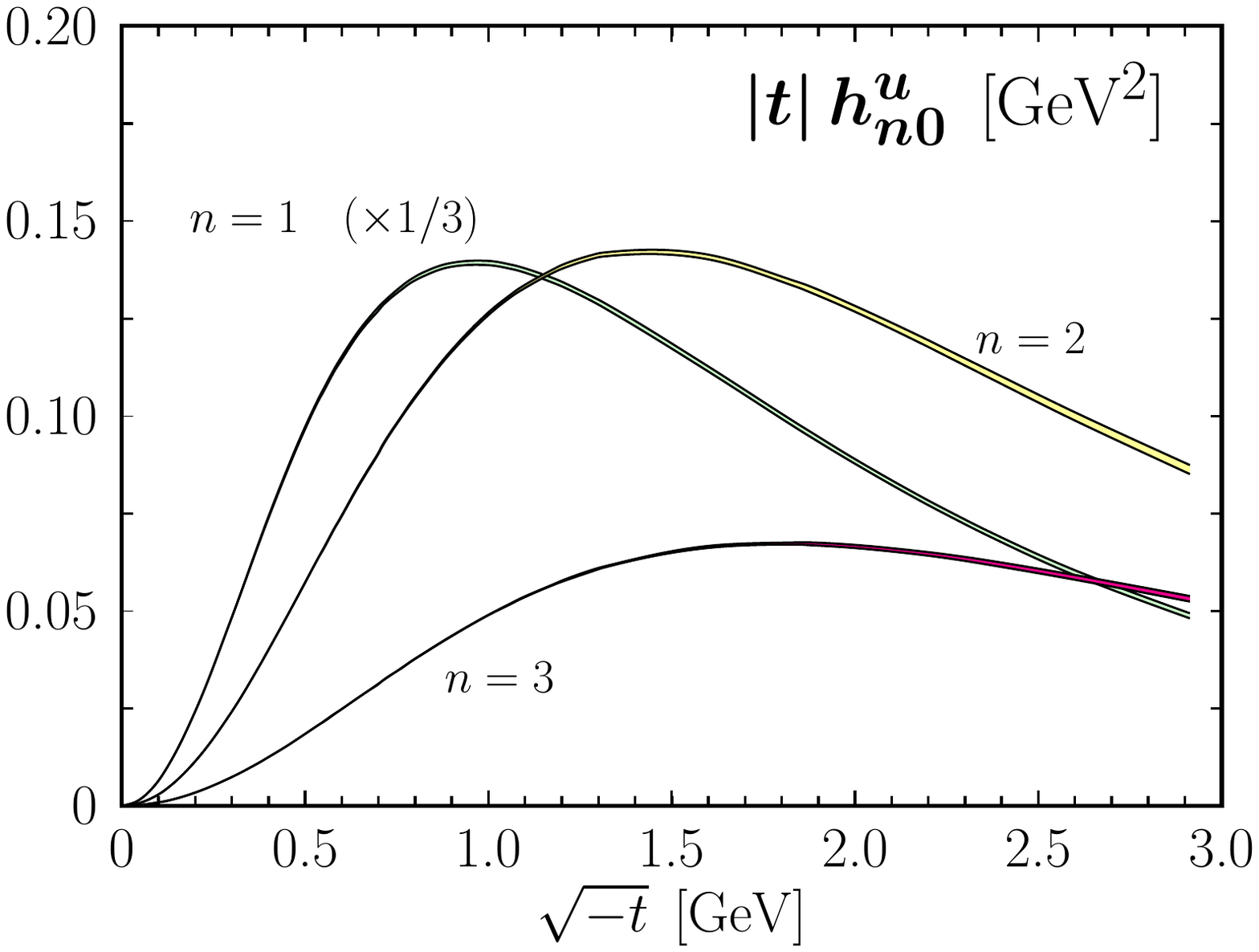}
\hspace{1.5em}
\includegraphics[width=0.45\textwidth,%
viewport=100 355 570 705]{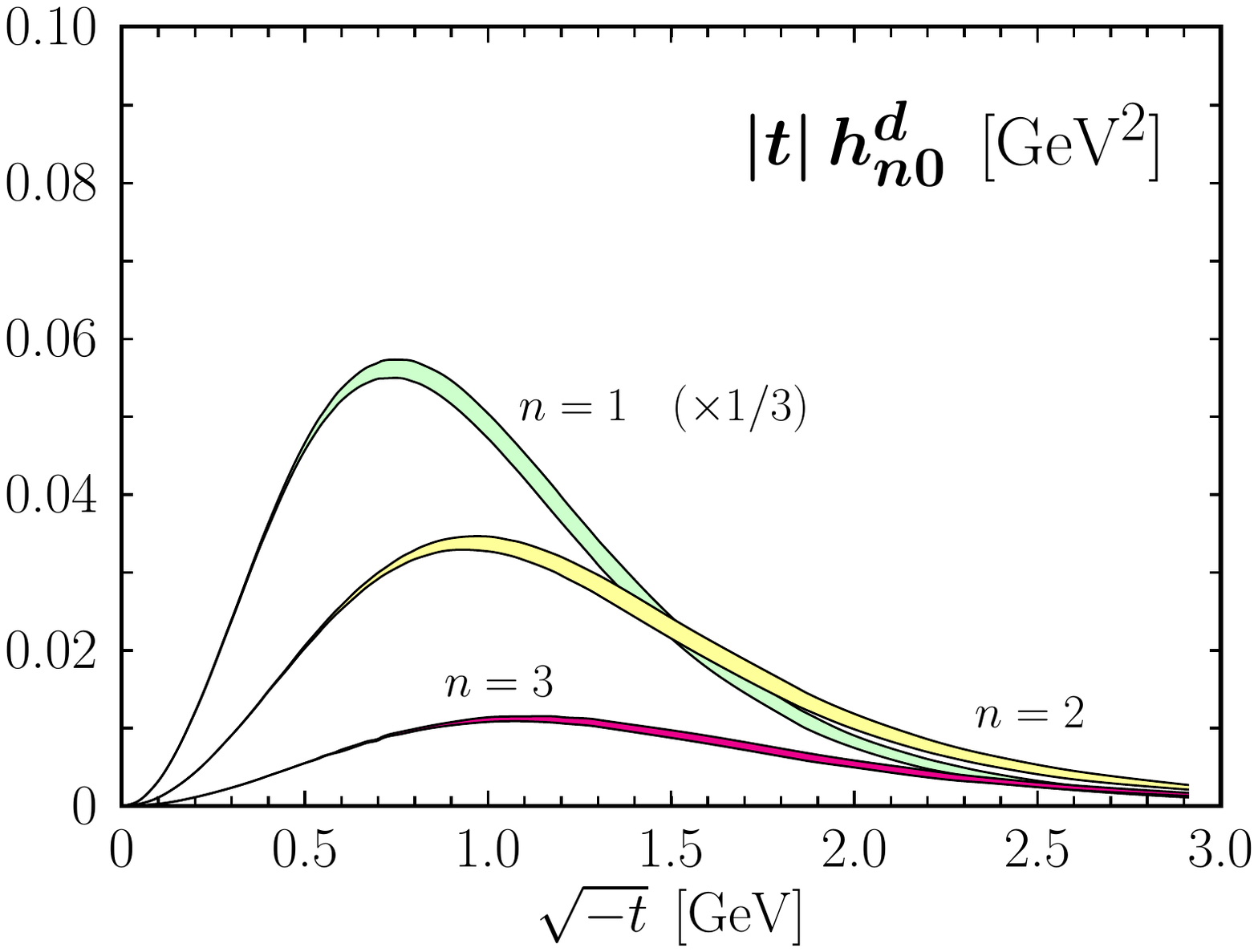} \\[1em]

\includegraphics[width=0.45\textwidth,%
viewport=105 305 570 655]{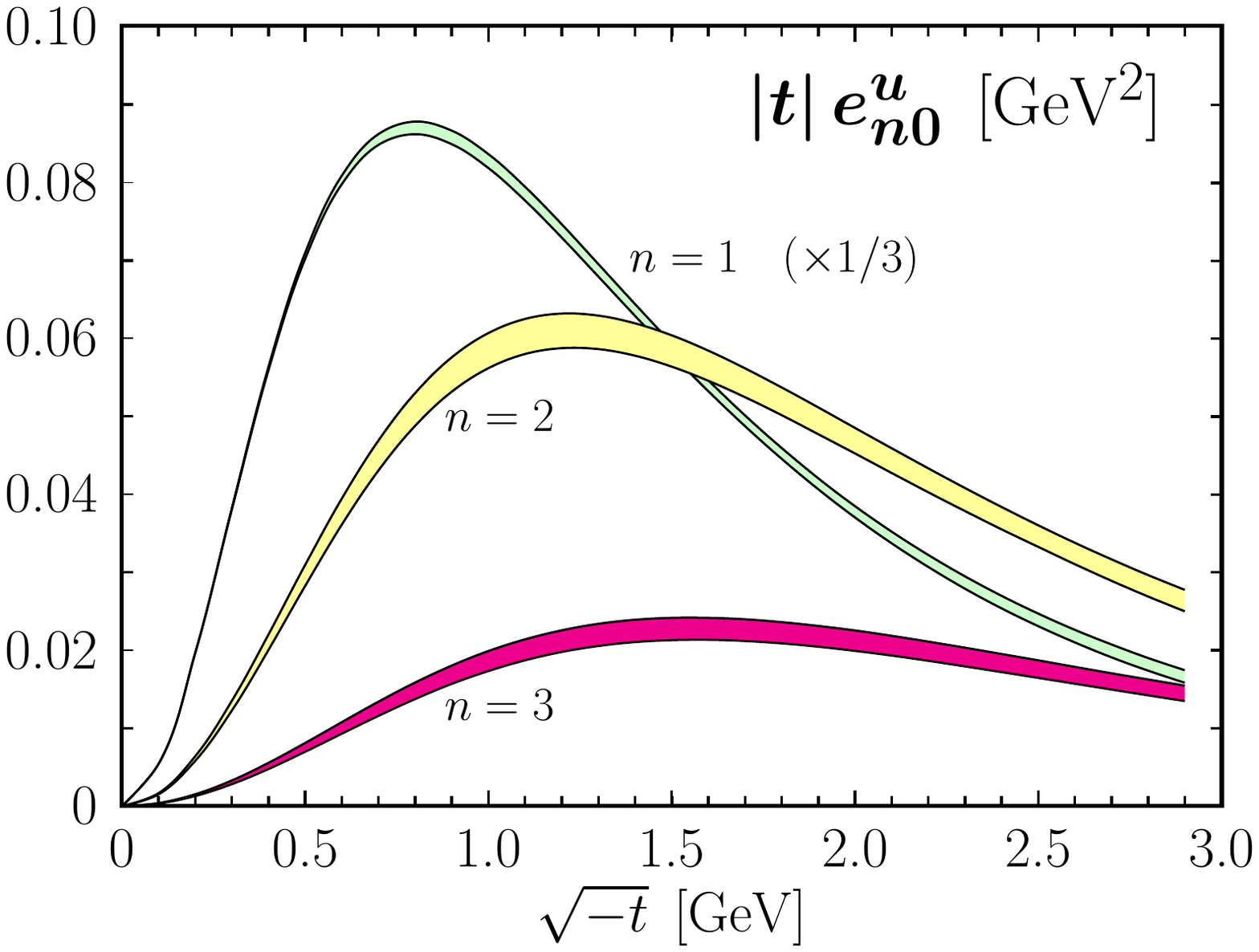}
\hspace{1.5em}
\includegraphics[width=0.45\textwidth,%
viewport=105 295 570 645]{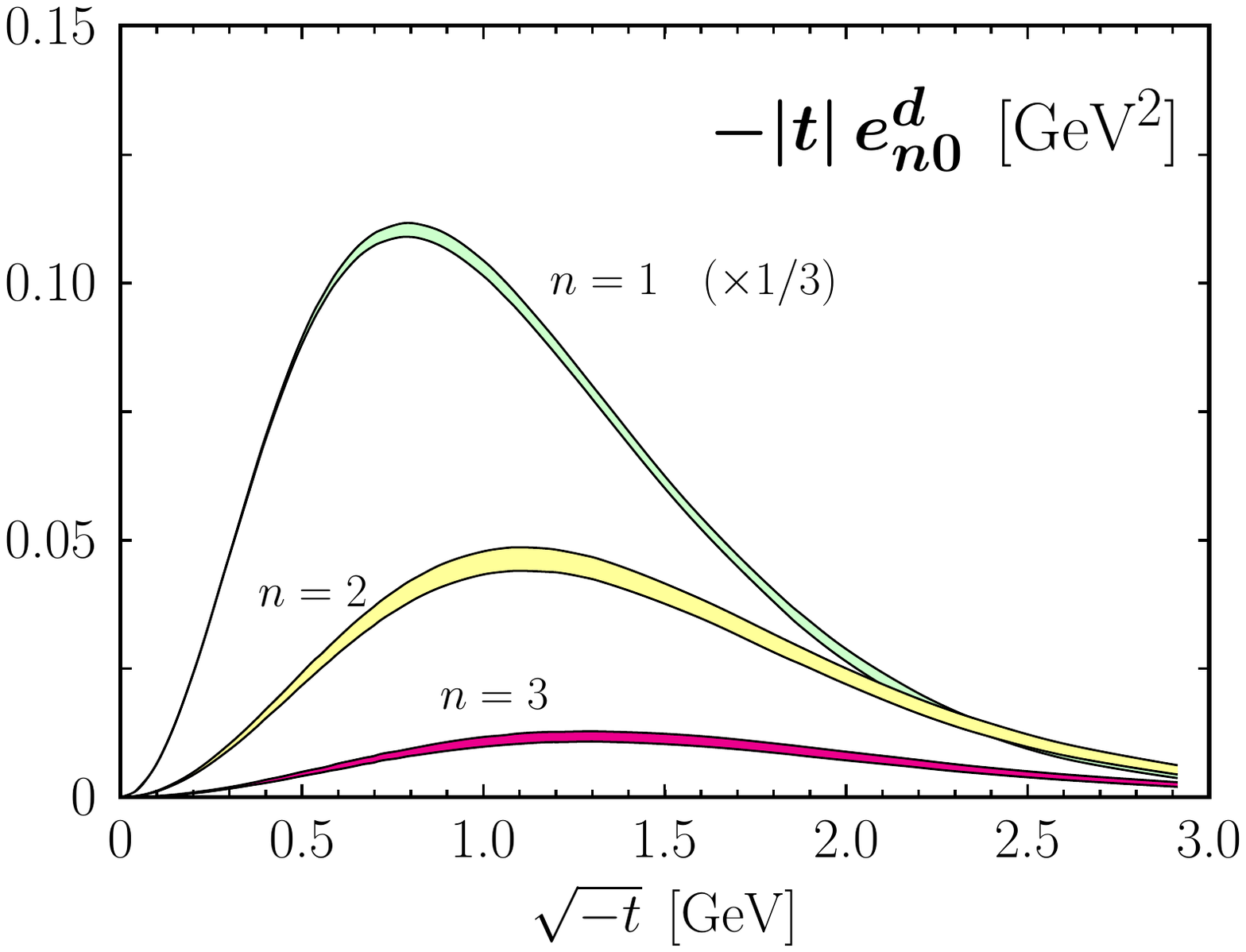}
\end{center}
\caption{\label{fig:moments} The first three moments of the GPDs obtained
  in our default fit ABM 1.  For better visibility of the higher moments,
  $h_{10}^q$ and $e_{10}^q$ have been scaled with $1/3$.}
\end{figure*}

We have determined the GPDs $H_v^q$ and $E_v^q$ by fitting their integrals
over $x$ to the form factors $F_1^q$ and $F_2^q$ of the electromagnetic
current.  We now use these GPDs to compute higher moments in $x$.  In
\fig{fig:moments} we show the moments
\begin{align}
h_{n0}^{q}(t) &= \int_0^1 \dd x\, x^{n-1} H_v^q(x,t) \,,
\nonumber \\
e_{n0}^{q}(t) &= \int_0^1 \dd x\, x^{n-1} E_v^q(x,t) \,,
\end{align}
for $n=1,2,3$. In keeping with a standard notation
\cite{Diehl:2003ny,Belitsky:2005qn}, the second subscript $0$ indicates
that the GPDs are evaluated at zero skewness.

The third moments $h^q_{30}$ and $e^q_{30}$ are form factors of a
twist-two operator containing two covariant derivatives
\cite{Diehl:2003ny,Belitsky:2005qn}. The second moments $h^q_{20}$ and
$e^q_{20}$ are not directly connected to local operators, since the form
factors of the relevant twist-two operator with one covariant derivative
correspond to the sum and not to the difference of quark and antiquark
contributions. In other words, the moments $h^q_{20}$ and $e^q_{20}$ give
the valence contributions (i.e.\ the difference of quark and antiquark
contributions) to the form factors of a twist-two operator, which happens
to be the quark part of the energy-momentum tensor.

\begin{table*}[p]
\begin{center}
\renewcommand{\arraystretch}{1.2}
\begin{tabular}{lccccccc} \hline
PDF     & $u-\bar{u}$ & $d-\bar{d}$ & $2 \bar{u}$ & $2 \bar{d}$
        & $2 (\bar{u} + \bar{d})$ & $s+\bar{s}$ & $s-\bar{s}$
\rule{0pt}{2.6ex} \\
\hline
ABM     & $0.297$ & $0.115$ & $0.062$ & $0.077$ & $0.139$ & $0.035$ & 0 \\ 
CT      & $0.287$ & $0.118$ & $0.058$ & $0.072$ & $0.130$ & $0.040$ & 0 \\
GJR     & $0.280$ & $0.116$ & $0.064$ & $0.080$ & $0.144$ & $0.021$ & 0 \\
HERAPDF & $0.284$ & $0.105$ & $0.074$ & $0.091$ & $0.165$ & $0.044$ & 0 \\
MSTW    & $0.282$ & $0.115$ & $0.064$ & $0.076$ & $0.140$ & $0.033$ 
        & $0.0019$ \\
NNPDF   & $0.290$ & $0.124$ & $0.059$ & $0.074$ & $0.133$ & $0.020$ 
        & $0.0029$ \\
\hline
\end{tabular}
\end{center}
\caption{\label{tab:2nd-mom-PDFs} Second moments of the parton densities
  specified in \protect\tab{tab:pdfs}, given at scale $\mu=2 \gev$.}
\end{table*}

\begin{table*}[p]
\begin{center}
\renewcommand{\arraystretch}{1.5}
\begin{tabular}{lcccc} \hline
fits     & $e_{20}^{u}(0)$   & $e_{20}^{d}(0)$ 
         & $e_{20}^{u+d}(0)$ & $e_{20}^{u-d}(0)$ \\
\hline
ABM 1    & $0.163 {}^{+0.010}_{-0.006}$ & $-0.122 {}^{+0.008}_{-0.006}$
         & $0.041 {}^{+0.008}_{-0.010}$ &  $0.284 {}^{+0.012}_{-0.012}$ \\
all fits & $0.163 {}^{+0.018}_{-0.032}$ & $-0.122 {}^{+0.028}_{-0.033}$
         & $0.041 {}^{+0.011}_{-0.053}$ &  $0.284 {}^{+0.040}_{-0.060}$ \\
\hline
ABM 0    & $0.204 {}^{+0.009}_{-0.010}$ & $-0.156 {}^{+0.009}_{-0.009}$ 
         & $0.048 {}^{+0.006}_{-0.010}$ &  $0.360 {}^{+0.017}_{-0.017}$ \\
\hline
\end{tabular}
\end{center}
\caption{\label{tab:2nd-mom-E} Results for second moments of $E_v^q$ at
  $t=0$.  The central values in the rows labeled ABM 1 and ``all fits''
  correspond to the default fit and refer to $\mu = 2\gev$, with two
  different error estimates as described in the text.  The last row gives
  the moments obtained from the fit ABM 0 and refers to $\mu=1 \gev$.}
\end{table*}

\begin{figure*}[p]
\begin{center}
\includegraphics[width=\textwidth,%
viewport=0 0 570 210]{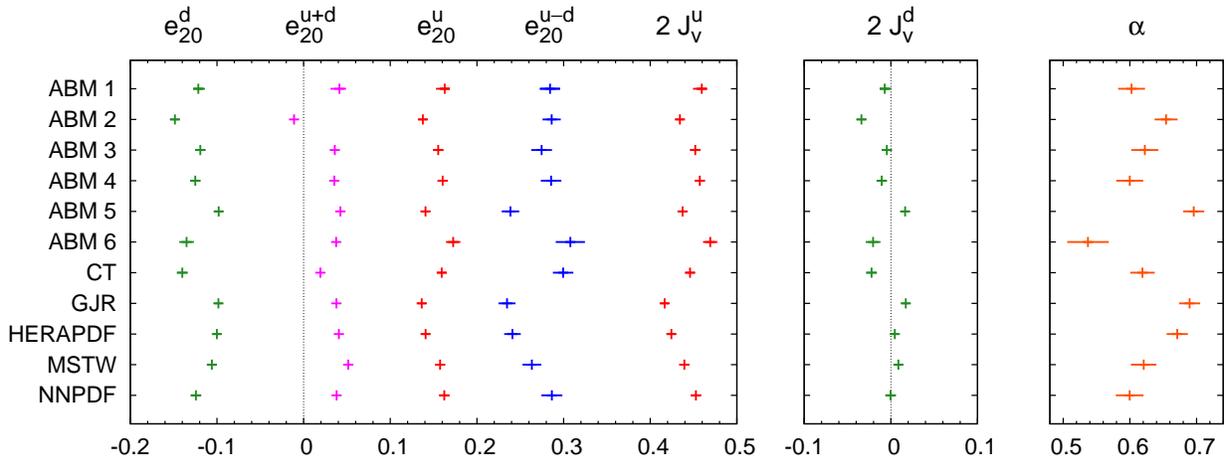}
\end{center}
\caption{\label{fig:2nd-mom} Results for the second moments of GPDs at
  $t=0$ and for the parameter $\alpha$ from the GPD fits specified in
  \tab{tab:fit-basics}.  Only the error bars for fit ABM 1 include the
  uncertainty that results from varying $\beta_u$ and $\beta_d$, which is
  computed as described in \protect\sect{sec:def-fit}.}
\end{figure*}

In \fig{fig:moments} we observe a strong decrease of the moments with the
index $n$, which can be understood from the rather strong decrease of the
GPDs with $x$. Moreover, we find that decrease of the moments with $-t$
becomes slower as $n$ increases. Within our ansatz, this is naturally
explained by the decrease of the profile functions with $x$, which results
in a weaker $t$ slope for the moments that are dominated by higher values
of $x$. The same trend has been observed in lattice calculations of GPD
moments, see e.g.\ \cite{Bratt:2010jn} and the review
\cite{Hagler:2009ni}. We refrain, however, from a quantitative comparison
with lattice results, since recent studies \cite{Bali:2012av,Green:2012ud}
suggest that in lattice computations of GPD moments it is more involved
than previously assumed to achieve full control over the extrapolation to
the physical pion mass and over the removal of contributions from excited
nucleon states.

We now take a closer look at the second moments at $t=0$. For $H_v^q$ they
just give the momentum fraction carried by the parton species in
question. The corresponding numbers for the PDF sets used in our analysis
are shown in \tab{tab:2nd-mom-PDFs}. We see that these momentum fractions
are known with reasonable accuracy, although there is a notable spread
between the different PDF sets. We can also assess the importance of sea
quark contributions in the momentum sum rule. The numbers in
\tab{tab:2nd-mom-PDFs} show that the second moment of the distribution $2
(\bar{u} + \bar{d})$, i.e.\ the momentum fraction carried by the
non-strange sea is comparable to the momentum fraction due to the valence
$d$ quark distribution.  The contribution from strange quarks and
antiquarks appears to be relatively unimportant.

The second moments of $E_v^q$ are given in \tab{tab:2nd-mom-E}. In the
first row we give the results of fit ABM 1 with errors computed as
described in \sect{sec:def-fit} (i.e.\ including the variation of
$\beta_u$ and $\beta_d$). A more conservative uncertainty estimate is
given in the second row, where we take the central values of fit ABM 1 and
determine the error from the spread of values from fits ABM 2 to ABM 6 and
from the fits with the alternative PDF sets in \tab{tab:fit-basics}. The
only notable source of uncertainty not included in this estimate is the
bias from our functional ansatz for the GPDs themselves, which we cannot
assess within our present study. A graphical display of the second moments
obtained in the different fits is shown in \fig{fig:2nd-mom}. We find that
$e_{20}^u(0)$ and $e_{20}^d(0)$ are determined rather well. Their sum is
found to be very small; within its error it may be positive or negative.

In the third row of \tab{tab:2nd-mom-E} we give the second moments from
fit ABM 0, which refers to the lower scale $\mu = 1 \gev$. The
uncertainties are computed as for the first row, including again the
variation of $\beta_u$ and $\beta_d$.  The moments $e_{20}^u(0)$ and
$e_{20}^d(0)$ are larger in absolute magnitude than at $\mu = 2 \gev$,
which is in line with what is expected from scale evolution. The
isosinglet combination $e_{20}^{u+d}(0)$ remains very small.

\begin{table}[b]
\begin{center}
\renewcommand{\arraystretch}{1.5}
\begin{tabular}{lcc} \hline
fits  & $2 J^{u}_v$ & $2 J^{d}_v$ \\
\hline
ABM 1    & $0.460 {}^{+0.006}_{-0.010}$ & $-0.007 {}^{+0.008}_{-0.006}$ \\
all fits & $0.460 {}^{+0.018}_{-0.048}$ & $-0.007 {}^{+0.021}_{-0.033}$ \\
\hline
ABM 0    & $0.560 {}^{+0.009}_{-0.010}$ & $-0.019 {}^{+0.009}_{-0.009}$ \\
\hline
\end{tabular}
\end{center}
\caption{\label{tab:Jq} Results for the total angular momentum of quarks
  minus the contribution from antiquarks according to Ji's sum rule.  For
  further explanations see \tab{tab:2nd-mom-E}.}
\end{table}

According to Ji \cite{Ji:1996ek}, the sum
\begin{equation}
  \label{Ji-sum-rule}
2 J_v^q = h_{20}^q(0) + e_{20}^q(0)
\end{equation}
of second moments gives two times the angular momentum carried by quarks
of flavor $q$, minus the corresponding antiquark contribution. Its values
for our different fits are shown in \fig{fig:2nd-mom} and listed in
\tab{tab:Jq} with error estimates analogous to those in
\tab{tab:2nd-mom-E}. We find that at $\mu = 2 \gev$ valence $u$ quarks
carry about half of the nucleon spin. By contrast, the contribution of
valence $d$ quarks is very small, even consistent with zero, as a result
of the cancellation between $h_{20}^u(0)$ and $e_{20}^u(0)$. Let us
emphasize that we cannot separately determine the contributions from
antiquarks to the nucleon spin in our analysis. If the second moments of
the PDFs are any guidance, then this contribution may not be negligible.

Within errors, our numbers for $J^{u}_v = J^{u-\bar{u}}$ and $J^{d}_v =
J^{d-\bar{d}}$ are consistent with the results in \cite{Bacchetta:2011gx},
where these quantities are estimated using a model dependent connection
between GPDs and Sivers distributions (see \sect{sec:lensing}).  Our
numbers are also close to the corresponding values of $J^{u+\bar{u}}$ and
$J^{d+\bar{d}}$ obtained in lattice calculations \cite{Hagler:2009ni},
although such a comparison should be taken with care given the comments we
made earlier in this section.  An overview of other determinations of
$J^{u+\bar{u}}$ and $J^{d+\bar{d}}$ can be found in
\cite{Bacchetta:2012xf}.

%%%%%%%%%%%%%%%%%%%%%%%%%%%%%%%%%%%%%

\subsection{Impact parameter distributions}
\label{sec:distance}

In \sect{sec:fit-ansatz} we explained how the GPDs $H_v^q(x,t)$ and
$E_v^q(x,t)$, after Fourier transformation to impact parameter space, give
information about the spatial distribution of partons with momentum
fraction $x$ in the impact parameter plane.  Let us now see how this plays
out quantitatively for our default fit ABM 1.

The impact parameter $\vbs$ introduced in \sect{sec:fit-ansatz} gives the
distance of a parton from the center of the proton, where the center is
determined by the transverse positions of all proton constituents,
weighted with their momentum fraction \cite{Burkardt:2002hr}.  The impact
parameter of a parton with $x$ close to $1$ thus tends to coincide with
the center of the proton, and a better quantity to assess the overall
proton size in that case is the distance $\vbs /(1-x)$ between the struck
parton and the center of all spectator partons in the transverse plane.
Following \cite{DFJK4}, we introduce the average
\begin{equation}
  \label{def-dq}
d_q(x) = \frac{\sqrt{\langle \vbs^2\rangle^q_x}}{1-x}
       = \frac{2 \sqrt{f_q(x)}}{1-x} \,.
\end{equation}
of this distance.  It is plausible to require that the proton remains of
finite transverse size for configurations where one parton has momentum
fraction $x\to 1$, which implies that $\langle \vbs^2 \rangle_x^q \sim
(1-x)^2$ in that limit as we anticipated in \sect{sec:fit-ansatz}.

\begin{figure}[t]
\begin{center}
\includegraphics[width=0.47\textwidth,%
viewport=70 160 520 500]{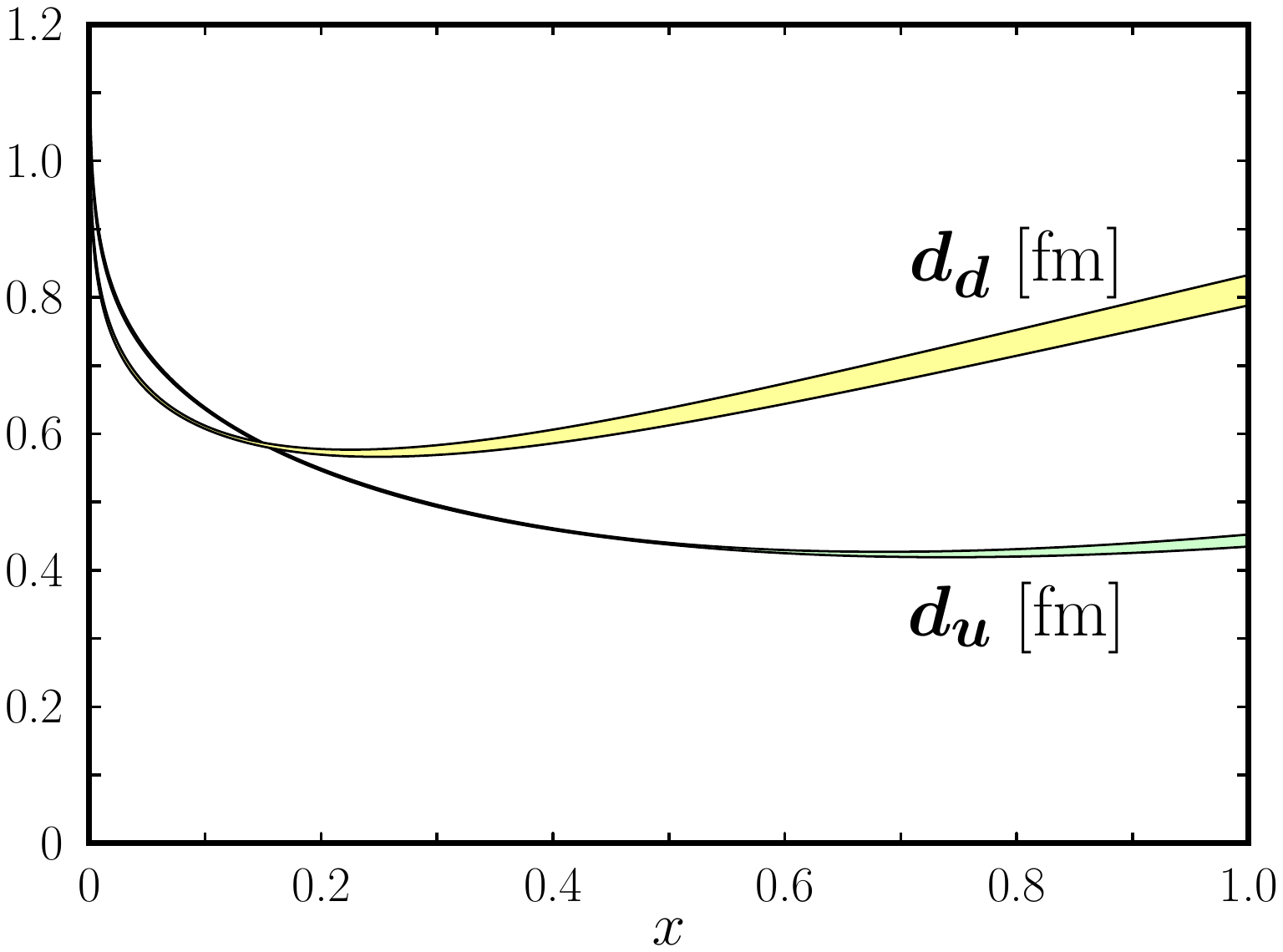} \\[1em]
\includegraphics[width=0.47\textwidth,%
viewport=150 315 595 645]{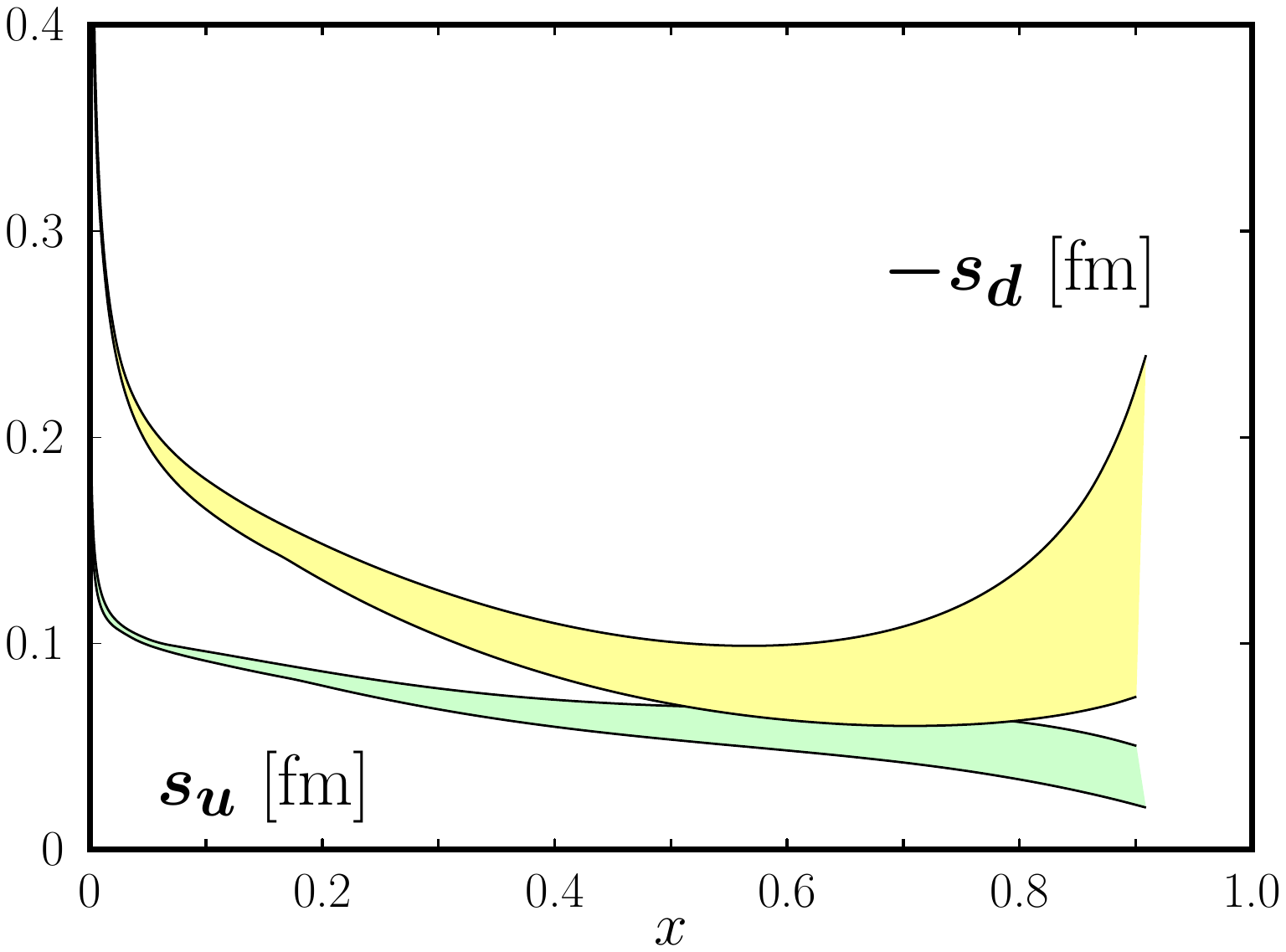}
\end{center}
\caption{\label{fig:distance} The average distance $d_q$ between the
  active quark and the cluster of spectators for $u$ and $d$ quarks, and
  the average shift $s_q$ of this distance along $y$ in a proton polarized
  along the $x$ direction.}
\end{figure}

In \fig{fig:distance} this distance is shown for $u$ and $d$ quarks.  The
values of $d_q$ appear reasonable for an estimate of the transverse size
of the proton.  In consistency with our earlier analysis \cite{DFJK4}, we
find that $d_d$ is clearly larger than $d_u$ for medium to large $x$.  The
comparison of the relevant parameters $A_u$ and $A_d$ in the profile
function (see \sect{sec:var-fits}) shows that this trend is seen in all
our GPD fits.  Due to the term with $\alpha'_q\ms \log(1/x)$ in the
profile function, the distance between the struck quark and the cluster of
spectators growths logarithmically when $x$ goes to zero.

In analogy to \eqref{def-dq} we introduce the average size
\begin{equation}
  \label{def-sq}
s_q(x) =\frac{\langle b^y\rangle_x^q}{1-x} =
        \frac{1}{2m}\,\frac{e_v^q(x)}{q_v(x)}\,\frac{1}{1-x}
\end{equation}
of the shift in the distance between the struck quark and the spectator
system that is induced by transverse proton polarization.  As is seen in
\fig{fig:distance}, this shift is significantly nonzero and larger in size
for $d$ quarks than for $u$ quarks.  For $x\to 1$ the numerical values we
find for $s_q$ are very uncertain, which is due to the corresponding
uncertainties in the parton densities we discussed in \sect{sec:pdfs}.  We
therefore limit the plot of $s_q$ to the region $x < 0.9$.

Observing that the r.h.s.\ of the positivity bound \eqref{bound-final}
assumes its maximum for $g_q(x) /f_q(x) = 3/4$ and omitting $\Delta
q_v(x)$, we obtain the bound
\begin{align}
  \label{distance-bound}
|s_q(x)| < 0.38\, d_q(x) \,.
\end{align}
We note that the weaker bound $|s_q(x)| \le 0.5\, d_q(x)$ can be derived
directly from the general form \eqref{bound-bspace}, independently of our
exponential ansatz \eqref{HE-ansatz} for the $t$ dependence of the GPDs,
see \cite{burkardt03}.  For our default fit we find that in the interval
$0.15 < x < 0.9$ the ratio $|s_q(x)| / d_q(x)$ is below 0.16 for $u$
quarks and below 0.27 for $d$ quarks, so that the bound
\eqref{distance-bound} on the displacement of quarks due to transverse
proton polarization is far from saturated in the $x$ region where it
should be applicable.
  
\begin{figure}[t]
\begin{center}
\includegraphics[width=0.48\textwidth,%
viewport=150 320 600 675]{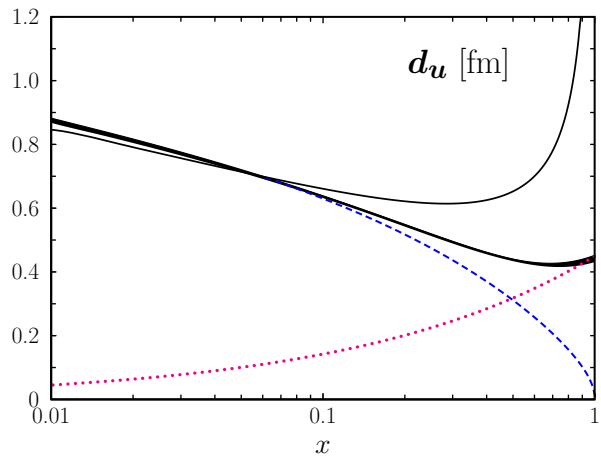}
\end{center}
\caption{\label{fig:distance-decomp} Comparison of the distance $d_u$
  evaluated with the Regge-type profile function
  \protect\eqref{eq:regge-profile} (thin solid line) and with $f_q$ in our
  default fit (thick solid line).  We also show the separate contributions
  from the Regge terms (dashed) and the large-$x$ term (dotted) in $f_q$,
  as specified in the text.  The Regge-type profile function is evaluated
  with the parameters $\alpha'_{R}=0.9 \gev^{-2}$ and $B_R^{} = 0$ taken
  from \protect\cite{GK3}.}
\end{figure}

In the phenomenology of deeply virtual Compton scattering (DVCS) and of
hard exclusive meson production, a Regge-type approximation of the profile
function $f_q$, namely
\begin{equation}
f_{R}(x) = \alpha'_{R} \log(1/x) + B_{R}^{}
\label{eq:regge-profile}
\end{equation}
is frequently used.  In \fig{fig:distance-decomp} we compare the distance
$d_u$ obtained with this approximation and its counterpart calculated from
the profile function \eqref{profile-ansatz} in our default fit.  We also
show the separate contributions of the Regge terms $\bigl[ \alpha_u'\ms
\log(1/x) + B_u \bigr] (1-x)^3$ and of the large-$x$ term $A_u\ms x
(1-x)^2$ in $f_u(x)$.  The Regge-type approximation $f_R$ agrees well with
$f_u$ at small $x$, which is expected from their definitions (small
deviations are due to the slightly different parameters in $f_R$ and our
fit).  For growing $x$ the Regge-type approximation increasingly deviates
from the distance evaluated with our profile function $f_u$, and in the
limit $x\to 1$ it increases as $1/(1-x)$, which is clearly an unphysical
behavior.  The form \eqref{eq:regge-profile} is thus suitable
at small $x$ but should not be used in the large-$x$ region.

%% file: applications.tex
\section{Applications}
\label{sec:applications}

\subsection{Axial form factor}
\label{sec:axialFF}

The axial form factor of the nucleon, $F_A(t)$, is related by the sum rule
\eqref{eq:FA} to the quark helicity dependent GPDs.
Data on the axial form factor are scarce (mainly limited to $-t < 1
\gev^2$) and show a considerable spread, see for instance the review
\cite{meissner}.  The measurement \cite{kitagaki}, which covers the
largest $t$-range, namely $0.1 \gev^2 < -t < 3 \gev^2$, is presented in
form of a dipole parameterization $F_A(t) = F_A(0) / (1-t/M_A^2)^2$ with
parameters $F_A(0)=1.23 \pm 0.01$ and $M_A = \bigl( 1.05^{+0.12}_{-0.16}
\bigr) \gev$.

In view of this situation we do not attempt to fit the GPDs for polarized
quarks to the data on the axial form factor.  Instead, we use the ansatz
\eqref{eq:htilde} for $\widetilde{H}_v^q(x,t)$ with the profile functions
$f_u(x)$ and $f_d(x)$ fixed by our fit of the unpolarized GPDs.  The
polarized parton distributions $\Delta q_v$ for valence quarks are taken
from the analysis of DSSV \cite{DSSV}, as we already did when evaluating
the positivity constraint \eqref{bound-final}.

\begin{figure}
\begin{center}
\includegraphics[width=0.48\textwidth,%
viewport=125 320 590 670]{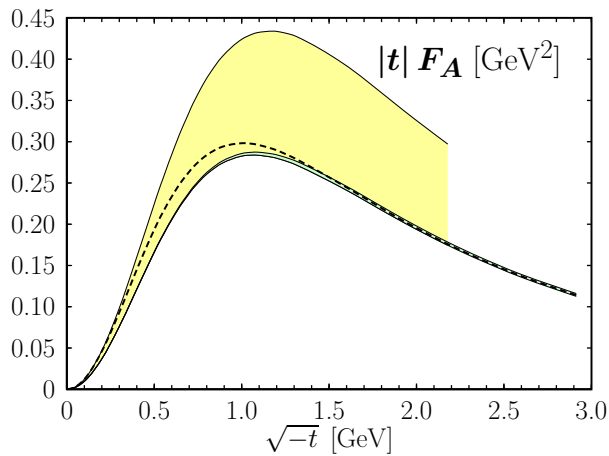}
\end{center}
\caption{\label{fig:FA} The axial form factor $F_A$ of the nucleon. The
  data \cite{kitagaki} are represented by their dipole parameterization as
  a yellow band. The GPD result is evaluated with the default fit for
  $f_u(x)$ and $f_d(x)$ and the polarized PDFs from \cite{DSSV}. The
  dashed line represents our result for $F_A$ including an estimate of the
  sea quarks, as explained in the text.}
\end{figure}

With this ansatz for $\widetilde{H}_v^q$ the axial form factor is
evaluated from the sum rule \eqref{eq:FA}, neglecting the antiquark
contribution for the time being.  The results are compared to the dipole
fit of \cite{kitagaki} in \fig{fig:FA}.  One sees that our simple model of
${\widetilde H}^q_v$ is compatible with experiment, although it is at the
lower edge of the large uncertainty band of the dipole fit.

In order to get a feeling for the possible size of the antiquark
contribution in the sum rule for $F_A$, we use the parameterization
\eqref{eq:htilde} also for the polarized sea quark distribution
$\widetilde{H}^{\bar{q}}$.  Admittedly, this is an ad hoc ansatz and
should not be over-interpreted.  Since the isotriplet combination
$\Delta\bar{u}(x)-\Delta\bar{d}(x)$ of forward densities is not very small
in the DSSV analysis, the effect of the sea quarks on the axial form
factor is noticeable for $-t$ below $1 \gev^2$.  The agreement with the
dipole fit improves within this estimate, as is seen in \fig{fig:FA}.

A more general ansatz than \eqref{eq:htilde} is the exponential form
$\widetilde{H}^q_v(x,t) = \Delta q_v(x)\, \exp\bigl[ t \ms \tilde{f}_q(x)
\bigr]$ with a profile function of its own.  The density interpretation of
the impact parameter distributions $q_v(x,\vbs^2) \pm \Delta
q_v(x,\vbs^2)$ for valence quarks with definite helicity implies the bound
$\tilde{f}_q(x) \le f_q(x)$ in the region where antiquarks can be
neglected.  Taking $\tilde{f}_q(x) < f_q(x)$ instead of $\tilde{f}_q(x) =
f_q(x)$ would increase the integral giving the axial form factor and could
thus improve the agreement with the dipole fit, especially at higher $t$.
We will, however, not pursue this possibility in the present work.

%%%%%%%%%%%%%%%%%%%%%%%%%%%%%%%%%%

\subsection{Compton form factors}
\label{sec:compton-scatt}

\begin{figure}
\begin{center}
\includegraphics[width=0.50\textwidth,%
viewport=100 310 560 660]{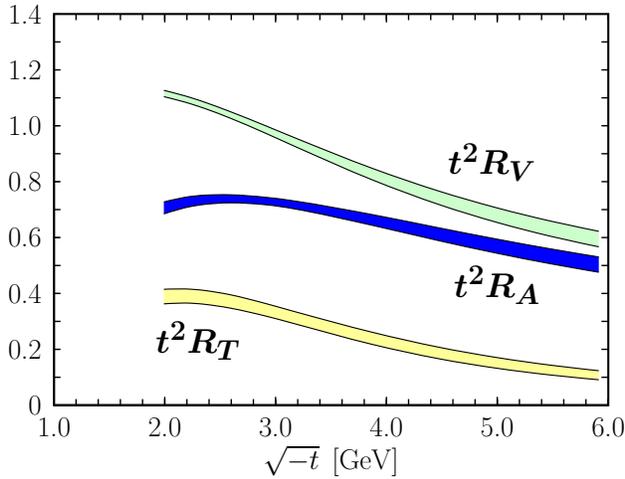}
\end{center}
\caption{\label{fig:Compton} The Compton form factors evaluated from our
  default fit at the scale $\mu=2 \gev$.  They are scaled by $t^2$ and
  shown in units of $\gev^4$.}
\end{figure}

As argued in \cite{rad98,DFJK1}, the amplitude for wide-angle Compton
scattering (i.e.\ Compton scattering at large values of the Mandelstam
variables $s$, $t$ and $u$) factorizes into a hard subprocesses $\gamma
q\to \gamma q$ and form factors given by the $1/x$ moments of GPDs:
\begin{align}
R_V
& = \sum_q e_q^2\, \int_0^1 \frac{\dd x}{x}\,
         \bigl[ H^q_v(x,t)+2 H^{\bar q}(x,t) \bigr] \,,
\nonumber\\
R_T
& = \sum_q e_q^2\, \int_0^1\frac{\dd x}{x}\,
         \bigl[ E^q_v(x,t)+2 E^{\bar q}(x,t) \bigr] \,,
\nonumber\\
R_A
& = \sum_q e_q^2\, \int_0^1\frac{\dd x}{x}\,
   \bigl[ \widetilde{H}^q_v(x,t)+ 2 \widetilde{H}^{\bar{q}}(x,t) \bigr] \,.
\end{align}
This factorization, which bears some similarity to the handbag
factorization of DVCS, is formulated in a symmetric frame where the
skewness $\xi$ is zero.  The form factors are mildly scale dependent as
discussed in \cite{DFJK4}.  We evaluate these Compton from factors from
our default fit of the valence-quark GPDs at the scale $\mu=2 \gev$,
taking again the ansatz \eqref{eq:htilde} for $\widetilde{H}_v^q$.  The
results and their parametric uncertainties are shown in \fig{fig:Compton}
for $-t > 2 \gev^2$.  They are rather similar to those obtained in our
previous work \cite{DFJK4}, except that $R_T$ in the present analysis is
somewhat larger at small $t$ and falls off slightly faster as $-t$
increases.  The sea quark contribution to these form factors can be safely
neglected since we are only interested in the large $t$ region (c.f.\ our
estimate of the sea quark contribution to the axial form factors in
\sect{sec:axialFF}).

\begin{figure}[t]
\begin{center}
\includegraphics[width=0.48\textwidth,%
viewport=175 235 570 665]{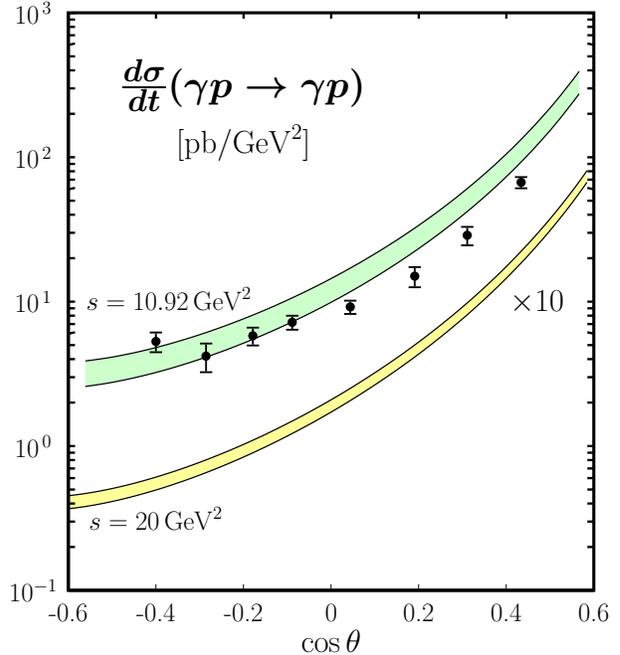}
\end{center}
\caption{\label{fig:ComptonCS} The Compton cross section at $s=10.92$ and
  $20 \gev^2$, evaluated from the Compton form factors shown in
  \protect\fig{fig:Compton}. The data points for $s=10.92 \gev^2$ are from
  \protect\cite{hall-a}.}
\end{figure}

In the handbag approach, the unpolarized cross section for wide-angle
Compton scattering reads \cite{DFJK1}
\begin{multline}
\frac{\dd \sigma}{\dd t} 
= \frac{\pi \alpha_{\text{em}}^2}{s^2}\, \frac{(s-u)^2}{-us} \,
   \biggl[ R_V^2(t) - \frac{t}{4m^2}\, R_T^2(t)
\\
  + \frac{t^2}{(s-u)^2}\, R_A^2(t) \biggr] \,.
\end{multline}
In \fig{fig:ComptonCS} we plot this quantity for $s = 10.92$ and $20
\gev^2$ and compare with the Hall A data \cite{hall-a} from JLab.  The
theoretical results include next-to-leading order QCD corrections
\cite{DFJK4,HKM} and an estimate of the uncertainties due to the finite
proton mass as specified in \cite{Diehl:2002ee}.  The latter uncertainties
are responsible for the error bands, because the parametric errors on the
Compton form factors resulting from our GPD fit are rather small.  Despite
the fact that the agreement with the Hall A data is not perfect, in
particular in the forward hemisphere, we consider our result as a
remarkable success: in a parameter-free calculation the deviations between
experiment and theory are less than about $30\%$.

Another interesting quantity is the correlation parameter $A_{LL}$
($K_{LL}$) between the helicities of the incoming photon and the incoming
(outgoing) proton \cite{HKM,DFJK2}.  As a consequence of the neglect of
quark masses, one has $A_{LL}=K_{LL}$ in the handbag approach.  To a good
approximation, the correlation parameters are given by the ratio of $R_A /
R_V$ times a known factor.  There is a measurement of $K_{LL}$ for $s=6.9
\gev^2$ and $-t = 4 \gev^2$ \cite{kll}, but we deem the corresponding
value $-u = 1.1 \gev^2$ to be too low for applying the handbag approach.
The situation should improve with future measurements of Compton
scattering at higher energies.

%%%%%%%%%%%%%%%%%%%%%%%%%

\subsection{Chromodynamic lensing and the Sivers distribution}
\label{sec:lensing}

The Sivers distribution quantifies the anisotropy in the
transverse-momentum distribution of unpolarized partons inside a proton
that is polarized in the transverse plane.  Its very existence offers deep
insights into the dynamics of partons in QCD: the distribution is naively
time-reversal odd and can only be nonzero due to interactions of the
spectator partons that single out a direction in time
\cite{Brodsky:2002cx}.  These interactions play a role in every type of
parton distribution; technically they are described by Wilson lines if one
works in a covariant gauge \cite{Collins:2002kn}.

We have already seen that in a transversely polarized proton there is a
sideways shift in the spatial distribution of partons, which is described
by the term with $e_v^q(x,\vbs^2)$ in \eqref{lateral-shift}.  The idea of
chromodynamic lensing \cite{Burkardt:2003uw} is that this anisotropy in
transverse \emph{space} induces an anisotropic distribution of the
transverse parton \emph{momentum} via the spectator interactions just
mentioned.  These interactions are of non-perturbative nature, and to date
the lensing effect cannot be computed from first principles in QCD.  It
can however be explored in simple model calculations, where typically the
spectator system is approximated by a diquark and its interactions are
treated as gluon exchange in perturbation theory, see e.g.\
\cite{Meissner:2007rx,Gamberg:2010xi}.  

In the recent study \cite{Bacchetta:2011gx} an ansatz for the lensing
effect was used to compute GPDs from a phenomenological extraction of the
Sivers distributions.  In this work we take the opposite approach and
combine a simple model for chromodynamic lensing with our fit result for
the distribution $e_v^q(x,\vbs)$, thus obtaining an estimate for the
Sivers distribution of valence $u$ and $d$ quarks.
A useful quantity for this purpose is the first $k_T^2$ moment of the
Sivers function, defined as
\begin{equation}
f_{1T}^{\perp (1),\, q-\bar{q}}(x) 
= \int \dd^2\tvec{k}\,
   \frac{\tvec{k}^2}{2 m^2}\,
   f_{1T}^{\perp,\, q-\bar{q}}(x,\tvec{k}{}^2) \,,
\end{equation}
where we abbreviate
\begin{equation}
f_{1T}^{\perp,\, q-\bar{q}}(x,\tvec{k}{}^2)
 = f_{1T}^{\perp,\, q}(x,\tvec{k}{}^2)
 - f_{1T}^{\perp,\, \bar{q}}(x,\tvec{k}{}^2)
\end{equation}
and use the standard definition of the Sivers distribution
\cite{Bacchetta:2004jz}.  We take a model that involves a scalar diquark
and perturbative one-gluon exchange \cite{Meissner:2007rx}, which gives
\begin{align}
  \label{lensing-model}
& f_{1T}^{\perp (1),\, q-\bar{q}}(x)
\nonumber \\
&\quad =
- \pi \alpha_s\ms C_F\, \frac{1}{2 m^2}
  \int \frac{\dd^2 \tvec{l}}{(2\pi)^2}\, 
  \frac{1}{\tvec{l}{}^2 + m_g^2}\,
\nonumber \\
&\qquad \times \frac{\tvec{l}^2}{1-x}\,
  E_v^q\biggl( x, - \frac{\tvec{l}^2}{(1-x)^2} \biggr)
\nonumber \\
&\quad =
- \int \dd^2 \tvec{b}\;  \frac{b^j}{2 m^2}\,
  I^j\biggl(\frac{\tvec{b}}{1-x} \biggr)\,
 \frac{\partial e_v^q(x, \vbs^2)}{\partial \vbs^2} \,.
\end{align}
The so-called lensing function thus reads
\begin{align}
I^j(\tvec{b}) &=
  2\pi \alpha_s\ms C_F\, 
  \frac{\partial}{\partial b^j}
  \int \frac{\dd^2 \tvec{l}}{(2\pi)^2}\,
  \frac{e^{-i \tvec{b}\ms \tvec{l}}}{\tvec{l}{}^2 + m_g^2}
\nonumber \\
 &= - \alpha_s\ms C_F\, \frac{b^j}{b^2}\,
      \bigl[ b m_g \ms K_1(b m_g) \bigr] \,,
\end{align}
where $b = \sqrt{\tvec{b}{}^{2}}$.  We note that the expression in squared
brackets tends to $1$ for $m_g\to 0$.  These results follow from
equations~(87), (89) and (90) in \cite{Meissner:2007rx} if we introduce a
mass $m_g$ in the gluon propagator and if we change $e_q\ms e_s\to -
g^2\ms C_F = - 4\pi \alpha_s\ms C_F$, as is appropriate when going from an
Abelian gluon model to color SU(3) with a color singlet target.  We allow
for a gluon mass in order to explore non-perturbative effects at least in
a very simple fashion.  With our exponential ansatz \eqref{HE-ansatz} for
the $t$-dependence of $E_v^q$, we obtain
\begin{multline}
  \label{lensing-result}
f_{1T}^{\perp (1),\, q-\bar{q}}(x)
\\
 = - \alpha_s\ms C_F\, \frac{(1-x)\ms e_v^q(x)}{8 m^2 g_q(x)}\,
     \chi\Biggl[ \frac{m_g^2\ms g_q^{}(x)}{(1-x)^2} \Biggr] \,,
\end{multline}
where the auxiliary function
\begin{equation}
\chi(z) = 1 - z\ms e^z \int_1^\infty \frac{du}{u}\, e^{-u z}
\end{equation}
behaves like $1 - z \log(1/z)$ for $z\to 0$ and like $1/z$ for $z\to
\infty$.  With the definition \eqref{def-dq} of the distance function
$d_q(x)$ this implies
\begin{align}
\bigl| f_{1T}^{\perp (1),\, q-\bar{q}}(x) \bigr|
 \, \le \, \frac{\alpha_s\ms C_F}{2 m^2 \ms d_q^{\ms 2}(x)}\,
           \frac{f_q(x)}{g_q(x)}\, \frac{e_v^q(x)}{1-x}\,.
\end{align}
In our numerical study we set $\alpha_s = 1$, bearing in mind that the
typical scale of the one-gluon exchange is non-perturbative, as can easily
be seen in \eqref{lensing-model}.  Even without a small $\alpha_s$, the
$k_T^2$ moment of the Sivers function is suppressed by the factor $1 \big/
\bigl[ 2 m^2\ms d_q^{\ms 2}(x) \bigr]$, given that the distance $d_q(x)$
is significantly larger than the Compton wavelength $1/m \approx 0.2 \fm$
of the proton (see \fig{fig:distance}).

\begin{figure}[!ht]
\begin{center}
\includegraphics[width=0.48\textwidth,%
viewport=100 315 555 650]{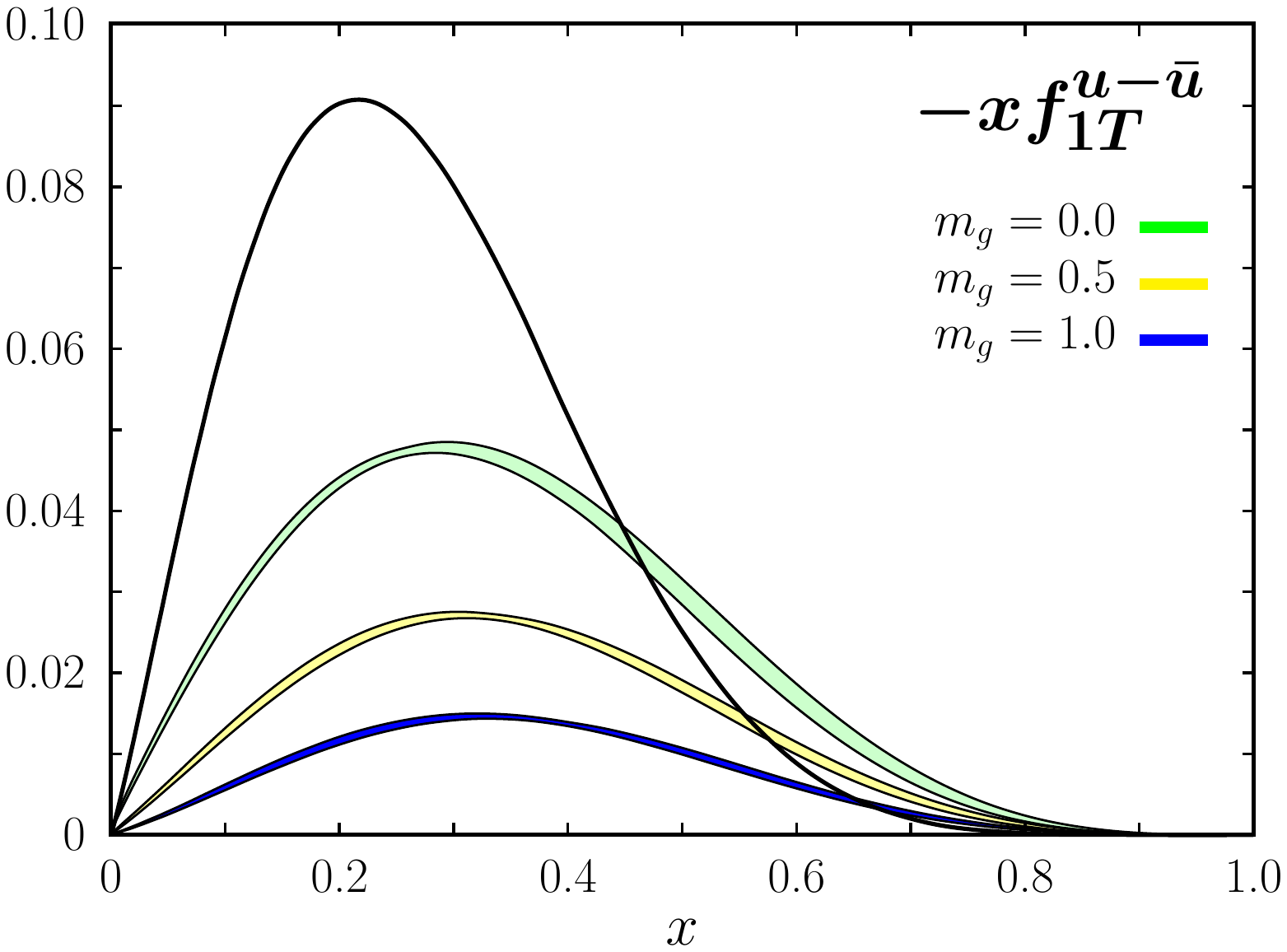} \\[1em]
\includegraphics[width=0.48\textwidth,%
viewport=100 300 555 635]{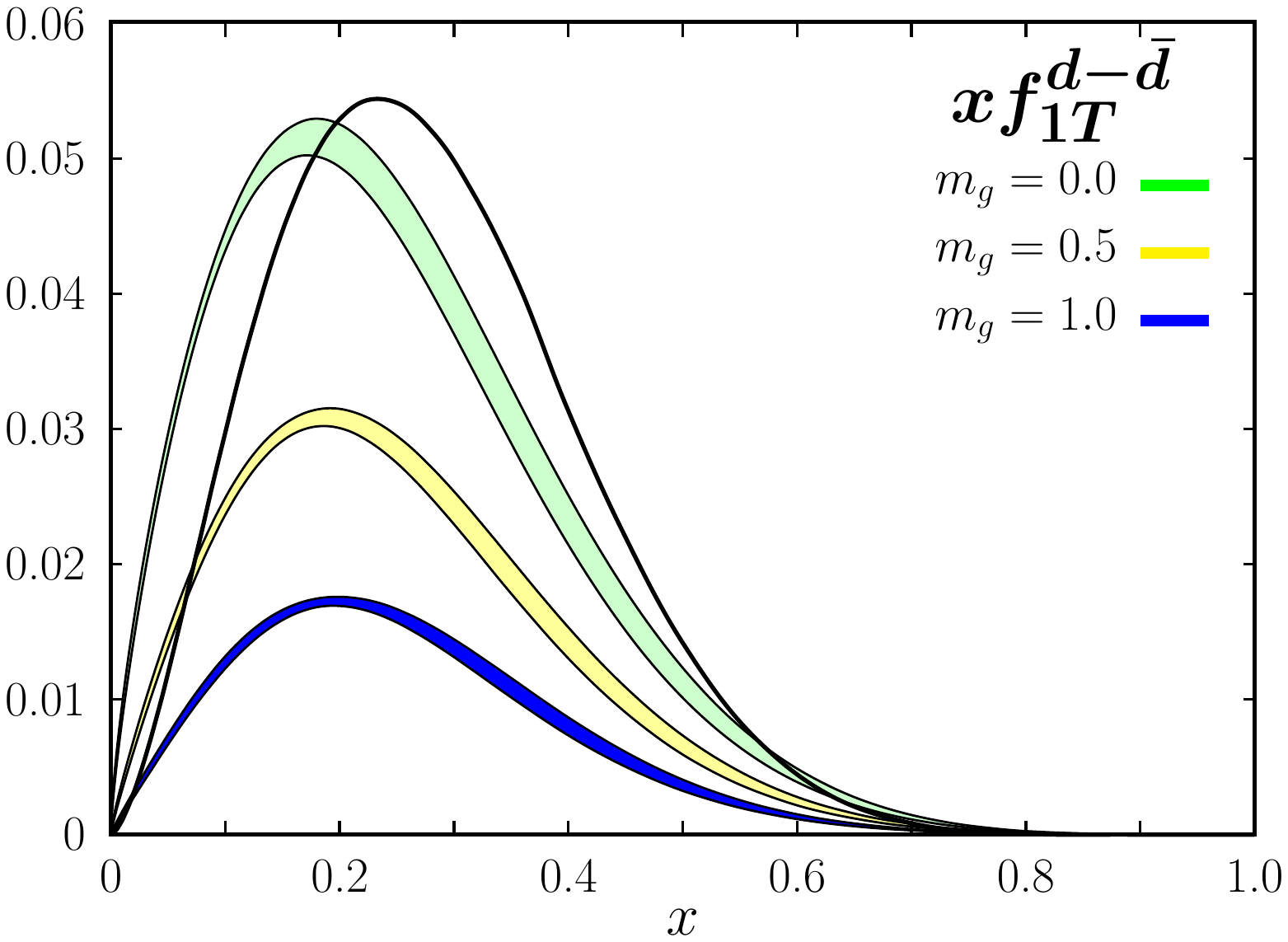}
\end{center}
\caption{\label{fig:sivers} The first $k_T^2$ moment of the Sivers
  function for valence $u$ and $d$ quarks, evaluated at the scale $\mu =
  1\gev$ and multiplied with $x$.  The bands represent the results of the
  model \protect\eqref{lensing-result} together with our fit ABM 0, and
  the solid lines show the recent phenomenological extraction
  \protect\cite{Anselmino:2012aa}.}
\end{figure}

The model leading to \eqref{lensing-result} does not include evolution
effects, neither for the Sivers function nor for $e_v^q$ (the two types of
function actually evolve quite differently).  We evaluate
\eqref{lensing-result} at scale $\mu = 1 \gev$ rather than at $\mu = 2
\gev$, so as to be closer to the non-perturbative region that is natural
for the model.  We hence use the distribution $e_v^q$ obtained in the fit
ABM 0 (see \tab{tab:fit-basics}) rather than the one from our default fit.
The result is shown in \fig{fig:sivers} for three representative values of
the gluon mass $m_g$.  It is compared with the recent extraction
\cite{Anselmino:2012aa} of the Sivers distribution from semi-inclusive
deep inelastic scattering data.  We observe overall agreement in sign and
order of magnitude between our model results and the phenomenological
extraction, but there are clear discrepancies as well, most notably for
the relative size of $u$ and $d$ quark distributions and for the detailed
shape in $x$.  As a word of caution we note that the extraction in
\cite{Anselmino:2012aa} (as well as any other current extraction of the
Sivers function) is subject to important theoretical uncertainties.  Our
overall assessment is however that the simple model for chromodynamic
lensing we have used cannot be expected to yield precise predictions.

%%%%%%%%%%%%%%%%%%%%%%%%%

\subsection{Skewness dependence}
\label{sec:skewness}

The sum rules relating GPDs to electromagnetic form factors do not contain
information about the dependence of the GPDs on the skewness parameter
$\xi$.  We have therefore used these sum rules at $\xi=0$, see
\eqref{eq:SumRule} and \eqref{valence-GPDs}, because at this point GPDs
admit a density interpretation after Fourier transform to impact parameter
space.  Furthermore, the $t=0$ limit of $H_v^q$ is then given by the usual
PDFs.  In this section we explore the GPDs that are obtained from our
default fit together with a model ansatz for the $\xi$ dependence.

As is well known \cite{mueller94,rad99}, GPDs admit an integral
representation in terms of double distributions
$k(\rho,\eta,t)$.\footnote{%
  It is common to write the arguments of the double distribution as
  $k(\beta,\alpha,t)$.  We changed notation here because $\alpha$ and
  $\beta$ are already used otherwise.}
For valence quarks one can write\footnote{%
  In general, the integration extends over $-1 < \rho < 1$ and $|\rho|-1 <
  \eta < 1-|\rho|$.  Its restriction to $\rho > 0$ for valence quarks has
  been proposed in \cite{Belitsky:2001ns}.}
\begin{multline}
K_v^q(x,\xi,t) 
  = \int_0^1 \dd\rho \int_{\rho-1}^{1-\rho} \dd\eta\, 
\\[0.2em]
\times
  \delta(\rho+\xi\eta-x)\, k_v^q(\rho,\eta,t) \,,
\label{eq:double-distribution}
\end{multline}
where $K_v^q = H_v^q, E_v^q$ as before.  A useful property of this
representation is that, without any restriction on the double
distributions, it ensures the polynomiality property of the resulting
GPDs, which is required by Lorentz covariance, see e.g.\
\cite{Diehl:2003ny,Belitsky:2005qn}.  From \eqref{eq:double-distribution}
it also follows that $K_v^q(x,\xi,t)=0$ for $x\leq -\xi$.  An often used
model for the double distribution, suggested long ago by Radyushkin and
Musatov\cite{rad-mus}, is to assume
\begin{equation}
k_v^q(\rho,\eta,t) = K_v^q(\rho,0,t)\, w(\rho,\eta)
\label{eq:ansatz}
\end{equation}
with a weight function that generates the $\xi$-dependence of the GPDs and
is normalized as
\begin{equation}
\int_{\rho -1}^{1-\rho} \dd\eta\, w(\rho,\eta) = 1 \,.
\end{equation}
The form
\begin{align}
  \label{weight-ansatz}
w(\rho,\eta) & =  \frac{3}{4}\,
   \frac{(1-\rho)^2-\eta^2}{(1-\rho)^{3}}
\end{align}
has been used in many phenomenologically analyses of DVCS and exclusive
meson production, see e.g.\ \cite{Goeke:2001tz,GK3,schaefer,GK4}.

With the help of the $\delta$ function we can perform the integral over
$\eta$ in \eqref{eq:double-distribution}.  Inserting the ansatz specified
by \eqref{eq:ansatz} and \eqref{weight-ansatz} we obtain
\begin{multline}
K_v^q(x,\xi,t) = \frac{3}{4 \xi^3}\,
   \int_{\rho_{\text{min}}}^{\rho_{\text{max}}}
   \frac{\dd\rho}{1 - \rho}\; K_v^q(\rho,0,t)
\\[0.2em]
\times \biggl( 1 + \xi - \frac{1-x}{1 - \rho} \biggr)
         \biggl( \frac{1-x}{1 - \rho} - 1 + \xi \biggr)
  \label{eq:dd2}
\end{multline}
with integration boundaries
\begin{align}
\rho_{\text{min}} &=
\begin{cases}
\, 0                              & \text{for}~ x \le \xi \,,
\\[0.2em]
\, x - \dfrac{\xi}{1-\xi}\, (1-x) & \text{for}~ x  >  \xi \,,
\end{cases}
\nonumber \\[0.2em]
\rho_{\text{max}} &= x + \dfrac{\xi}{1+\xi}\, (1-x)  \,.
\end{align}
The expression in the second line of \eqref{eq:dd2} is zero at the
integration boundaries and has its maximum at $\rho = x$.

\begin{figure}[t]
\begin{center}
\includegraphics[width=0.46\textwidth,%
viewport=135 200 470 660]{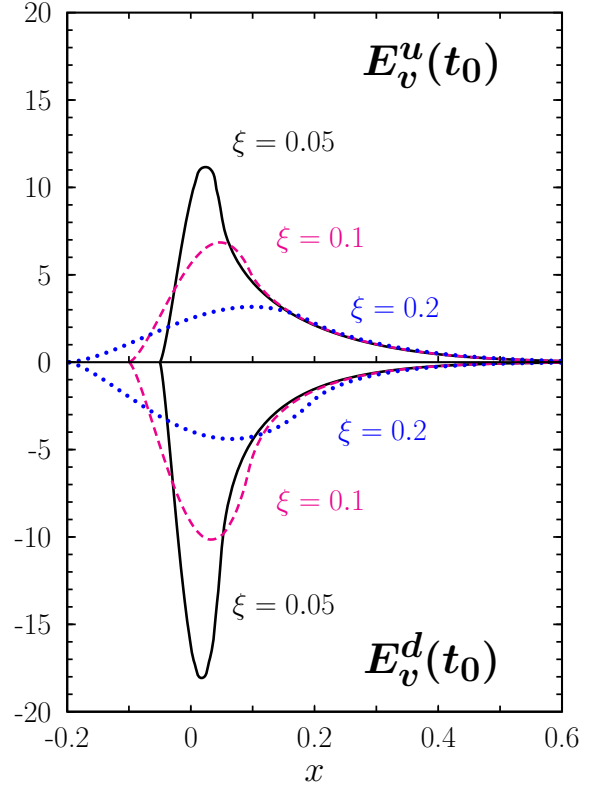}
\end{center}
\caption{\label{fig:GPDs-skewness} The valence GPD $E_v^q$ at the scale
  $\mu=2 \gev$ for selected values of $\xi$.  We set $-t$ to its minimum
  value $-t_0 = 4 \xi^2 m^2 /(1-\xi^2)$ that can be achieved in a physical
  process at given $\xi$.}
\end{figure}

In \fig{fig:GPDs-skewness} we show $E_v^q$ obtained in the model just
described at various values of the skewness $\xi$.  The GPD $H_v^q$ looks
similar in shape.  We see in the figure that the valence-quark GPDs
exhibit a pronounced maximum (or minimum) at a value of $x$ between 0 and
$\xi$.  Note that in this region the integral in \eqref{eq:dd2} extends
down to $\rho=0$, where $K_v^q(\rho,0,t)$ is singular for small enough $t$
according to our discussion in \sect{sec:xmin-xmax}.  With this behavior
of the GPDs, the dominant contribution to the convolution integrals in the
amplitudes of DVCS and exclusive meson production is accumulated in the
vicinity of $x \approx \xi$.

Let us now discuss the behavior of the GPDs in the limit $x\to 1$ at fixed
$\xi$ and $t$.  For small enough $(1-x)$ the integration in \eqref{eq:dd2}
extends over a narrow interval around $\rho=x$, so that we need to know
the behavior of the zero-skewness GPDs for $x\to 1$ at fixed $t$.  With
our ansatz in \sect{sec:fit-ansatz} we have $K_v^q(\rho,0,t) \approx
K_v^q(\rho,0,0) \approx c\, (1-\rho)^\beta$ in that limit.  Inserting this
approximation in \eqref{eq:dd2}, we can easily perform the $\rho$ integral
and obtain
\begin{align}
  \label{eq:GPD-limit}
& K^q_v(x,\xi,t) \approx c\, (1-x)^\beta\;
      \frac{3}{2\beta(\beta-1)(\beta-2)}
\nonumber \\[0.3em]
& \quad \times \frac{1}{\xi^3}\,
  \biggl[ \frac{1 + \xi(\beta-1)}{(1+\xi)^{\beta-1}}
        - \frac{1 - \xi(\beta-1)}{(1-\xi)^{\beta-1}} \biggr]  \,.
\end{align}
For $\xi\ll 1$ this simplifies to
\begin{multline}
K^q_v(x,\xi,t) \approx c\, (1-x)^\beta \; 
\\[0.1em]
\times \Bigl[ \ms 1 + \tfrac{1}{10}\ms (\beta+1) (\beta+2)\, \xi^2
       + \mathcal{O}(\xi^4) \ms \Bigr] \,,
\end{multline}
i.e.\ the GPD at finite $\xi$ is larger in absolute value than in the
forward limit, which it approaches smoothly.  We see that in the limit
$x\to 1$ our ansatz yields a $t$-independent GPD for all values of $\xi$,
which (as in the case of zero skewness) is necessary for having a finite
average size of the proton.  Note that with the Regge-type approximation
\eqref{eq:regge-profile} for the profile function, the GPD at nonzero
skewness contains a factor $\exp[t B_R]$ and thus remains $t$-dependent
for $x\to 1$.

In the limit $x\to 1$ at fixed $\xi$ and $t$ there is a prediction that
the dominant graphs describing the GPDs are such that, starting from a
three-quark configuration of the proton, all longitudinal momentum is
transferred to a single quark by perturbative gluon exchange
\cite{yuan03}.  This generalizes the corresponding statement for the $x\to
1$ limit of the ordinary PDFs \cite{Mueller:1981sg}.  The analysis in
\cite{yuan03} predicts a power behavior
\begin{align}
  \label{large-x-perturb}
H^q_v(x,\xi,t) &\approx (1-x)^3\, a_v^q(\xi)
\nonumber \\
E^q_v(x,\xi,t) &\approx (1-x)^5\, b_v^q(\xi)
\end{align}
with a dependence on $\xi$ but no dependence on~$t$.  The
$t$-independence in this prediction is consistent with the finite-size
requirement discussed above.  As we see in \eqref{eq:GPD-limit}, the
behavior \eqref{large-x-perturb} follows if $q_v(x) \sim (1-x)^{3}$ and
$e_v^q(x) \sim (1-x)^5$ for $x\to 1$.  For the valence PDF $q_v(x)$ this
is just the behavior predicted by dimensional counting.

The prediction in \eqref{large-x-perturb} must be taken with due caution,
since among other things it does not take into account the change of the
large-$x$ behavior induced by DGLAP evolution; for a more detailed
discussion we refer to section~3.5 in \cite{DFJK4}.  We saw in
\sect{sec:pdfs} that the phenomenologically extracted valence PDFs
approximately follow a power law in $(1-x)$ at large $x$.  As discussed
there, we cannot determine the corresponding power for the mathematical
limit $x\to 1$, but we can instead fit such a power for small but finite
$(1-x)$.  The resulting powers for $u$ quarks are typically between $3$
and $4$ at the scale $\mu = 2 \gev$ and thus not very far from the above
prediction.  By contrast, the effective large-$x$ power for $d$ quarks is
significantly larger (except for the CT parton densities).  As for
$e_v^q(x)$, we find from \tab{tab:fit-basics} that the large-$x$ powers
obtained in our GPD fits range from $4$ to $6$.  All in all, we do not
find strong evidence for the quantitative validity of the prediction
\eqref{large-x-perturb}, at least not at $\mu = 2 \gev$.

%% file: summary.tex
\section{Summary}
\label{sec:sum}

We have performed a detailed review of the data on electromagnetic nucleon
form factors and---with an ansatz for the functional dependence---used
them to determine the GPDs $H_v^q$ and $E_v^q$ for unpolarized valence
quarks.

We find that the quality of the experimental data is clearly important for
the quantitative extraction of GPDs.  A resolution of the inconsistencies
between several data sets on the ratio $R^p = \mu_p G_E^p / G_M^p$ at low
$t$ (all dating from after 2000) is of highest urgency in this respect.
We also point out inconsistencies in the data on $G_M^n$ between
\cite{Anklin:1998ae,Kubon:2001rj} and more recent measurements at
Jefferson Lab.  We decided to discard the data points in
\cite{Anklin:1998ae,Kubon:2001rj} that are in conflict with the more
recent results and caution against the use of parameterizations that fit
the older data.  Finally, our fits show that the precision of $G_M^p$, the
best known among all form factors, is of great importance for the
determination of GPDs.  Quantitative control over two-photon exchange
contributions is crucial in this respect, and measurements of the ratio of
elastic $e^+p$ and $e^-p$ cross sections
\cite{Arrington:2004hk,JLab:E07-005,Kohl:2011zz} will hopefully bring
further progress to this area.  We find that the inclusion of the very
precisely measured squared neutron charge radius has a clear impact on
fits of the form factors, whereas our global fits of the nucleon form
factors in the full available $t$ range are not well suited to determine
with precision the charge or magnetic radii, i.e.\ the derivatives of form
factors at $t=0$.  A resolution of the discrepancies between the
determinations of the proton charge radius from the Lamb shift in either
electronic or muonic hydrogen would be highly welcome, as it would allow
one to include this radius as an extra constraint in the determination of
the form factors.

The electromagnetic nucleon form factors receive contributions from
strange quarks and antiquarks, which are small but remain poorly known.
Estimating their size, we find that they start to be quantitatively
relevant at the present level of precision, but that their neglect does
not change the qualitative picture.  Experimentally, the strange form
factors can be determined in parity violating electron-nucleon scattering,
but it is not clear to us whether in the foreseeable future this can be
brought to a sufficient level of precision and be extended over a
significant range in $t$.  Perhaps more help will come from determinations
in lattice QCD, once the present systematic uncertainties in those
extractions can be brought under quantitative control.  If one is willing
to make a model ansatz for the GPDs, one can connect the strange Dirac
form factor $F_1^s$ with the difference $s(x) - \bar{s}(x)$ of parton
distributions.  This difference is poorly known at present, but
significant progress in their determination can be expected in forthcoming
years from the production of $W$ and $Z$ bosons at the LHC.

We have used the available data on the nucleon form factors to extract an
interpolated set of data for the Dirac and Pauli form factors in the quark
flavor basis, i.e.\ $F_1^u, F_1^d, F_2^u$ and$ F_2^d$.  The $t$-range over
which all form factors can be determined is currently limited by the
measurement of the electric neutron form factor or of the ratio $R^n =
\mu_n G_E^n / G_M^n$.  Measurements for $-t$ above $3.4 \gev^2$ after the
$12 \gev$ upgrade at Jefferson Lab will be of immediate interest in this
respect.  They might in particular reveal whether any of the isosinglet
combinations $F_2^p + F_2^{n\phantom{p}\!\!\!}$ or $G_E^p +
G_E^{n\phantom{p}}$ has a zero crossing at $-t$ around $4 \gev^2$.  Our
fits strongly suggest that $G_M^p$ is dominated by $F_1^u$ at large $t$,
so that $F_1^u$ may be regarded as reasonably well known up to $-t$ of
order $30 \gev^2$.

The flavor form factors we have extracted exhibit several clear trends.
$F_1^d$ decreases significantly faster with $-t$ than $F_1^u$, a behavior
that is naturally explained by the Feynman mechanism, given that the PDF
for $d$ quarks decreases significantly faster with $x$ than the one for
$u$ quarks.  Interestingly, we do not see the same trend for the Pauli
form factors: the ratio $F_2^d / F_2^u$ remains quite flat in the region
of $-t$ up to $3.4 \gev^2$, where we have information from the data.  It
will be most interesting to see in future data whether this trend
continues for higher $-t$.

Using our interpolated data set, we have investigated several one- and
two-parameter fits to individual form factors and their linear
combinations.  We find that a satisfactory description of all form factors
up to $-t = 3.4 \gev^2$ is possible with a power-law $F(t)/F(0) = (1 -
t/M^2)^{-p}$ in the flavor basis of the Dirac and Pauli form factors.  A
dipole form works only well in a few cases.  Using the product
\eqref{global-power-fit-fct} of two power laws for each flavor form
factor, we obtain an excellent global fit to all form factor data in their
available $t$ range.

The current form factor data, together with a slight extension of the
ansatz for GPDs developed in \cite{DFJK4}, allow for a significant advance
in the determination of GPDs compared with our previous work.  As we
already observed in \cite{DFJK4}, the fit of the GPD is significantly
constrained by the positivity requirements that arise from the relation
between zero-skewness GPDs and parton densities in impact parameter space.
We perform a global fit of all form factor data and find significant
correlations between the different parameters describing the GPDs.
Independent information about the distributions $E_v^u$ and $E_v^d$, whose
forward limits are unknown, would be most helpful.  In the future this
might be provided by the determination of higher moments of GPDs in
lattice QCD, and by measurements of exclusive processes (which are,
however, described by GPDs at finite skewness and thus add a further
kinematic dependence that needs to be modeled successfully).  We note that
the electromagnetic form factors provide indirect constraints on GPDs at
high values of $t$, which will conceivably never be accessible in hard
exclusive scattering processes.

We performed fits of the GPDs using several PDF sets for the forward limit
of $H_v^q$ and with several settings regarding parameters and the data
selection.  The spread of the GPDs due to these variations is more
important than the parametric errors of the individual fits.  An improved
knowledge of the valence PDFs, especially at small and at large $x$, will
be of great use and can be expected from forthcoming measurements at the
LHC or at Jefferson Lab.

Despite the present uncertainties, certain features are common to all our
fits and can hence be regarded as firm results under the hypothesis that
our basic ansatz for the functional form of the GPDs is adequate.  In
particular, we find that $E_v^q$ decreases faster with $x$ than $H_v^q$,
although not very much for $d$ quarks.  The data on $G_E^n$ and the
associated squared radius $r_{nE}^2$ favor a small splitting between the
effective Regge slopes $\alpha'_q$ for $u$ and $d$ quarks, which we take
to be equal in $H_v^q$ and $E_v^q$.  We also find that the effective Regge
intercept $\alpha_q$ is larger in $E_v^q$ than in $H_v^q$ in all our fits.
By contrast, we cannot conclude whether the data prefers different
parameters $\alpha_q$ in $E_v^q$ for $u$ and $d$ quarks.

From the $t$-dependence of the fitted GPDs we can compute the average
squared impact parameter of valence quarks at given $x$ and find it to be
clearly larger for $d$ quarks than for $u$ quarks when $x$ is above $0.3$.
The sideways shift $s_q(x)$ in the average impact parameter induced by
transverse proton polarization (which is proportional to $E_v^q$ at $t=0$)
is found to be $0.1 \fm$ for $u$ quarks and $-0.17 \fm$ for $d$ quarks at
$x=0.1$ .  It decreases in absolute size as $x$ increases.

Evaluating Ji's sum rule with out fitted GPDs, we can extract the total
angular momentum carried by quarks of a given flavor minus the
corresponding contribution from antiquarks.  We find
\begin{align}
J_v^u &=  0.230^{+ 0.009}_{- 0.024} \,,
&
J_v^d &= -0.004^{+ 0.010}_{- 0.016}
\end{align}
at a scale $\mu = 2 \gev$, where the uncertainties reflect all variations
we allowed in our fits.  The total angular momentum carried by sea quarks
cannot be determined with information from the electromagnetic form
factors.  We note that from current PDF determinations one finds that the
non-strange sea carries about $15\%$ of the longitudinal proton momentum
at $\mu = 2 \gev$.

We have used our fitted valence GPDs for unpolarized quarks to study a
variety of related quantities.  Assuming a simple model for the GPDs of
longitudinally polarized quarks, where their $t$-dependence is taken to be
equal to the unpolarized case, we can evaluate the isotriplet axial form
factor $F_A(t)$ of the nucleon and find it to be at the lower edge of the
uncertainty band from the experimental extraction of this quantity.  A
slower decrease with $t$ of the polarized GPDs, as well as the
contribution of polarized sea quarks, could increase $F_A(t)$ and improve
the agreement with the data, which themselves remain rather imprecise and
limited in $t$.  Also for this quantity, advances in lattice computations
will be of great interest.

Evaluating the form factors that describe wide-angle Compton scattering in
the handbag approach, we obtain agreement with the data at the 30\% level
for $s \approx 11 \gev^2$, which we consider to be a success for a
parameter free calculation, given the uncertainties of the handbag
approach itself in this kinematic region.

Combining our GPDs with a simple model for chromodynamic lensing, we can
estimate the Sivers distributions for valence $u$ and $d$ quarks.
Comparing the results with a recent phenomenological determination of
these distributions, we find overall agreement in sign and magnitude but
clear differences in the details.  We think that the model of the lensing
effect, which uses a simple quark-diquark picture and one-gluon exchange,
is the most important source of uncertainty in this comparison.

Finally, we have combined our fitted zero-skewness GPDs with the double
distribution ansatz of Musatov and Radyushkin in order to compute the
valence GPDs at finite skewness.  These results can be used as an input
for phenomenological analyses of hard exclusive scattering processes, in
particular for the kinematics of current and future fixed-target
experiments.

In conclusion, the improvements in the measurement of the electromagnetic
form factors over the last decade has significantly advanced our ability
to perform a model-dependent extraction of the GPDs for unpolarized
valence quarks.  Further significant progress can be expected for the next
decade from the measurement of form factors, parton densities, and hard
exclusive processes and from lattice calculations.  This will hopefully
also enable us to get a firmer grip on the model dependence due to
assuming a functional form of the GPDs.

%% file: appendices.tex
\section{Tables of form factors}
\label{app:tables}

In \tab{tab:Rn-values} we give the values and errors of $R^n$ for the data
set we have selected.  Except for the entries from Plaster 05, Geis 08 and
Riordan 10, which directly quote results on $R^n$, we have computed this
ratio from $G_E^n$ and the assumed value of $G_M^n$ for the reasons
explained in \sect{sec:neutron-data}.

In \tab{tab:flavorFF} we list the results for the flavor form factors we
have extracted from our default data set as explained in
\sect{sec:interpol-procedure}.

\begin{table*}[p]
\begin{center}
\renewcommand{\arraystretch}{1.2}
\begin{tabular}{ccccl}
\hline
$-t \, [\gev^2]$ & $R^n$ & \multicolumn{2}{c}{error} & reference \\
               &       & total & stat. \\
\hline
$0.21$~ & $0.1108$ & $0.0260$ & $0.0252$   & Passchier 99 \\
\hline
$0.495$ & $0.1334$ & $0.0201$ & $0.0177$   & Zhu 01 \\
\hline
$0.5$~~ & $0.1517$ & $0.0121$ & $0.0095$   & Warren 03 \\
$1.0$~~ & $0.2457$ & $0.0350$ & $0.0292$ \\
\hline
$0.15$~ & $0.0706$ & $0.0122$ & $0.0095$   & Herberg 99 \\
$0.34$~ & $0.1485$ & $0.0212$ \\
\hline
$0.30$~ & $0.1140$ & $0.0132$ & $0.0126$   & Glazier 04$^{\,a}$ \\
$0.59$~ & $0.1566$ & $0.0240$ & $0.0230$ \\
$0.79$~ & $0.1992$ & $0.0395$ & $0.0383$ \\
\hline
$0.447$ & $0.1444$ & $0.0175$ & $0.0170$   & Plaster 05$^{\,b}$ \\
$1.132$ & $0.2506$ & $0.0218$ & $0.0210$ \\
$1.450$ & $0.3616$ & $0.0353$ & $0.0344$ \\
\hline
$0.35$~ & $0.1112$ & $0.0083$ &           & Rohe 05 \\
\hline
$0.67$~ & $0.1763$ & $0.0258$ &           & Bermuth 03 \\
\hline
$0.142$ & $0.0505$ & $0.0078$ & $0.0072$  & Geis 08 \\
$0.203$ & $0.0695$ & $0.0093$ & $0.0084$ \\
$0.291$ & $0.1022$ & $0.0135$ & $0.0127$ \\
$0.415$ & $0.1171$ & $0.0189$ & $0.0182$ \\
\hline
$1.72$~ & $0.273$~ & $0.0361$ & $0.020$~   & Riordan 10 \\
$2.48$~ & $0.412$~ & $0.0600$ & $0.048$~ \\
$3.41$~ & $0.496$~ & $0.0813$ & $0.067$~ \\
\hline
\multicolumn{5}{l}{${}^{\,a}$ gives asymmetric errors; we have 
  symmetrized by} \\[-0.3em]
\multicolumn{5}{l}{${}^{~~}$ taking the maximum of the two errors} \\[-0.3em]
\multicolumn{5}{l}{${}^{\,b}$ values from Table IX, ``FSI+MEC+IC+RC''}
\end{tabular}
\end{center}
\caption{\label{tab:Rn-values} Values of $R^n$ measured in polarization
  experiments on deuterium or ${}^3\mathrm{He}$.  References for the data
  sets are given in \tab{tab:Rn-data}.}
\end{table*}

\begin{table*}[p]
\renewcommand{\arraystretch}{1.15}
\begin{center}
\begin{tabular}{ccccc}
\hline
$-t [\gev^2]$ & $F_1^u$ & $F_1^d$ & $F_2^u$ & $F_2^d$ \\
\hline
$0.039$ & $1.810 \pm 0.039$ & $0.908 \pm 0.034$ & $1.471 \pm 0.059$ & $-1.833 \pm 0.056$ \\ 
$0.088$ & $1.610 \pm 0.032$ & $0.815 \pm 0.048$ & $1.251 \pm 0.044$ & $-1.565 \pm 0.071$ \\ 
$0.142$ & $1.414 \pm 0.027$ & $0.681 \pm 0.017$ & $1.104 \pm 0.036$ & $-1.352 \pm 0.043$ \\ 
$0.156$ & $1.392 \pm 0.037$ & $0.705 \pm 0.056$ & $1.044 \pm 0.045$ & $-1.347 \pm 0.070$ \\ 
$0.203$ & $1.241 \pm 0.026$ & $0.588 \pm 0.016$ & $0.972 \pm 0.033$ & $-1.133 \pm 0.038$ \\ 
$0.243$ & $1.148 \pm 0.036$ & $0.560 \pm 0.048$ & $0.857 \pm 0.040$ & $-1.092 \pm 0.052$ \\ 
$0.291$ & $1.023 \pm 0.016$ & $0.475 \pm 0.014$ & $0.798 \pm 0.030$ & $-0.948 \pm 0.049$ \\ 
$0.300$ & $1.020 \pm 0.016$ & $0.478 \pm 0.014$ & $0.780 \pm 0.054$ & $-0.937 \pm 0.103$ \\ 
$0.340$ & $0.979 \pm 0.018$ & $0.468 \pm 0.019$ & $0.696 \pm 0.039$ & $-0.890 \pm 0.068$ \\ 
$0.350$ & $0.924 \pm 0.051$ & $0.416 \pm 0.026$ & $0.719 \pm 0.061$ & $-0.832 \pm 0.065$ \\ 
$0.415$ & $0.862 \pm 0.014$ & $0.375 \pm 0.015$ & $0.626 \pm 0.028$ & $-0.740 \pm 0.046$ \\ 
$0.447$ & $0.831 \pm 0.012$ & $0.368 \pm 0.013$ & $0.577 \pm 0.025$ & $-0.711 \pm 0.041$ \\ 
$0.495$ & $0.759 \pm 0.011$ & $0.320 \pm 0.013$ & $0.523 \pm 0.021$ & $-0.643 \pm 0.036$ \\ 
$0.500$ & $0.758 \pm 0.010$ & $0.327 \pm 0.009$ & $0.513 \pm 0.020$ & $-0.648 \pm 0.035$ \\ 
$0.590$ & $0.668 \pm 0.010$ & $0.272 \pm 0.013$ & $0.445 \pm 0.018$ & $-0.564 \pm 0.033$ \\ 
$0.623$ & $0.640 \pm 0.010$ & $0.259 \pm 0.014$ & $0.413 \pm 0.016$ & $-0.546 \pm 0.028$ \\ 
$0.670$ & $0.612 \pm 0.010$ & $0.243 \pm 0.012$ & $0.378 \pm 0.016$ & $-0.513 \pm 0.026$ \\ 
$0.790$ & $0.524 \pm 0.010$ & $0.198 \pm 0.015$ & $0.326 \pm 0.014$ & $-0.421 \pm 0.023$ \\ 
$1.000$ & $0.438 \pm 0.014$ & $0.160 \pm 0.012$ & $0.244 \pm 0.016$ & $-0.310 \pm 0.018$ \\ 
$1.132$ & $0.372 \pm 0.008$ & $0.124 \pm 0.007$ & $0.223 \pm 0.010$ & $-0.255 \pm 0.014$ \\ 
$1.177$ & $0.372 \pm 0.009$ & $0.132 \pm 0.015$ & $0.191 \pm 0.010$ & $-0.262 \pm 0.018$ \\ 
$1.401$ & $0.279 \pm 0.012$ & $0.080 \pm 0.017$ & $0.170 \pm 0.013$ & $-0.188 \pm 0.018$ \\ 
$1.450$ & $0.279 \pm 0.008$ & $0.090 \pm 0.007$ & $0.155 \pm 0.009$ & $-0.189 \pm 0.010$ \\ 
$1.644$ & $0.247 \pm 0.012$ & $0.071 \pm 0.018$ & $0.125 \pm 0.012$ & $-0.149 \pm 0.018$ \\ 
$1.720$ & $0.231 \pm 0.011$ & $0.058 \pm 0.007$ & $0.118 \pm 0.011$ & $-0.131 \pm 0.008$ \\ 
$2.480$ & $0.139 \pm 0.003$ & $0.028 \pm 0.004$ & $0.062 \pm 0.004$ & $-0.072 \pm 0.005$ \\ 
$3.410$ & $0.087 \pm 0.002$ & $0.013 \pm 0.003$ & $0.035 \pm 0.002$ & $-0.037 \pm 0.003$ \\ 
\hline
\end{tabular}
\end{center}
\caption{\label{tab:flavorFF} The flavor form factors we obtain by
  interpolation of the data, as explained in
  \protect\sect{sec:interpol-procedure}.}
\end{table*}

%%%%%%%%%%%%%%%%%%%%%%%%%%%%%%%%%%%%%%%%%%%%%%%%%%%%%%%

\section{Matrices for computing fit errors}
\label{app:matrices}

In this appendix we give the information that is needed to compute
parametric errors for our default fit ABM 1 and for the power-law fit of
\sect{sec:power-law-fit}.  A convenient proceduce to propagate errors is
the so-called Hessian method used in modern PDF determinations, see e.g.\
\cite{Martin:2009iq,Pumplin:2001ct}.  We briefly describe this method and
then list the relevant matrices.

Let us introduce the column vector $\boldsymbol{p}$ of the $n$ original
fit parameters, as well as the vector of transformed parameters
$\boldsymbol{z}$ defined by
\begin{equation}
\boldsymbol{p} - \boldsymbol{p}_0 = E\ms \boldsymbol{z} \,,
\end{equation}
where $\boldsymbol{p}_0$ is the set of parameters that minimizes $\chi^2$.
The matrix $E$ satisfies
\begin{equation}
E\, E^{T} = V
\end{equation}
with the standard covariance matrix $V$ for the parameters
$\boldsymbol{p}$.
The deviation of $\chi^2$ from its minimum value is then given by
\begin{equation}
\Delta \chi^2 = (\boldsymbol{p} - \boldsymbol{p}_0)^T\, V^{-1}\,
                ( \boldsymbol{p} - \boldsymbol{p}_0)
              = \boldsymbol{z}^T\ms \boldsymbol{z} \,,
\end{equation}
The error on a function $f$ of the parameters, as given by linear error
propagation, can be written as
\begin{align}
  \label{error-result}
\Delta f &= \sqrt{ \sum_{i=1}^n
    \biggl[ \frac{\partial f(\boldsymbol{p})}{\partial z_i}
    \biggr]^2_{\boldsymbol{p} \, = \boldsymbol{p}_0} }
\nonumber \\[0.1em]
 &= \sqrt{ \sum_{i=1}^n \biggl[ \frac{f(\boldsymbol{p}_i^+) 
                      - f(\boldsymbol{p}_i^-)}{2} \biggr]^2 } \,,
\end{align}
where the parameter set $\boldsymbol{p}_i^{\pm}$ is specified by the
condition
\begin{equation}
( \boldsymbol{p}_i^{\pm} - \boldsymbol{p}_0 )_j^{}
 = \pm E_{ji}^{} \,.
\end{equation}
In the second step of \eqref{error-result} we have approximated the
derivative by a difference quotient, which is consistent in the region
where linear error propagation is adequate.  The vector given by the $i$th
column of the matrix $E$ thus gives the amount by which the central values
of the parameters need to be shifted to obtain a set of parameters on the
$\Delta \chi^2 = 1$ contour.

In \tab{tab:errormats} we give the matrix $E_D$ for the default GPD fit
(ABM 1) and the matrix $E_P$ for the power law fit of
\sect{sec:power-law-fit}.  The order of entries in the matrices
corresponds to the following vectors of parameters:
\begin{align}
\boldsymbol{p}_D &=
\begin{pmatrix}
A_u \\ A_d \\ B_u \\ B_d \\
C_u \\ C_d \\ D_u \\ D_d \\
\alpha'_d \\ \alpha
\end{pmatrix} \,,
&
\boldsymbol{p}_P &=
\begin{pmatrix}
a_1^u+b_1^u \\ b_1^u \\ a_1^d+b_1^d \\ b_1^d \\
a_2^u+b_2^u \\ b_2^u \\ a_2^d+b_2^d \\ b_2^d \\
p_1^u \\ p_1^d \\ p_2^u \\ p_2^d \\ q_1^u-p_1^u
\end{pmatrix} \,.
\end{align}
The central values of the fit parameters are given in
\eqref{eq:def-fit-pars} and \tab{tab:def-fit-pars} for the GPD fit, and
those of the power-law fit are given in \tab{tab:global-power-fit-pars}.

\newpage
\onecolumn

\begin{sidewaystable}
{\small
\begin{align*}
E_D &= 10^{-3} \times
\left( \begin{array}{rrrrrrrrrr}
0.060 & 1.201 & -4.092 & -3.574 & 2.970 & 10.026 & 11.518 & -41.918 &
0.907 & -22.550 \\ -0.005 & -0.093 & 0.024 & -1.106 & -0.821 & -0.066 &
3.009 & 17.045 & -159.989 & -165.471 \\ 0.104 & 2.164 & -1.558 & 3.194 &
-1.961 & -1.283 & -25.959 & 51.507 & -8.565 & 21.610 \\ -0.015 & -0.371 &
0.347 & -8.188 & 0.839 & 1.139 & -29.116 & 48.996 & 12.208 & 43.508 \\
0.036 & 0.177 & -2.366 & -1.876 & 1.400 & -9.399 & 51.439 & 39.615 &
-3.476 & 56.682 \\ 0.018 & 0.031 & -0.310 & 0.081 & 4.304 & -2.954 &
-1.180 & 19.597 & 107.532 & -224.070 \\ 0.092 & 0.036 & -2.379 & -1.919 &
-2.562 & -13.660 & -26.668 & -55.362 & -8.389 & -13.809 \\ 0.084 & -0.124
& 0.265 & 1.643 & 13.882 & -2.601 & -14.377 & -3.909 & -45.726 & 56.891 \\
0.433 & 1.096 & 4.812 & -2.455 & -0.164 & -1.406 & 13.764 & -19.246 &
3.122 & -9.610 \\ 0.540 & -1.440 & -2.562 & 1.711 & -1.754 & 3.737 & -4.755 & 18.333 & 3.021 & 2.667
\end{array} \right)
\\[2em]
E_P &= 10^{-3} \times
\left( \begin{array}{rrrrrrrrrrrrr}
-0.05 & -0.15 & -0.69 & 4.00 & 5.69 & 0.32 & -2.88 & 20.44 & -13.86 &
-33.85 & 83.24 & 15.29 & 38.31 \\ -0.23 & -0.32 & -0.52 & 0.31 & -0.38 &
5.46 & -7.55 & -26.34 & 36.87 & -30.86 & 46.64 & 27.83 & 64.55 \\ 0.01 &
0.05 & 0.48 & -3.60 & -5.21 & 3.40 & -3.14 & 23.20 & -14.61 & -32.92 &
83.24 & 16.18 & 39.48 \\ 0.03 & 0.06 & 1.09 & -1.66 & 2.87 & 1.05 & -1.79
& 4.79 & -3.55 & 48.92 & -47.60 & 135.29 & 187.97 \\ -0.04 & 0.48 & -0.36
& -0.24 & 1.39 & 7.68 & 15.91 & 5.88 & 11.62 & -45.60 & -29.22 & 1.08 &
15.42 \\ -0.22 & 1.61 & 2.51 & 2.89 & -1.98 & -1.59 & -2.48 & 7.00 & 17.33
& -12.97 & -13.26 & -13.57 & 15.41 \\ -0.01 & 0.13 & -0.33 & -0.40 & 0.55
& 0.81 & 3.69 & 11.13 & 39.94 & 114.51 & 72.94 & -42.97 & -14.98 \\ -0.05
& 0.57 & -1.15 & -1.11 & 0.73 & -0.41 & -1.59 & -4.10 & -16.28 & 4.80 &
-12.59 & -180.39 & 172.11 \\ -0.63 & -0.87 & 0.11 & -1.02 & 0.40 & -0.33 &
0.83 & 10.82 & -12.92 & 19.88 & -38.70 & -15.18 & -36.65 \\ 0.06 & 0.14 &
2.20 & -3.72 & 6.00 & 2.44 & -5.15 & -1.94 & 4.01 & -18.25 & 6.27 & -66.43
& -96.37 \\ -0.09 & 0.71 & 0.44 & 0.88 & -0.12 & 4.37 & 2.14 & -21.02 &
-51.70 & 56.49 & 54.09 & 27.26 & -46.10 \\ -0.13 & 1.47 & -2.70 & -2.75 &
1.70 & -2.68 & -3.39 & 2.24 & 4.26 & -9.88 & -4.27 & 75.45 & -64.94 \\
-0.11 & -0.09 & 0.97 & -1.88 & 1.08 & -7.89 & 10.23 & -14.50 & 2.68 &
-32.51 & 82.55 & 23.19 & 57.91
\end{array} \right)
\end{align*}
}
\caption{\label{tab:errormats} The matrix $E_D$ for the default GPD fit
  ABM 1 and the matrix $E_P$ for the power-law fit of
  \protect\sect{sec:power-law-fit}.  Entries are given in units of
  $\gev^{-2}$ for rows 1 to 9 in $E_D$ and for rows 1 to 8 in $E_P$.}
\end{sidewaystable}